\documentclass[11pt]{article}
\pdfoutput=1 
\usepackage[utf8]{inputenc}
\usepackage{caption}
\usepackage{braket,slashed,bm}
\usepackage{jheppub}
\usepackage{array,multirow}
\usepackage{amsmath}
\usepackage[normalem]{ulem}
\usepackage{calligra}
\usepackage{floatrow}
\DeclareMathAlphabet{\mathpzc}{OT1}{pzc}{m}{it}
\usepackage{mathrsfs}
\usepackage{tikz}
\usetikzlibrary{arrows,decorations.pathmorphing,backgrounds,positioning,fit,petri,automata,shadows,calendar,mindmap,
decorations.markings,calc,hobby,shapes.misc}

\usepackage{subcaption}
\usepackage{lscape}
\usepackage{hyperref}
\usepackage{graphicx}
\usepackage{booktabs}
\newfloatcommand{capbtabbox}{table}[][\FBwidth]
\usepackage{feynmp-auto}
\usepackage{pgfplots}
   \pgfplotsset{compat=1.5}
   \usepgfplotslibrary{fillbetween}

\graphicspath{{./Figures/}}
\newcommand{\titlemath}{\texorpdfstring}
\def\beq{\begin{equation}}
\def\eeq{\end{equation}}
\def\bea{\begin{eqnarray}}
\def\eea{\end{eqnarray}}
\def\nn{\nonumber \\}
\def\hyp{\mathsf{y}}
\renewcommand{\a}{\alpha}

\renewcommand{\d}{\delta}
\newcommand{\g}{\gamma}

\newcommand{\hst}{s_{\hat\theta}}
\newcommand{\hct}{c_{\hat\theta}}
\newcommand{\hsdt}{s_{2\hat\theta}}
\newcommand{\hcdt}{c_{2\hat\theta}}
\newcommand{\hsw}{s_{\hat \theta}}
\newcommand{\hcw}{c_{\hat \theta}}
\newcommand{\dsw}{\d s_\theta}
\newcommand{\gsm}[2]{g^{#1,SM}_{#2}}
\def\gcb{{\overline g_{1}}}
\def\gcw{{\overline g_{2}}}
\def\gcg{{\overline g_{3}}}

\def\tc{{\overline \theta}}
\def\ckin{C_{H,\text{kin}}}
\newcommand{\Lagr}{\mathcal{L}}
\renewcommand{\to}{\rightarrow}
\newcommand{\MG}{\textsc{MadGraph5}}

\DeclareMathOperator{\re}{Re}
\newcommand{\hc}{\text{h.c.}}

\title{The Higgs width in the SMEFT}

\author[a,b]{Ilaria Brivio,}

\author[a]{Tyler Corbett,}

\author[a]{Michael Trott}

\affiliation[a]{Niels Bohr Institute \& Discovery Center, University of Copenhagen,
Blegdamsvej 17, DK-2100, Copenhagen, Denmark}
\affiliation[b]{Institut f\"{u}r Theoretische Physik, Universit\"{a}t Heidelberg, Philosophenweg 16, DE-69120 Heidelberg, Germany}

\abstract{We calculate the total and partial inclusive Higgs widths at leading order in the Standard Model Effective Field Theory (SMEFT). We report results
incorporating SMEFT corrections for two and four body Higgs decays through vector currents in this limit.
The narrow width approximation is avoided and all phase space integrals are directly evaluated.  We explain why the narrow width approximation fails more significantly in the SMEFT compared
to the SM, despite the narrowness of the observed $\rm SU(2) \times U(1)$ bosons in both theories.
Our results are presented in a manner that allows various input parameter schemes to be used, and
they allow the inclusive branching ratios and decay widths of the Higgs to be numerically determined
without a Monte Carlo generation of phase space for each Wilson coefficient value chosen.}

\begin{document}
\maketitle
\section{Introduction}
In the Standard Model (SM) the width of the Higgs is  small ($\sim 4 \, {\rm MeV}$) compared to the Higgs mass of $m_h \sim 125 \, {\rm GeV}$.
The width is known to high accuracy in terms of the parameters of the SM, and this makes it interesting to study perturbations due to physics beyond the SM on the total and partial widths.
Although difficult to directly measure, the Higgs width is essential to inferring the full set of partial widths from the observed branching ratios
--which match well with SM predictions for the Higgs at the $\sim 10\%$ level.
Precise knowledge of the Higgs width is a key requirement to accurately interpreting experimental results on Higgs decays now and in the future.
This remains true when the SM is extended into the Standard Model Effective Field theory (SMEFT).

The SMEFT is defined
under the assumptions that: physics beyond the SM is present at scales $\Lambda > \sqrt{2 \, \langle H^\dagger H\rangle} = \bar{v}_T$,
no light ($m \ll \bar{v}_T$) hidden states are lurking in the particle spectrum with couplings
to the SM, and a $\rm SU(2)_L$ scalar doublet with hypercharge $\hyp_h = 1/2$ is present in the low energy limit defining the EFT.\footnote{More precisely
the direct meaning of this standard assumption is that the local operators are {\it analytic} functions of the field $H$ in the SMEFT. The
analyticity of the local contact operators making up the SMEFT is a basic feature of this theory.
This basic EFT point was discussed in the recent SMEFT review \cite{Brivio:2017vri}.}
The SMEFT extends the SM with operators $\mathcal{Q}_i^{(\rm{d})}$ of mass dimension $\rm{d}$
\bea\label{smeftdefinition}
	\Lagr_{\textrm{SMEFT}} &=& \Lagr_{\textrm{SM}} + \Lagr^{(5)}+\Lagr^{(6)} +
	\Lagr^{(7)} + \dots,  \\ \nonumber
  \Lagr^{(\rm{d})} &=& \sum_i \frac{C_i^{(\rm{d})}}{\Lambda^{\rm{d}-4}}\mathcal{Q}_i^{(\rm{d})}
	\textrm{ for } \rm{d}>4.
\eea
The operators $Q_i^{(\rm{d})}$ are suppressed by $\rm{d}-4$ powers of the cut-off scale $\Lambda$ and
the $C_i^{(\rm{d})}$ are the Wilson coefficients. In this work we use the non-redundant $\mathcal{L}^{(6)}$ Warsaw basis \cite{Grzadkowski:2010es}.
This basis removed some residual redundancies (see also \cite{AguilarSaavedra:2010zi,Alonso:2014zka}) in the over-complete basis of Ref.~\cite{Buchmuller:1985jz}.
We often use the notation $\tilde{C}= C v_T^2/\Lambda^2$ for dimensionless rescaled Wilson coefficients.
In this work we report the corrections to the two and four body decay of the  Higgs width through vector currents,
i.e the interference effects of $\Lagr^{(6)}$ with the SM prediction of the Higgs Width. We neglect odd dimension
operator effects from $\Lagr^{(5)}$ as this operator violates lepton number and does not interfere in the processes
that we calculate at tree level. This same reasoning applies to neglecting $	\Lagr^{(7)}$ corrections.
We neglect corrections due to $\Lagr^{(8)}$, as including a consistent and complete set of such
corrections is beyond the scope of this work.

A key strength of a SMEFT analysis of experimental data is
that it represents a consistent general low energy (or infrared - IR) limit of physics beyond the SM, so long as its defining assumptions are satisfied, and all operators
at each order in the power counting of the theory are
retained.  This is the approach we adopt in this paper.
A further strength of the SMEFT is that it addresses a key challenge to the program of studying the Higgs precisely to look for deviations in its properties as a sign of physics beyond the SM.
The difficulty of directly measuring the Higgs width experimentally (model independently) in the LHC environment is well known. For some related results see Refs.~\cite{Sirunyan:2019twz,Bredenstein:2006rh,Kauer:2012hd,Caola:2013yja,Campbell:2013una}. This fact is also relevant when
considering successor machines for a future precision Higgs phenomenology program. It is important to stress that the perturbations to the Higgs width
are systematically calculable and of a limited form in the SMEFT, when the assumptions of this theoretical framework are adopted. Due to this, even when the Higgs width is difficult to directly measure,
it is possible to bound it indirectly due to calculating directly its allowed perturbations in the SMEFT.

In this paper, we report a consistent calculation of the width of the Higgs to order $1/\Lambda^2$ for a set of two and four body decays (through vector currents)
in the SMEFT.\footnote{Four body decays where a vector is emitted off the fermion pair produced by the Higgs is considered beyond the scope of this work.
Such corrections are suppressed by small yukawa couplings, and also (generally) kinematically suppressed. These results, as well as a set of
other interference effects that are also omitted here, will be included in a follow up work.}
 Our results are presented in a semi-analytic fashion, with inclusive phase space integrals explicitly evaluated and reported.
 Our results allow the total inclusive width, partial widths and branching ratios to be determined as a function of the Wilson coefficients
without a Monte Carlo generator being run. This allows the Wilson coefficient space of the SMEFT to be sampled efficiently in global studies of the properties of the Higgs, and
combined with other particle physics experimental results. We believe this is of some value going forward in the LHC experimental program.

A key observation feeding into the important impact of the calculation reported here is the relative success of the narrow width approximation in the SM and the SMEFT.
The narrow width approximation in the SM relies on the fact that SM interactions are of limited mass dimension ($d \leq 4$) for its numerical adequacy in predicting many experimental results
This is the case as renormalizability leads to $h \gamma \gamma$ and $h \gamma Z$ effective vertices being one loop effects.
In the SMEFT, the presence of interaction
terms of mass dimension $d > 4$ leads to a more serious breakdown of the narrow width approximation, primarily due to neglected interference effects using this approximation.
This is despite the fact that the
$\rm SU(2) \times U(1)$ gauge bosons remain narrow, with $\Gamma/M \ll 1$.
In this work we incorporate off-shell effects neglected in the narrow width approximation, and a consistent set of interference effects present in the SMEFT at LO for the processes we calculate, to address this issue.

The outline of this paper is as follows. In Section~\ref{inputschemes} we define how the
$\{\hat{\alpha}_{ew}, \hat{m}_Z, \hat{G}_F, \hat{M}_h \}$
and $\{\hat{m}_{W}, \hat{m}_Z, \hat{G}_F, \hat{M}_h \}$ electroweak parameter input schemes are related to Lagrangian parameters.
In Section~\ref{commonshifts} we define some common Lagrangian parameter shifts, including vertex corrections, widths, and shifts to the propagators
as combinations of Wilson coefficients. In Section~\ref{correctionstowidths} we define the consistent leading order results for the SMEFT corrections to a critical set of two and four body decays of the Higgs.
This includes an extensive discussion of the results for four body Higgs decays, and the required determination of the phase space integrations over four body phase space. In Section~\ref{sec:Pheno} we discuss the numerical results and quantify the impact of different contributions, with special attention to the terms that are usually neglected when using the narrow width approximation for the $W,Z$ bosons.
Finally in Section~\ref{conclusions} we conclude.

\section{SM and SMEFT theoretical conventions}
The SM Lagrangian \cite{Glashow:1961tr,Weinberg:1967tq,Salam:1968rm} notation is fixed to be
\bea\label{sm1}
\mathcal{L} _{\rm SM} &=& -\frac14 G_{\mu \nu}^A G^{A\mu \nu}-\frac14 W_{\mu \nu}^I W^{I \mu \nu} -\frac14 B_{\mu \nu} B^{\mu \nu}  + \sum_{\psi} \overline \psi\, i \slashed{D} \, \psi, \\
&\,&\hspace{-0.75cm} + (D_\mu H)^\dagger(D^\mu H) -  \lambda \left(H^\dagger H -\frac12 v^2\right)^2 -  \left[H^{\dagger j} \, \overline d\, Y_d\, q_{j}
+ \widetilde H^{\dagger j} \overline u\, Y_u\, q_{j} + H^{\dagger j} \overline e\, Y_e\,  \ell_{j} + \hbox{h.c.}\right]. \nonumber
\eea
The chiral projectors have the convention $\psi_{L/R} = P_{L/R} \, \psi$ where
$P_{R} = \left(1 + \gamma_5 \right)/2$, and
the gauge covariant derivative is defined with a positive sign convention
\bea
D_\mu = \partial_\mu + i g_3 T^A A^A_\mu + i g_2  \tau^I W^I_\mu/2 + i g_1 {\bf \hyp_i} B_\mu,
\eea
with $I=\{1,2,3\}$, $A=\{1\dots 8\}$ , $\tau^I$ denotes the Pauli matrices and
${\bf \hyp_i}$ the $\rm U_Y(1)$ hypercharge generator with charge normalization ${\bf \hyp_i}= \{1/6,2/3,-1/3,-1/2,-1,1/2\}$ for $i =\{q,u,d,\ell,e,H \}$.
Notation for $\Lagr^{(6)}$ largely descends from Ref.~\cite{Grzadkowski:2010es} with $\phi$ replaced by $H$ for the Higgs $\rm SU(2)_L$ field. We use
the Hermitian derivative conventions
\bea
H^\dagger \, i\overleftrightarrow D_\mu H &=& i H^\dagger (D_\mu H) - i (D_\mu H)^\dagger H, \\
H^\dagger \, i\overleftrightarrow D_\mu^I H &=& i H^\dagger \tau^I (D_\mu H) - i (D_\mu H)^\dagger \tau^I H.
\eea
The normalization of $\tau^I$ is such that $\rm{tr}[\tau^I \tau^J] = 2 \, \delta^{IJ}$. Our conventions are consistent with Ref.~\cite{Brivio:2017vri}, and we refer the reader to this work for more notational details.
We use the notation $k_{ij}^\alpha=(k_i+k_j)^\alpha$ and $k_{ij}^2=(k_i+k_j)^2$  for the Lorentz invariant four vector  and its square, with final state spinor pairs produced from the decay of a vector boson. For example, in the massless fermions limit, pairs $(\bar{u}(k_i),v(k_j))$, $(\bar{u}(k_k),v(k_l))$ can be produced by vectors carrying four momentum $k_{ij}^2= 2 k_i\cdot k_j$, $k_{kl}^2=2k_k\cdot k_l$.

\section{Input schemes and analytical results}\label{inputschemes}

Operators in $\Lagr^{(6)}$ can have a significant impact on the determination of Lagrangian parameters
from experimental imputs. The SMEFT has a significant input parameter scheme dependence of this form.
An input parameter scheme is an (informed) choice, with no scheme carrying unique benefits.
In any case, scheme dependence cancels out when experimental measurements are directly related to one
another, by-passing Lagrangian parameters. As generally this is not done,
in this work we present the Higgs width in two input parameter schemes, to avoid drawing overly scheme dependent
conclusions. For more discussion on the benefits of the schemes used see Ref.~\cite{Berthier:2015oma,Brivio:2017bnu}

For the $\{\hat{\alpha}_{ew}, \hat{m}_Z, \hat{G}_F \}$ input parameter scheme many of these results are summarized in Ref.~\cite{Brivio:2017vri}, which in turn is based on \cite{Grinstein:1991cd,Alonso:2013hga,Berthier:2015oma,Berthier:2015gja,Bjorn:2016zlr,Berthier:2016tkq}. The corresponding results in the
$\{\hat{m}_{W}, \hat{m}_Z, \hat{G}_F \}$ input parameter scheme largely descend from Ref.~\cite{Brivio:2017bnu}. Here we collect and complete the theoretical results used for
a self contained presentation, and to define a consistent LO set of analytic results of the SMEFT. These results are then used to consistently define the Higgs width with leading SMEFT corrections.

Our notation follows the "hat-bar" convention of Refs.~\cite{Brivio:2017vri,Alonso:2013hga,Berthier:2015oma}.
Lagrangian parameters directly determined from the measured input parameters are defined as having hat superscripts.
Lagrangian parameters in the canonically normalized SMEFT Lagrangian are indicated with
bar superscripts. The differences between these parameters come about due to the SMEFT perturbations of the SM.
With this convention, a leading order
shift in a SM Lagrangian parameter ($P$) due to the SMEFT, when such a parameter is determined from an input parameter set, is given by
\bea
\delta P = \bar{P} - \hat{P}.
\eea
Note the sign convention applied to these shift definitions, and that in the SM limit ($C_i\to 0$) hatted and bar quantities coincide and the SM inference from experimental results (at tree level)
is recovered. The implementation of this convention has some historical legacies. $\delta G_F$ is dimensionless while $G_F$ has mass dimension minus two
requiring a further dimensionful rescaling from a naive implementation of this convention.

In unitary gauge, the Higgs doublet is expanded as
\begin{align}
H &= \frac{1}{\sqrt 2} \left(\begin{array}{c}
0 \\
 \left[ 1+ \ckin \right]  h + \bar{v}_T
 \end{array}\right), & \quad  \ckin &\equiv \left(\tilde C_{H\Box}-\frac14 \tilde C_{HD}\right),
 \label{Hvev}
\end{align}
to obtain a canonical normalization. Here $\langle H^\dagger H \rangle $ has been defined to include corrections due to $\Lagr^{(6)}$
so that $\bar{v}_T \equiv ( 1+ 3 C_H \bar{v}^2/8 \lambda\Lambda^2 ) \bar{v}$ where $\sqrt{2 \langle H^\dagger H \rangle_{SM}} \equiv \bar{v}$.
Below, we include cross terms in theoretical predictions, where higher order SM perturbative corrections interfere with the $\Lagr^{(6)}$ corrections.
We note that the total contribution to $S$ matrix elements is gauge invariant order by order in the SMEFT power counting expansion; i.e.
the $A_{SM}$ amplitude contributing to an $S$ matrix element through $A_{SM} \times A^{(6)}/\Lambda^2$ is separately gauge invariant, as is $A^{(6)}/\Lambda^2$ alone.

The gauge fields are redefined into script fields to canonically normalize the SMEFT, including $\Lagr^{(6)}$ corrections, as
\begin{align}\label{canonical1}
G_\mu^A &= \mathcal{G}_\mu^A \left(1 + \tilde C_{HG} \right), &
W^I_\mu  &=  \mathcal{W}^I_\mu \left(1 + \tilde C_{HW} \right), &
B_\mu  &=  \mathcal{B}_\mu \left(1 + \tilde C_{HB} \right).
\end{align}
The modified coupling constants are simultaneously redefined
\begin{align}\label{canonicalagain}
\gcg &= g_3 \left(1 + \tilde C_{HG} \right), & \gcw &= g_2 \left(1 + \tilde C_{HW} \right), & \gcb &= g_1 \left(1 +\tilde  C_{HB} \right),
\end{align}
so that the products $g_3 G_\mu^A=\gcg \mathcal{G}_\mu^A$, etc.\ are unchanged.

The rotated script field eigenstate basis for $\{\mathcal{W}^3_\mu,\mathcal{B}_\mu\}$ in the SMEFT to $\Lagr^{(6)}$ is given by~\cite{Grinstein:1991cd,Alonso:2013hga}
\begin{align}\label{ZArotation}
\left[ \begin{array}{cc}  \mathcal{W}_\mu^3 \\ \mathcal{B}_\mu \end{array} \right]
&=
\left[ \begin{array}{cc}  1   &  -  \frac{1}{2} \,   \tilde C_{HWB} \\
- \frac{1}{2} \,  \tilde   C_{HWB} & 1 \end{array} \right] \, \left[ \begin{array}{cc} \cos \tc  &  \sin \tc \\
-\sin \tc & \cos \tc \end{array} \right] \left[ \begin{array}{cc}  \mathcal{Z}_\mu \\ \mathcal{A}_\mu \end{array} \right].
\end{align}
The $\Lagr_{SM} +\Lagr^{(6)}$ rotation angle is
\bea
\tan \bar{\theta} = \frac{\bar g_1}{\bar g_2} + \frac{\tilde  C_{HWB}}{2} \left(1 - \frac{\bar g_1^2}{\bar g_2^2}\right).
\eea
The mass eigenstate fields of the SM $\{Z_\mu,A_\mu \}$ are defined via the $C_{HWB} \rightarrow 0$ and $\{\cos \tc, \sin \tc\} \rightarrow \{\cos \theta, \sin \theta\}$
limit of Eq.~\eqref{ZArotation} where $c_\theta = \cos \theta = g_2/\sqrt{g_1^2 + g_2^2}$, $s_\theta = \sin \theta = g_1/\sqrt{g_1^2 + g_2^2}$.
The relation between the
mass eigenstate fields in $\Lagr_{SM}$ and $\Lagr_{SM} +\Lagr^{(6)}$ is explicitly \cite{Brivio:2017btx}
\begin{align}
  Z_\mu &=\mathcal Z_\mu\left( 1+ \hsw^2 \tilde C_{HB} + \hcw^2 \tilde  C_{HW} + \hsw \hcw \tilde C_{HWB}
  \right), \nn
  &+ \mathcal A_\mu\left( \hsw \hcw (\tilde C_{HW}-\tilde C_{HB}) - \left(\frac12-  \hsw^2\right)\tilde  C_{HWB}+\frac{\dsw^2}{2\hsw\hcw}
  \right),\\
  A_\mu &=\mathcal A_\mu\left( 1+ \hcw^2 \tilde C_{HB} + \hsw^2 \tilde C_{HW} - \hsw \hcw \tilde C_{HWB}
  \right), \nn
  &+ \mathcal Z_\mu\left( \hsw \hcw (\tilde C_{HW}-\tilde C_{HB}) - \left(\frac12-  \hsw^2\right) \tilde C_{HWB}-\frac{\dsw^2}{2\hsw\hcw}
  \right).
\end{align}
These expressions hold in both input parameter schemes using notation defined in the following section.
$\Lagr^{(8)}$ corrections to this formalism where recently reported in Ref.~\cite{Hays:2018zze}

In addition to the electroweak input parameters we discuss below in detail, we also require experimental inputs to fix $\{m_t, \alpha_s, m_c, m_b, m_\tau,V_{CKM}^{ij}, \Delta \alpha_{had}^{(5)}, \cdots \}$.
Barred mass parameters are generally defined to be the pole masses in $\Lagr_{SMEFT}$, including $\Lagr^{(6)}$ corrections. For recent discussion and results on CKM parameters in the SMEFT from an input parameter
perspective, see Refs.~\cite{Brivio:2017btx,Descotes-Genon:2018foz}.
Note that we generally neglect terms in the SMEFT corrections to SM results relatively suppressed by small quark masses.

\subsection{\titlemath{$\{\hat{\alpha}_{ew}, \hat{M}_Z, \hat{G}_F, \hat{M}_h \}$}{(aEM, MZ, GF, Mh)} input parameter scheme}
For the $\{\hat{\alpha}_{ew}, \hat{M}_Z, \hat{G}_F, \hat{M}_h \}$ input parameter scheme, in unitary gauge, we define
\begin{equation}
 \begin{aligned}
  \hat e &=\sqrt{4\pi\hat{\a}_{ew}}, \quad
 & \hat v_T &= \frac{1}{2^{1/4}\sqrt{\hat G_F}}, \quad
 &  \hst^2 &= \frac{1}{2}\left[1-\sqrt{1-\frac{4\pi\hat{\a}}{\sqrt2\hat G_F \hat M_Z^2}}\right],\quad
  &
 \hat M_W^2 &= \hat M_Z^2\hct^2,
 \\
 \hat g_1 &= \frac{\hat{e}}{\hct},
 &
 \hat g_2 &= \frac{\hat{e}}{\hst}, \quad
 &  \hat{g}_Z &= -   \frac{\hat g_2}{\hct}, \nonumber
 \end{aligned}
\end{equation}
and $ \hct^2 \equiv 1- \hst^2$. It is convenient to define
\begin{align}\label{gfmwshift}
 \d G_F &= \frac{1}{\sqrt2} \left(\tilde C^{(3)}_{\substack{Hl \\ee}}+\tilde C^{(3)}_{\substack{Hl \\ \mu \mu}} - \frac{1}{2}(\tilde C'_{\substack{ll \\ \mu ee \mu}}+\tilde C'_{\substack{ll \\ e \mu \mu e}})\right),  \\
\frac{\d  M_W^2}{\hat{M}_W^2} &=  -\frac{\hsdt}{4  \, \hcdt}\left( \frac{\hct}{\hst} \tilde C_{HD} +\frac{\hst}{\hct} 2\sqrt2\, \d G_F
+ 4 \tilde C_{HWB}\right), \label{eq:dMW}\\
  \d s_\theta^2  &=\frac{\hsdt}{8\hcdt}\left[\hsdt\left(\tilde C_{HD}+2\sqrt2 \, \d G_F \right)+4 \tilde C_{HWB}\right], \\
\frac{\d e}{\hat{e}} &= 0.
 \end{align}
 The $\rm U(3)^5$ limit used here treats the two flavour contractions of $\mathcal{Q}_{ll}$ as independent \cite{Cirigliano:2009wk}
\bea
(C_{\substack{ll \\  mnop}} \delta_{mn} \, \delta_{op} + C'_{\substack{ll \\  mnop}}  \delta_{mp} \, \delta_{no})(\bar l_m \gamma_\mu l_n)(\bar l_o \gamma^\mu l_p).
\eea
 We also define corrections to the $Z$ and $h$ mass parameters even though the corresponding input parameter $\hat{M}_Z,  \hat{M}_h$ fix the
 location of the propagator pole, i.e. by definition
a pole in a resonance scan is such that $\delta M_Z^2 = \bar{M}_Z^2 - \hat{M}_Z^2 \equiv 0$ and $\delta M_h^2 = \bar{M}_h^2 - \hat{M}_h^2 \equiv 0$.

We define shifts to the $Z$, $h$ masses as a convenient shorthand notation for common combinations
of Lagrangian parameters in $\Lagr_{\textrm{SMEFT}}$. We are then faced with a notational conundrum, as the natural notational choice in each case
is zero by definition. We overcome this challenge with a slight modification of notation compared to Ref.~\cite{Brivio:2017vri} by defining
\bea
 \d m_Z^2 &=  \dfrac{\hat{M}_Z^2}{2} \tilde C_{HD} + \dfrac{2^{3/4} \sqrt{\pi \hat{\alpha}} \, \hat{M}_Z}{\hat{G}_F^{1/2}} \tilde C_{HWB},\\
 \d m_h^2 &=  \hat M_h^2\left(-\dfrac{3\tilde C_H }{2\lambda}+2 \tilde C_{H\square}-\dfrac{\tilde C_{HD}}{2}\right), \label{mhinput}
\eea
where the lowercase $m$ takes on an meaning distinguishing it from the uppercase $M_{Z,h}$ resonance pole mass, whose shift vanishes by definition.
One should note this notational refinement when comparing to past works. See Ref.~\cite{Brivio:2017vri} for more details.

\subsection{\titlemath{$\{\hat{M}_{W}, \hat{M}_Z, \hat{G}_F, \hat{M}_h \}$}{(MW, MZ, GF, Mh)} input parameter scheme}
In this scheme
\begin{align}
\hat{e} &=  2\cdot 2^{1/4} \hat{M}_W\sqrt{\hat{G}_F} s_{\hat\theta},  \quad
& \hat v_T &= \frac{1}{2^{1/4}\sqrt{\hat G_F}},  \quad
& s^2_{\hat\theta} &= 1-\frac{\hat{M}_W^2}{\hat{M}_Z^2}, \nonumber \\
\hat{g}_1 &= 2\cdot 2^{1/4}\hat{M}_Z\sqrt{\hat{G}_F \left(1 -\frac{\hat{M}_W^2}{\hat{M}_Z^2}\right)}, & \quad
\hat{g}_2 &= 2\cdot 2^{1/4}\hat{M}_W\sqrt{\hat{G}_F}, & \quad   \hat{g}_Z &= -   \frac{\hat g_2}{\hct}. \nonumber
\end{align}
$ \d G_F, \d m_h^2$ are unchanged from the expressions in Eqs.~\eqref{gfmwshift},~\eqref{mhinput} and
\begin{align}
 \d m_Z^2 &=  \frac{\hat{M}_Z^2}{2} \tilde C_{HD} + 2 \, \hat{M}_Z \, \hat{M}_W \sqrt{1 - \frac{\hat{M}_W^2}{\hat{M}_Z^2}}\tilde C_{HWB},\\
 \d m_W^2 &= 0, \\
 \d m_h^2 &=  \hat M_h^2\left(-\frac{3\tilde C_H }{2\lambda}+2 \tilde C_{H\square}-\frac{\tilde C_{HD}}{2}\right), \\
\d s^2_{\theta} &= -\frac{\hat{M}_W^2}{2\hat{M}_Z^2}\tilde C_{HD}-\frac{\hat{M}_W}{\hat{M}_Z}\sqrt{1-\frac{\hat{M}_W^2}{\hat{M}_Z^2}} \tilde C_{HWB}, \\
\frac{\delta e}{\hat{e}} &\equiv \frac{\delta \alpha}{2 \, \hat{\alpha}} = -\frac{\delta G_F}{\sqrt{2}} + \dfrac{\d m_Z^2}{\hat{M}_Z^2} \frac{\hat{M}_W^2}{2 \, (\hat{M}_W^2 - \hat{M}_Z^2)}
- \tilde C_{HWB} \frac{\hat{M}_W}{\hat{M}_Z} \, \sqrt{1-\frac{\hat{M}_W^2}{\hat{M}_Z^2}}.
\end{align}

\section{Preliminaries: some common parameter shifts}\label{commonshifts}
For each input parameter scheme, the expression for a physical observable depends (in part) on the shift in the usual SM Lagrangian parameters
through the formulae in the previous two sections. Here we give a common set of such shifts. The $\delta P$ are a useful short hand notation that can be used at times in a specific gauge,
but do not span, and are not equivalent to, a complete and well defined gauge independent operator basis for $\Lagr^{(6)}$ in the SMEFT.
The remaining SMEFT corrections to physical observables appear through the direct dependence on
the operators in calculated amplitudes, and through the expansion of the $W$ pole mass in the $\{\hat{\alpha}_{ew}, \hat{M}_Z, \hat{G}_F, \hat{M}_h \}$
input parameter scheme.

\subsection{Effective \titlemath{$\mathcal A^{\mu} \bar{\psi}\gamma_\mu \psi$}{Aff} couplings}
In either input parameter scheme we can define the $\mathcal A^{\mu}$ effective couplings as
\bea
 \Lagr_{\mathcal A,eff} = - \hat{e} \,  \left[Q_\psi (1+ \delta e/ \hat{e}) \, J_{\psi}^{\mathcal A^\mu}  \right] \mathcal A^{\mu}, \quad J_\psi^{\mathcal A^\mu} = \bar{\psi} \, \gamma_\mu \, \psi,
 \eea
where $Q_\psi= \{2/3,-1/3, -1 \}$ for $\psi = \{u,d,e\}$. As class seven operators in the Warsaw basis are of the form $H^\dagger \, i\overleftrightarrow D_\mu H \bar{\psi} \gamma_\mu \psi$
and the Higgs is uncharged  under $\rm U(1)_{em}$, further flavour non-universal contact operator contributions due to expanding out these operators are not present. Chirality flipping dipole operators
generate effective couplings of the photon field to $\rm U(1)_{em}$ charged fermions at $\mathcal{L}_6$. However, as these contributions interfere with the SM amplitudes proportional to quark masses,
even if the Wilson coefficient is not assumed proportional to the Yukawa matrix to impose a controlled breaking of flavour symmetry; these contributions are neglected.

\subsection{Effective \titlemath{$\mathcal Z^{\mu} \bar{\psi}\gamma_\mu \psi$}{Zff}  and \titlemath{$h \, \mathcal Z^{\mu} \bar{\psi}\gamma_\mu \psi$}{hZff}  couplings}
The $\mathcal Z$ couplings are modified as
\begin{equation}\label{def_Zcurrent}
 \mathcal{L}_{\mathcal Z,{\rm eff}}  =  \hat g_{Z}  \,  J_{\psi^{pr}}^{\mathcal Z^\mu} \mathcal Z_\mu, \quad J_{\psi^{pr}}^{\mathcal Z^\mu} = \bar{\psi}_p \, \gamma_\mu \left[(\bar{g}^{\psi}_V)_{pr}- (\bar{g}^{\psi}_A)_{pr} \, \gamma_5 \right] \psi_r
\end{equation}
where $\psi = \{u,\nu, d, e\}$ with normalization
$\gsm{\psi}{V} = T_3/2 - Q_\psi  s_{\hat{\theta}}^2$, $\gsm{\psi}{A} = T_3/2$ and $2 \, T_3(\psi) = \{1,1,-1,-1\}$ while  $F[\tilde C_1,\tilde C_2,\tilde C_3  \cdots]_{pr} \equiv (\tilde C_{\substack{1 \\ pr}} + \tilde C_{\substack{2 \\ pr}} + \tilde C_{\substack{3 \\ pr}} + \cdots)/4$ yielding
\begin{align}
\delta (g^{\ell}_V)_{pr}&=\delta \bar{g}_Z \, (\gsm{\ell}{V})_{pr} - F[\tilde C_{\substack{H e}},\tilde C^{(1)}_{\substack{H \ell}},\tilde C^{(3)}_{\substack{H \ell}}]_{pr} - Q_\ell \, \delta_{pr} \, \delta s_\theta^2, \\
\delta(g^{\ell}_A)_{pr}&=\delta \bar{g}_Z \, (\gsm{\ell}{A})_{pr} -
F[-\tilde C_{\substack{H e}},\tilde C^{(1)}_{\substack{H \ell}},\tilde C^{(3)}_{\substack{H \ell}}]_{pr},  \\
\delta(g^{\nu}_{A/V})_{pr}&=\delta \bar{g}_Z \, (\gsm{\nu}{A/V})_{pr} -
F[\tilde C^{(1)}_{\substack{H \ell}},-\tilde C^{(3)}_{\substack{H \ell}}]_{pr},
\\
\delta (g^{u}_V)_{pr}&=\delta \bar{g}_Z \, (\gsm{u}{V})_{pr}  -
F[\tilde C_{\substack{H u}}, \tilde C_{\substack{H q}}^{(1)},-\tilde C^{(3)}_{\substack{H q}}]_{pr} - Q_u \, \delta_{pr} \, \delta s_\theta^2,\\
\delta(g^{u}_A)_{pr}&=\delta \bar{g}_Z \, (\gsm{u}{A})_{pr}
+F[\tilde C_{\substack{H u}}, -\tilde C_{\substack{H q}}^{(1)},\tilde C^{(3)}_{\substack{H q}}]_{pr},  \\
\delta (g^{d}_V)_{pr}&=\delta \bar{g}_Z \,(\gsm{d}{V})_{pr}
-F[\tilde C_{\substack{H d}},\tilde C_{\substack{H q}}^{(1)},\tilde C^{(3)}_{\substack{H q}}]_{pr} - Q_d \, \delta_{pr} \, \delta s_\theta^2, \\
\delta(g^{d}_A)_{pr}&=\delta \bar{g}_Z \,(\gsm{d}{A})_{pr}
-F[-\tilde C_{\substack{H d}},\tilde C_{\substack{H q}}^{(1)},\tilde C^{(3)}_{\substack{H q}}]_{pr},
\end{align}
where $\hat g_{Z} = - \hat{g}_2/c_{\hat{\theta}} = - 2 \, 2^{1/4} \sqrt{\hat{G}_F} \hat{M}_Z = - \sqrt{\hat{g}_1^2 + \hat{g}_2^2}$ and
\begin{equation}
 \delta \bar{g}_Z =- \frac{\delta G_F}{\sqrt{2}} - \frac{\delta m_Z^2}{2\hat{M}_Z^2} + s_{\hat{\theta}} \, c_{\hat{\theta}} \, \tilde C_{HWB}.
\end{equation}

The SMEFT introduces $h \, \mathcal Z^{\mu} \bar{\psi}\gamma_\mu \psi$ couplings that are forbidden in the SM due to it being limited to $d\leq 4$ interactions.
We define these couplings as
\begin{equation}\label{def_Zhcurrent}
 \mathcal{L}_{\mathcal Zh,{\rm eff}}  = \frac{2 \, \hat g_{Z}}{\hat{v}_T} \,  \mathcal Z^\mu \, h \, \bar{\psi}_p \, \gamma_\mu \left[\delta \bar{C}_{\substack{\psi\\pr}}^{hV}- \delta \bar{C}_{\substack{\psi\\pr}}^{hA} \, \gamma_5 \right] \psi_r,
\end{equation}
where as above
 \begin{align}
\delta \bar{C}_{\substack{\ell\\pr}}^{hV}&= -  F[\tilde C_{\substack{H e}},\tilde C^{(1)}_{\substack{H \ell}},\tilde C^{(3)}_{\substack{H \ell}}]_{pr}, \quad &
\delta \bar{C}_{\substack{\ell\\pr}}^{hA}&=  - F[-\tilde C_{\substack{H e}},\tilde C^{(1)}_{\substack{H \ell}},\tilde C^{(3)}_{\substack{H \ell}}]_{pr}, \\
\delta \bar{C}_{\substack{\nu\\pr}}^{hV}&= - F[\tilde C^{(1)}_{\substack{H \ell}},-\tilde C^{(3)}_{\substack{H \ell}}]_{pr}, \quad &
\delta \bar{C}_{\substack{\nu\\pr}}^{hA}&= -  F[\tilde C^{(1)}_{\substack{H \ell}},-\tilde C^{(3)}_{\substack{H \ell}}]_{pr}, \\
\delta \bar{C}_{\substack{u\\pr}}^{hV}&= - F[\tilde C_{\substack{H u}},\tilde C^{(1)}_{\substack{H q}},-C^{(3)}_{\substack{H q}}]_{pr},\quad &
\delta \bar{C}_{\substack{u\\pr}}^{hA}&= F[\tilde C_{\substack{H u}}, -\tilde C_{\substack{H q}}^{(1)},\tilde C^{(3)}_{\substack{H q}}]_{pr}, \\
\delta \bar{C}_{\substack{d\\pr}}^{hV}&= - F[\tilde C_{\substack{H d}},\tilde C^{(1)}_{\substack{H q}},\tilde C^{(3)}_{\substack{H q}}]_{pr},\quad &
\delta \bar{C}_{\substack{d\\pr}}^{hA}&= - F[-\tilde C_{\substack{H d}},\tilde C^{(1)}_{\substack{H q}},\tilde C^{(3)}_{\substack{H q}}]_{pr}.
\end{align}

In the results that follow, we calculate in the limit that final state fermions are neglected. Using chiral eigenstates of the fermions is advantageous in some
results, and we note that the left and right handed SM couplings follow in the standard manner. The chiral SMEFT corrections are
\begin{align}
\delta g_{\substack{L \\ pr}}^\psi &= \delta g_{\substack{V \\ pr}}^\psi + \delta g_{\substack{A \\ pr}}^\psi,& \quad  \delta g_{\substack{R \\ pr}}^\psi &= \delta g_{\substack{V \\ pr}}^\psi - \delta g_{\substack{A \\ pr}}^\psi, \\
\delta C_{\substack{HL \\pr}}^{\psi} &= \delta \bar{C}_{\substack{\psi\\pr}}^{hV} +\delta \bar{C}_{\substack{\psi\\pr}}^{hA}, & \quad  \delta C_{\substack{HR\\ pr}}^{\psi}  &= \delta \bar{C}_{\substack{\psi\\pr}}^{hV} -\delta \bar{C}_{\substack{\psi\\pr}}^{hA} .
\end{align}
We introduce the convenient notation $(\bar{g}_{\pm}^{\psi_a})^2 = |\bar{g}_{L}^{\psi_a}|^2 \pm |\bar{g}_{R}^{\psi_a}|^2$ for some common combinations of the $\mathcal{Z}$ boson couplings that appear.

\subsection{Effective \titlemath{$\mathcal W^{\mu} \bar{\psi}_L\gamma_\mu \psi_L$}{Wff} and \titlemath{$\mathcal W^{\mu} h \bar{\psi}_L\gamma_\mu \psi_L$}{hWff} couplings}
In the case of the $\mathcal W$ effective couplings we define
\begin{equation}\label{Wcouplings.def}
 \Lagr_{\mathcal W,eff} =  - \frac{\hat{g}_2}{\sqrt{2}}  \mathcal W^+_\mu \, J_{\psi^{pr}}^{\mathcal W^{+,\mu}} + h.c.,
 \end{equation}
\begin{equation}
 J_{\ell^{pr}}^{\mathcal W^{+,\mu}} =\bar{\nu}_p \g^\mu \left[(g_V^{W_+,\ell})_{pr} - (g_A^{W_+,\ell})_{pr} \g_5\right]e_r, \quad
J_{q^{pr}}^{\mathcal W^{+,\mu}} = \bar{u}_p \, \gamma_\mu  \left[(g_V^{W_+,q})_{pr} - (g_A^{W_+,q})_{pr} \g_5\right] d_r. \nonumber
\end{equation}
In the SM
\begin{equation}
 (\bar g_{V}^{W_+,\ell})^{SM}_{pr}= (\bar g_A^{W_+,\ell})^{SM}_{pr} = \frac{(U_{PMNS}^\dagger)_{pr}}{2},
 \qquad\qquad
  (\bar g_V^{W_+q})^{SM}_{pr}=(\bar g_A^{W_+,q})^{SM}_{pr} = \frac{V_{pr}}{2},
\end{equation}
where $V = U(u,L)^\dagger U(d,L)$ is the CKM matrix, $U = U(e,L)^\dagger U(\nu,L)$ the PMNS matrix, with $U(\psi,L/R)$ the rotation matrix between the weak and mass eigenstates. For flavour diagonal components
\begin{align}
 \d (g_V^{W_\pm,\psi})_{rr}= \d (g_A^{W_\pm,\psi})_{rr} &= \frac{1}{2} \, \tilde C^{(3)}_{\substack{H \psi \\ rr}} - \frac{ \delta G_F}{2 \, \sqrt{2}} \delta_{rr},
\end{align}
for $\psi = \{q,\ell\}$ in the $\{\hat{M}_{W}, \hat{M}_Z, \hat{G}_F, \hat{M}_h \}$ input scheme and
\begin{align}
\delta(g^{W_{\pm},\psi}_V)_{rr} = \delta(g^{W_{\pm},\psi}_A)_{rr}  &= \frac{1}{2} \left(\tilde C^{(3)}_{\substack{H \psi \\ rr}} + \frac{1}{2} \frac{\hct}{\hst} \, \tilde C_{HWB} \delta_{rr} \right)
- \frac{1}{4} \frac{\delta s_\theta^2}{s^2_{\hat{\theta}}} \delta_{rr}.\label{delta_gW_quarks}
\end{align}
in the $\{\hat{\alpha}_{ew}, \hat{M}_Z, \hat{G}_F, \hat{M}_h \}$ scheme.

The SMEFT introduces $h \, \mathcal W_{+}^{\mu} \bar{\psi}T_+ \gamma_\mu P_L \psi, h \, \mathcal W_{-}^{\mu} \bar{\psi}T_- \gamma_\mu P_L \psi$ couplings as
\bea\label{def_Whcurrent}
 \mathcal{L}_{\mathcal Wh,{\rm eff}}  &=& - \frac{\sqrt{2} \hat g_{2}}{\hat{v}_T}   \, h \, \bar{\ell}_p \,  \left[ \mathcal W^+_\mu \, T^+ \, \tilde{C}^{(3)}_{\substack{H \ell \\pr}}
 +  \mathcal W^-_\mu T^-  \, \tilde{C}^{(3)}_{\substack{H \ell \\pr}} \right]  \gamma_\mu  \ell_r, \nonumber  \\
 &-& \frac{\sqrt{2} \hat g_{2}}{\hat{v}_T}  \, h \, \bar{q}_p \,  \left[ \mathcal W^+_\mu T^+  \tilde{C}^{(3)}_{\substack{H q \\pr}}
 +  \mathcal W^-_\mu T^-   \tilde{C}^{(3)}_{\substack{H q \\pr}} \right]  \gamma_\mu  q_r,
\eea
where $2  \, T^+ = \tau_1 + i \tau_2, \, 2 \, T^- = \tau_1 - i \tau_2$.
The off diagonal terms trivially follow.
Again, we note that the left and right handed $ \delta g^{W,\psi}_{L,R}$ SMEFT corrections are the sum and
difference of the vector and axial $W$ shifts respectively.

\subsection{Effective \titlemath{$h \bar{\psi} \psi$}{hff} couplings}
The pole masses of quarks and leptons inferred from experimental results can define input parameters $\hat{m}_\psi$.
These inputs also determine the Yukawa couplings through the definition
\begin{equation}\label{Yhat}
 \hat Y_\psi = 2^{3/4}\hat M_\psi \sqrt{\hat G_F}\,
\end{equation}
with normalization
\begin{equation}\label{hcouplings.def}
 \Lagr_{h,eff} =  - g_{\substack{h \psi \\ pr}} \, h \, \bar{\psi}_{\substack{R\\p}} \psi_{\substack{L\\r}} + h.c.
\end{equation}
In the SM  $g_{\substack{h \psi\\ pp}}^{SM} =\hat  Y_{\substack{\psi\\pp}}/\sqrt2$, and in the SMEFT \cite{Alonso:2013hga}
\begin{equation}
 \d g_{\substack{h \psi\\pr}} =\frac{\hat Y_{\substack{\psi \\pr}}}{\sqrt{2}}\left[\ckin - \frac{\d G_F}{\sqrt{2}} \right]- \frac{1}{\sqrt2 } \,\tilde C^*_{\substack{\psi H \\ pr}} \,.
\end{equation}
Note that in the $\rm U(3)^5$ limit $\tilde C^*_{\substack{\psi H \\ pr}}$ is proportional to $Y_{\substack{\psi\\ pr}}$.

\subsection{Massive boson propagator and width shifts}
For a consistent treatment of the SMEFT corrections to the SM, the propagators need to be expanded up to linear order in the Wilson coefficients, when a massive vector boson mediates an experimental measurement~\cite{Berthier:2016tkq}.
For a massive boson $B =\{\mathcal Z,\mathcal W,h\}$ we define
\begin{equation}
 D^B(k^2_{ij}) = \frac{1}{k^2_{ij}-\bar M_B^2+i\bar \Gamma_B \bar M_B+i\epsilon}\left[1+\d D^B(k^2_{ij})\right].
\end{equation}
The propagator in unitary gauge is then
\bea
\frac{-i}{k_{ij}^2 - \hat{M}_{B}^2 + i \, \hat{\Gamma}_{B} \, \hat{M}_{B}+i\epsilon} \left[g_{\mu \nu} - \frac{k_\nu k_\mu}{\hat{M}^2_{B}} \left(1- \frac{\delta M^2_{B}}{\hat{M}^2_{B}}\right)\right]\left[1+\d D^B(k^2_{ij})\right].
\eea
Note that as we calculate in the massless limit of the final state fermions, the longitudinal term $\propto k_\nu k_\mu$ vanishes in this limit.
The shift in the propagator is given by
\begin{align}
 \d D^B(k_{ij}^2) &= \frac{1}{k^2_{ij}-\hat M_B^2+i\hat \Gamma_B \hat M_B}\left[\left(1-\frac{i\hat \Gamma_B}{2\hat M_B}\right)\,\d M_B^2-i\hat M_B\d\Gamma_B\right]\,\label{def_Ds}.
\end{align}
A useful result is
\begin{align}
 2 {\rm Re}\left[ \d D^B(k^2_{ij})\right] &= \frac{2 (k^2_{ij} - \hat M_B^2) \delta M_B^2 - \hat{\Gamma}_B (\hat{\Gamma}_B \delta M_B^2 + 2 \hat{M}_B^2 \, \d \Gamma_B)}{(k^2_{ij} - \hat M_B^2)^2 + \hat{M}_B^2 \hat{\Gamma}_B^2 },
\end{align}
which can be directly used if $V_{CKM},U_{PMNS}$ phases are neglected and one considers a CP conserving set of Wilson coefficients in  $\Lagr^{(6)}$.
In a near on-shell region of phase space $k_{ij}^2 \simeq  \hat m_B^2$
\begin{align}\label{nearonshell}
 2 {\rm Re}\left[ \d D^B(\hat{m}_B^2)\right] &\simeq - \frac{\delta M_B^2}{\hat{M}_B^2} - 2 \frac{\d \Gamma_B}{\hat{\Gamma}_B}.
\end{align}

In the $\{\hat{\alpha}_{ew}, \hat{M}_Z, \hat{G}_F, \hat{M}_h \}$ scheme $\d M_Z^2 =\d M_h^2 = 0$, while in the $\{\hat{M}_{W}, \hat{M}_Z, \hat{G}_F, \hat{M}_h \}$ scheme $\d M_Z^2 = \d M_W^2 = \d M_h^2 = 0$.
The width shift should be included when studying experimental results in any scheme for a consistent SMEFT analysis. This can be done by expanding in the correction to the width, linearizing the dependence on the SMEFT correction in the final result
for an observable. This procedure is difficult to directly carry out interfacing SMEFTsim \cite{Brivio:2017btx} with \MG \, due to the implementation of widths in \MG. Our results present this result for inclusive quantities in
a semi-analytic form, determining this correction using direct numerical integration. This makes the dependence on the total width clear in the case of inclusive quantities, and at least clearer when considering non-inclusive quantities.

The decay of a Higgs boson to four fermion final states occurs through physical phase space where some of the intermediate propagators are necessarily off-shell.
The width of the unstable SM Bosons remains parametrically  less than the boson mass in the SMEFT, as the corrections introduced to the widths and masses are a small perturbative correction. As such, one can still expand
in the small ratios $\Gamma_B/M_B$, $\delta \Gamma_B/M_B$ the modified propagator of the SMEFT, finding
\begin{align}
2 {\rm Re} \left[\d D^B(k^2_{ij}) \right]  &=   2 \frac{\delta M_B^2}{\hat{M}_B^2} \frac{\hat{M}_B^2}{k^2_{ij} - \hat{M}_B^2} - 2 \frac{\delta \Gamma_B}{\Gamma_B}  \frac{\hat{M}_B^2 \hat{\Gamma}_B^2}{(k^2_{ij} - \hat{M}_B^2)^2}
+ \frac{\delta M_B^2}{\hat{M}_B^2} \frac{\hat{M}_B^2 \hat{\Gamma}_B^2 (k^2_{ij} + \hat{M}_B^2)}{(k^2_{ij} - \hat{M}_B^2)^3} + \cdots
\end{align}
The off shell region of phase space where $(k^2_{ij} - \hat{M}_B^2) \simeq \hat{\Gamma}_B \hat{M}_B$, and when $k^2_{ij}$ takes on other values, is averaged over in a four body decay of the Higgs through intermediate vector bosons $\mathcal{V} =\{\mathcal{W},\mathcal{Z},\mathcal{A}\}$.
The effect of this averaging in the SMEFT, compared to the SM, modifies the coefficient for $\delta \Gamma_B/\hat{\Gamma}_B$ in an $\mathcal{O}(1)$  manner, and this deviation from a naive expectation formed
using Eq.~\eqref{nearonshell} is included in our results, see Sec. \ref{propcorrections}.

 \subsubsection{\titlemath{$\delta \Gamma_{\mathcal{Z}}$}{Corrections to the Z width}}
At tree level corrections to $\mathcal{Z}$ partial and total widths due to $\Lagr^{(6)}$ are
\begin{align}
{\Gamma} \left(\mathcal{Z} \to \psi \bar{\psi} \right) &= \frac{ \, \sqrt{2} \, \bar{G}_F \bar{M}_Z^3 \, N_C^\psi}{3 \pi} \left( |\bar{g}^{\psi}_V|^2 + |\bar{g}^{\psi}_A|^2 \right), \\
 \delta {\Gamma} \left(\mathcal{Z} \to \psi \bar{\psi} \right) &= \frac{ \, \sqrt{2} \, \hat{G}_F \hat{M}_Z^3 \, N_C^\psi}{3 \pi} \left( 2 \gsm{\psi}{V} \delta g^{\psi}_{V} + 2 \gsm{\psi}{A} \delta g^{\psi}_{A} \right), \\
 \d \Gamma_{\mathcal{Z} \to {\rm Had}} &= 2 \, \delta \bar{\Gamma} \left(\mathcal{Z} \to u \bar{u} \right)+ 3  \, \delta \bar{\Gamma} \left(\mathcal{Z} \to d \bar{d} \right) \label{Zhadronicwidth}, \\
 \d \Gamma_{\mathcal{Z}}  &= 3\d \Gamma_{\mathcal{Z} \to \ell \bar{\ell}} + 3 \d \Gamma_{\mathcal{Z} \to \nu \bar{\nu}} +\d \Gamma_{\mathcal{Z} \to {\rm Had}} \label{Zwidth},
 \end{align}
 $N_C^\psi$ depends on the $\rm SU(3)_c$ representation of $\psi$.
 Off diagonal corrections due to local contact operators are neglected, as they interfere with SM contributions that have a significant numerical suppression. This reasoning is used
 in part to define a "pole parameter" set of SMEFT Wilson coefficients in Ref.~\cite{Brivio:2017btx}, and our results are consistent with this reasoning.\footnote{The neglect of flavour violating effects
 for the $h$ interactions with fermions also follows from the structure of flavour changing effects in the SM.}
Similarly corrections due to four fermion operators
 modify the inference of a partial $\mathcal Z$ width from an experimental cross section, with an intermediate $\mathcal Z$ boson. We also neglect these corrections
 as they are kinematically suppressed beyond the power counting suppression.

 \subsubsection{\titlemath{$\delta \Gamma_{\mathcal{W}}$}{Corrections to the W width}}
At tree level corrections to $\mathcal{W}$ partial and total widths due to $\Lagr^{(6)}$ are~\cite{Berthier:2016tkq}
 \begin{align}
 \Gamma^{SM}_{\mathcal{W}} &= \frac{3 \, \hat G_F \, \hat{M}_W^3}{2 \sqrt{2} \pi}, \\
\d\Gamma_{\mathcal{W}} &= \Gamma^{SM}_{\mathcal{W}}\left(\frac{4}{3}\delta g_{V/A}^{W, \ell} + \frac{8}{3}\delta  g_{V/A}^{W, q} +\frac{\delta M_W^2}{2 \hat{M}_W^2}\right), \\
  \delta\Gamma_{\mathcal{W} \rightarrow \bar \psi_p \psi_r} &= \frac{2 \, N_C \, \Gamma^{SM}_W}{9} \left(V^\psi_{pr} \delta g_{\substack{V/A\\pr}}^{W, \psi}+ V_{pr}^{\psi,\dagger} \delta (g_{\substack{V/A\\pr}}^{W, \psi})^\dagger + |V^\psi_{pr}|^2 \frac{\delta M_W^2}{4 \hat{M}_W^2}\right).
  \end{align}
 $V^\psi$ corresponds to the CKM ($\psi=q$) or Hermitian conjugate of the PMNS ($\psi=\ell$) matrix.
 in the $\{\hat{M}_{W}, \hat{M}_Z, \hat{G}_F, \hat{M}_h \}$ scheme  recall that $\delta M_W^2/ \hat{M}_W^2 =0$.

 \section{Corrections to the partial and total Higgs decay widths}\label{correctionstowidths}
 The total and partial Higgs width is also corrected in the SMEFT as follows.

 \subsection{\titlemath{$\delta \Gamma_{h \rightarrow \bar{\psi} \psi }$}{h > ff}}
 The decays to $\psi = \{u,c,d,s,b,e,\mu,\tau\}$ are each modified as
  \begin{align}
\bar{\Gamma} \left(h \to  \bar{\psi} \psi \right) &= \frac{|g_{h \psi}^{SM}|^2}{8 \pi} \, N_C^\psi \, \bar{M}_h\left(1-\dfrac{4\bar M_\psi^2}{\bar M_h^2}\right)^{3/2}, \quad &
 \delta {\Gamma}_{h \to  \bar{\psi} \psi} &= \frac{g_{h \psi}^{SM} \, {\rm Re}(\delta g_{h \psi})}{4 \pi} \, N_C^\psi \, \hat{M}_h \left(1-\dfrac{4\hat M_\psi^2}{\hat M_h^2}\right)^{3/2}.
  \end{align}

\subsection{\titlemath{$\delta \Gamma_{h \rightarrow \mathcal A \mathcal A }$}{h > AA}}\label{AAsection}
 The leading order SM prediction of $\Gamma(h \rightarrow \mathcal A \mathcal A)$ is \cite{Ellis:1975ap,Shifman:1979eb,Bergstrom:1985hp}
\bea
\bar{\Gamma}_{SM} \left(h \to  \mathcal A \mathcal A \right) &=& \frac{\bar{\alpha}^2 \, \bar{G}_F}{128 \sqrt{2} \pi^3} \, \bar{M}_h^3
\left| \sum_\psi N_C^\psi \, Q_\psi^2 \, F_{1/2} (\tau_\psi) + F_1(\tau_W) \right|^2,
\eea
where $\tau_W = 4 \bar{M}_W^2/\bar{M}_h^2,\tau_{ZW} = \bar{M}_Z^2/(4 \bar{M}_W^2)$ and $\tau_\psi = 4 \bar{M}_\psi^2/\bar{M}_h^2$, while
\bea
F_{1/2}(\tau_\psi) &=& - 2 \tau_\psi - 2 \tau_\psi (1- \tau_\psi)f(\tau_\psi), \\
F_{1}(\tau_W) &=&  2 + 3 \, \tau_W \left[1 + (2 - \tau_W) \, f(\tau_W)\right], \\
f (\tau_p) &=&
\begin{cases}
   \arcsin^2 \sqrt{1/\tau_p} ,&  \tau_p \ge 1\\
    - \frac{1}{4} \left[\ln \frac{1 + \sqrt{1- \tau_p}}{1 - \sqrt{1- \tau_p}} - i \pi \right]^2,              & \tau_p <1.
\end{cases}
\eea
The correction in the SMEFT due to $\Lagr^{(6)}$ is given by
\bea\label{gamgam}
\delta \Gamma_{h \to  \mathcal A \mathcal A} &=&  \hat{\Gamma}_{SM} \left(h \to  \mathcal A \mathcal A\right)   \left(2 \, \frac{\delta \alpha}{\hat \alpha}  - \sqrt{2} \delta G_F\right)
\nn
&
+& \frac{\hat{\alpha}^2 \,  \hat{G}_F\, \hat{M}_h^3 }{128 \sqrt{2} \pi^3}  \, \delta  \left| \sum_\psi N_C^\psi \, Q_\psi^2 \, F_{1/2} (\tau_\psi) + F_1(\tau_W) \right|^2 \nn
&-& \frac{\hat{\alpha}^2 \, \hat G_F\,\hat{M}_h^3}{2\sqrt2 \, \pi} {\rm Re} \left[ \sum_\psi N_C^\psi \, Q_\psi^2 \, F_{1/2} (\tau_\psi) + F_1(\tau_W) \right] \tilde C_{\mathcal A \mathcal A},
\eea
where $\tilde C_{\mathcal A \mathcal A} = \tilde C_{HW}/\hat{g}_2^2 + \tilde C_{HB}/\hat{g}_1^2 - \tilde C_{HWB}/\hat{g}_1 \, \hat{g}_2$.
Here the first line indicates the scheme dependent linear expansion in SMEFT corrections feeding into the SM loop diagrams. The second term indicates a shift due to a possible shift in the $W$ mass, while the fermion mass
inputs are assumed to be pole masses.
One loop calculations of this process in the SMEFT were reported in Refs.~\cite{Hartmann:2015oia,Ghezzi:2015vva,Hartmann:2015aia,Dedes:2018seb,Dawson:2018liq}.
Such corrections are important, but they the same order (in the SMEFT expansion and the loop expansion) as the scheme dependent corrections to the SM
results, which introduces scheme dependence that is only removed once a full one loop improvement of SMEFT predictions is obtained.
We consistently drop such loop suppressed SMEFT effects in this work and only retain the contribution from the third line of Eq.~\eqref{gamgam}.

\subsection{\titlemath{$\delta \Gamma_{h \rightarrow gg }$}{h > gg}}
The LO SM result for $\Gamma(h \rightarrow gg)$ is \cite{Georgi:1977gs,Ellis:1979jy}
\bea
\bar{\Gamma}_{SM} \left(h \to  gg \right) &=& \frac{\bar{\alpha}_s^2 \, \bar{G}_F}{64 \sqrt{2} \pi^3} \, \bar{M}_h^3
\left| \sum_\psi \, F_{1/2} (\tau_\psi) \right|^2.
\eea
The correction in the SMEFT due to $\Lagr^{(6)}$ is given by
\bea\label{ggresult}
\delta \Gamma_{h \to  gg} &=&  \hat{\Gamma}_{SM} \left(h \to  gg \right)   \left(2 \, \frac{\delta \alpha_s}{\hat \alpha_s}  - \sqrt{2} \delta G_F\right)
- \frac{\hat{\alpha}_s\,\hat G_F}{2\sqrt2 \pi^3} \, \hat{M}_h^3  \,{\rm Re} \left[  \sum_\psi \, F_{1/2} (\tau_\psi) \right] \,\tilde C_{HG}.
\eea

\subsection{\titlemath{$\delta \Gamma_{h \rightarrow \mathcal Z \mathcal A }$}{h > ZA}}\label{ZAsection}
The leading order SM prediction of $\Gamma(h \rightarrow \mathcal Z \mathcal A)$ is given by \cite{Bergstrom:1985hp,Manohar:2006gz}
\bea
\bar{\Gamma}_{SM} \left(h \to  \mathcal Z \mathcal A \right) &=& \frac{\bar{\alpha}^2 \, \bar{G}_F}{64 \sqrt{2} \pi^3} \, \bar{M}_h^3 \left(1- \frac{\bar{M}_Z^2}{\bar{M}_h^2}\right)^3
\left| \sum_\psi N_C^\psi \, Q_\psi \,  \frac{2 \, \bar{g}_V^{\psi}}{s_{2 \theta}} I_\psi (\tau_\psi, \tau_{Zt}) + I_W^Z(1/\tau_W,\tau_{ZW}) \right|^2, \nonumber
\eea
where
\bea
I_\psi (a, b) = -4 \int_{0}^1 dx \int_0^{1-x} dy \frac{1- 4 x y}{1 - 4 (a-b) x y - 4 b y (1-y) - i 0^+},
\eea
and $\tau_{Zt} = \bar{m}_Z^2/4 \bar{m}_t^2$ and $I_\psi (\tau_\psi,  0) = F_{1/2}(\tau_\psi)$. The remaining loop function is  \cite{Bergstrom:1985hp,Manohar:2006gz}
\bea
I_W^Z(a,b) = \frac{-4}{t_{\bar{\theta}}} \, \int_0^1 dx \int_0^1 dy \frac{(5 - t_{\bar{\theta}}^2 + 2 a (1- t_{\bar{\theta}}^2)) x y - (3 - t_{\bar{\theta}}^2)}{1 - 4(a-b) x y - 4 b y (1-y) - i 0^+}.
\eea
The correction in the SMEFT due to $\Lagr^{(6)}$ is given by
\bea
\delta \bar{\Gamma}_{h \to  \mathcal Z \mathcal A} &=& \frac{\hat{\alpha}^2 \, \hat{G}_F}{64 \sqrt{2} \pi^3} \, \hat{M}_h^3 \left(1- \frac{\hat{M}_Z^2}{\hat{M}_h^2}\right) \delta
\left| \sum_\psi N_C^\psi \, Q_\psi \,  \frac{2 \, \bar{g}_V^{\psi}}{s_{2 \theta}} I_\psi (\tau_\psi, \tau_{Zt}) + I_W^Z(1/\tau_W,\tau_{ZW}) \right|^2, \\
&-& {\frac{\hat{\alpha}^2 \,\hat G_F\, \hat{M}_h^3}{\sqrt2 \,\pi}} \left(1-\frac{\hat{M}_Z^2}{\hat{M}_h^2}\right)^3 \tilde C_{\mathcal{A} \mathcal{Z}} \, {\rm Re}
\left[\sum_\psi N_C^\psi \, Q_\psi \,  \frac{2 \, \bar{g}_V^{\psi}}{s_{2 \theta}} I_\psi (\tau_\psi, \tau_{Zt}) + I_W^Z(1/\tau_W,\tau_{ZW}) \right], \nonumber \\
&+& \hat{\Gamma}_{SM} \left(2 \, \frac{\delta \alpha}{\hat \alpha}  - \sqrt{2} \delta G_F\right), \nonumber
\eea
where $\tilde C_{\mathcal{A} \mathcal{Z}} = \tilde C_{HW}/(\hat{g}_1 \hat{g}_2) - \tilde C_{HB}/(\hat{g}_1 \hat{g}_2) -\tilde C_{HWB}(\hat{g}_2^2 - \hat{g}_1^2)/(2 \hat{g}_1^2 \hat{g}_2^2)$.

\subsection{Four fermion decays \titlemath{$\delta \Gamma_{h \rightarrow \bar{\psi} \psi \bar{\psi} \psi }$}{h > ffff}}
Some of the largest partial widths that remain are due to $h \rightarrow {\mathcal{V} \, \mathcal{V}^\star} \rightarrow \bar{\psi} \psi \, \bar{\psi} \psi$,
through combinations of vector bosons $\mathcal V =\{\mathcal W^\pm,\mathcal Z, \mathcal A \}$. These calculations, when the intermediate gauge bosons are allowed to be off-shell, have been developed
for the SM in Refs.~\cite{Kniehl:1990yb,Cahn:1988ru,Grau:1990uu,Gross:1994fv,Bredenstein:2006rh,Bredenstein:2006ha}. Here we extend this approach to the SMEFT,
avoiding an on-shell assumption and narrow width approximation to ensure the consistency of the SMEFT corrections included in a leading order analysis.

To define these corrections, it is useful to introduce the notation
\begin{align*}
[k_{ij}^2,\mathcal{V}] &= (k_{ij}^2 - m_\mathcal{V}^2) + i \Gamma_\mathcal{V} M_\mathcal{V}, &
[k_{ij}^2,\mathcal{V}^\dagger] &= [k^2_{ij},\mathcal{V}]^\star,
\end{align*}
and $J_{\psi_a^{pr}}^{\mathcal V^\nu}(k^2_{ij})$ when the
gauge boson $\mathcal V$ coupling to the current  producing the final states labeled with $\psi_a^{p,r}$ (of flavours $p,r$) carries four momentum squared $k^2_{ij}$. The definition of the propagator
has assumed a width prescription that is consistent with the implementation of widths for unstable states in \MG.\footnote{A generalization of the results to a different width prescription and the complex mass scheme is clearly also of interest, but is beyond the scope of this work.}
Further notation is defined as follows
\bea
T^{\prod_i \! \mathcal{V}_i^{\alpha_i}}_{\prod_i \psi_i^{r_is_i}}(k^2_{ij}) = \frac{\sum_s \left[ \prod_i (J_{\psi_i^{pr}}^{\mathcal V_i^{\alpha_i}}) \right]}{\prod_i [k^2_{ij},{\mathcal{V}_i}^2]},
\eea
for example
\bea
T_{\psi_1^{pr} \psi_1^{pr}}^{\mathcal V_1^{\mu} (\mathcal V_1^{\nu})^\dagger}(k^2_{ij},k^2_{ij}) &=& \frac{\sum_s \left[(J_{\psi_1^{pr}}^{\mathcal V_1^\mu})(J_{\psi_1^{pr}}^{\mathcal V_1^\nu})^\dagger \right]}
{|[k^2_{ij},\mathcal{V}_1]|^2}.
\eea
The flavour, colour and spin sums (denoted $\sum_s$) in each case are restricted to the allowed final states. Note that we are not using a convention that repeated indicies are always summed.\footnote{At times an Einstein summation convention is in place, particularly for flavour indicies for brevity of notation. The presence of a summation or not is believed to be clear from the physics in each case.} At times the momentum dependence of the $T$ are suppressed.
Determining the partial width from the expressions that follow is defined as
\bea
\hat{\Gamma} = \frac{1}{2 \hat{M}_h} \, \int  {\it d ps} \, |A|^2, \quad \int  {\it d ps}  = \int (2 \pi)^4 \delta^4(P_h -\sum_i k_i) \prod_{k_i} \frac{d^3 k_i}{(2 \pi)^3 2 E_{i}},
\eea
for a $4$ body phase space element with $k_i$ denoting the momentum of each final state spinor. In the Appendix we transform this four body phase space into
a form where Lorentz invariants are integrated over in the phase space volume, to allow direct numerical evaluations.
A convenient trace product to define for a compact presentation is
\bea
L_{k_1,k_2}^{\alpha \beta} &=& {\rm Tr} \left[ \slashed{k_1} \gamma^\alpha  \slashed{k_2} \gamma^{\beta} P_L\right].
\eea

\subsubsection{\titlemath{$\delta \Gamma_{h \rightarrow \mathcal W \, \mathcal W^\star \rightarrow \bar{\psi}_1 \psi_1' \, \bar{\psi}_2 \psi_2'}$}{h > WW* > ffff}}
\begin{figure}[h!]
\includegraphics[width=0.25\textwidth]{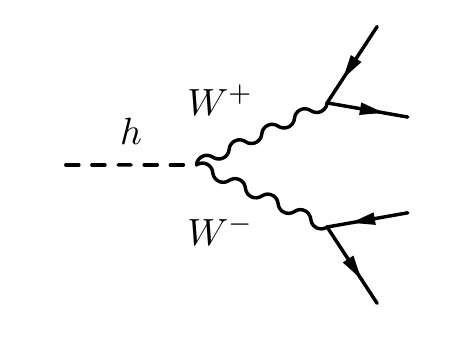}
\includegraphics[width=0.25\textwidth]{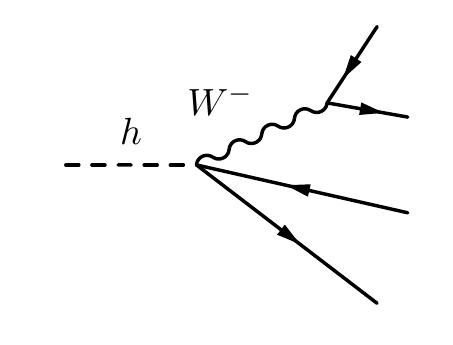}
\includegraphics[width=0.25\textwidth]{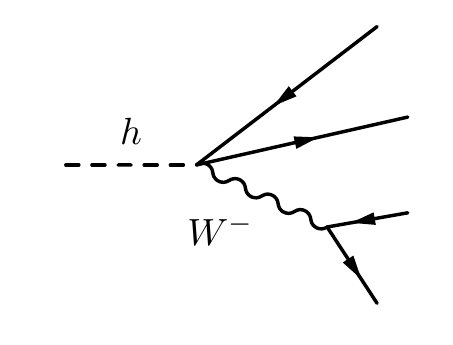}
\caption{Charged current contributions to $h \rightarrow 4 \psi$. The SM diagram also represents $\mathcal{O}(\bar{v}_T^2/\Lambda^2)$ operator insertions perturbing the SM prediction through two charged current exchanges. \label{fig:pureWW}} .
\end{figure}

The diagrams for pure charged current (CC) interference effects are shown in  Fig.~\ref{fig:pureWW}

We label $h \rightarrow \mathcal W \, \mathcal W^\star \rightarrow \bar{\psi}_1 \psi_1' \, \bar{\psi}_2 \psi_2'$
as $h \rightarrow F_1({\psi_1^{rs},\psi_2^{tu}})$.
In the SM the corresponding leading order result is
\bea
|A_{{\psi_1^{rs},\psi_2^{tu}}}^{\mathcal W \mathcal W^\dagger}|_{SM}^2 &=& |A^{\mathcal W^{\alpha \beta} (\mathcal W^{\gamma \delta})^\dagger}_{\psi_1^{rs},\psi_2^{tu}}|^2  g_{\alpha \gamma} g_{\beta \delta}, \nn
 |A^{\mathcal W^{\alpha \beta} (\mathcal W^{\gamma \delta})^\dagger}_{\psi_1^{rs},\psi_2^{tu}}|^2   &=&  \frac{\hat{g}_2^4 \bar{g}_2^4 \bar{v}^2_T}{16} T_{\psi_1^{rs}}^{\mathcal W^{\alpha} ( \mathcal W^{\beta})^\dagger}(k^2_{ij},k^2_{ij})\, T_{\psi_2^{tu}}^{ \mathcal  W^{\gamma} ( \mathcal W^{\delta})^\dagger} (k^2_{kl},k^2_{kl}).
\eea
Here $k_{ij,kl}$ are the momentum carried by the $\mathcal{W}$ propagators associated with the spinor pairs $(\bar{u}(k_i),v(k_j))$, $(\bar{u}(k_k),v(k_l))$.
The couplings in this expression are $\hat{g}_2^4 \, \bar{g}_2^4$ as the $\hat{g}_2$ couplings are defined to be those that couple the vector to $J^\mathcal{W}$, while the remaining dependence descends from the $h \mathcal{W}^2$ vertex.

The partial SM decay width at leading order is constructed from the one "kinematic number"
\bea
N^\mathcal{W W}_1 &=& \int  {\it d ps} \frac{k_{i}\cdot k_{l} \, k_{j} \cdot k_{k} }{|[k^2_{ij},\mathcal{W}]|^2 |[k^2_{kl},\mathcal{W}] |^2} \simeq 1.28 \times 10^{-6}.
\eea
We extract this number, and similar numbers below using various techniques to cross check results.
When considering SMEFT corrections,
novel kinematic numbers result from the novel populations of phase space due to the presence of local contact operators of mass dimension $d>4$.
An extended set of such kinematic numbers is required for describing four fermion Higgs decays in the SMEFT.
Such corrections are a key difference from the $\kappa$ formalism developed in Refs.
\cite{LHCHiggsCrossSectionWorkingGroup:2012nn,Zeppenfeld:2000td,Duhrssen:2003tba,Duhrssen:2004cv,Lafaye:2009vr,Espinosa:2012ir,Carmi:2012yp,Azatov:2012bz}.

We determine a kinematic number in a process using the Vegas algorithm and  CUBA numerical integration package \cite{Hahn:2004fe} primarily.
These results are also cross checked in  \MG  \, \cite{Alwall:2014hca} from leading order SM results. We have also (when possible) cross checked these results with an independent evaluation using the RAMBO algorithm \cite{Kleiss:1985gy} to directly determine the phase space integrals. In some cases, for phase space integrals that are highly singular,
or in the presence of multiple poles, the Vegas numerical approach was considered an essential step to obtain a reliable determination
of sufficient numerical accuracy. Some details on these approaches are given in the Appendix.

The SM result  for the pure charged current partial width is
\bea
\Gamma(h \rightarrow F_1({\psi_1^{rs},\psi_2^{tu}}))_{SM} =
\frac{8 \, N_C^{\psi_1} \, N_C^{\psi_2}  \,  \hat{g}_2^4  \bar{M}_W^4}{\hat{M}_h \, \bar{v}_T^2}  |(\bar g_{L,rs}^{W_\pm,\psi_1})^{SM}|^2 |(\bar g_{L,tu}^{W_\pm,\psi_2})^{SM}|^2 \, N^\mathcal{W W}_1.
\eea
The $\mathcal{L}^{(6)}$ SMEFT corrections can be classified by
the phase space integrations they multiply. The partial width corrections that simply perturb the SM prediction proportional to $N^\mathcal{W W}_1$  are\footnote{Note the hat notation on the predicted observable is again used here.}
\bea
\frac{\delta \Gamma(N^\mathcal{W W}_1)}{(\hat{\Gamma}(h \rightarrow F_1({\psi_1^{rs},\psi_2^{tu}}))_{SM}} =
2 \, \frac{{\rm Re}[ \d  (g_{L,rs}^{\mathcal W^+,\psi_{1}})]}{{\rm Re}[(g_{L,rs}^{\mathcal W^+,\psi_{1}})^{SM}]}+ 2 \, \frac{{\rm Re}[ \d  (g_{L,tu}^{\mathcal W^-,\psi_{2}})]}{{\rm Re}[(g_{L,tu}^{\mathcal W^-,\psi_{2}})^{SM}]}
+ 2 \left[\frac{ \delta M_W^2}{\hat{M}^2_W} - \frac{\delta G_F}{\sqrt{2}} + \ckin \right],\nn
\eea
in the limit that we neglect phases in the CKM and PMNS matrices. The generalization to the case where SM phases are not neglected is via
\bea
\frac{{\rm Re}[ \d  (g_{L,rs}^{\mathcal W^+,\psi_{1}})]}{{\rm Re}[(g_{L,rs}^{\mathcal W^+,\psi_{1}})^{SM}]} \rightarrow  \frac{{\rm Re}[ \d  (g_{L,rs}^{\mathcal W^+,\psi_{1}})] {\rm Re}[(g_{L,rs}^{\mathcal W^+,\psi_{1}})^{SM}]}{|(g_{L,rs}^{\mathcal W^+,\psi_{1}})^{SM}|^2}+ \frac{{\rm Im}[ \d  (g_{L,rs}^{\mathcal W^+,\psi_{1}})] {\rm Im}[(g_{L,rs}^{\mathcal W^+,\psi_{1}})^{SM}]}{|(g_{L,rs}^{\mathcal W^+,\psi_{1}})^{SM}|^2}.
\eea
The remaining corrections lead to non-SM phase space integrations due to the local contact operators present in the SMEFT, and are given by $2 \hat{M}_h \delta \Gamma_{h \rightarrow  F_1({\psi_1^{rs},\psi_2^{tu}})}$
which is
\bea
&=&  \sum_{n={ij},{kl}}  \int  {\it d ps} |A_{{\psi_1^{rs},\psi_2^{tu}}}^{\mathcal W \mathcal W^\dagger}|_{SM}^2 \left[
\frac{8 \, C_{\substack{H \psi_{n} \\ op}}^{(3)} (k_n^2 - \hat{M}_W^2)}{\hat{g}_2^2 \, (g_{L,op}^{\mathcal W,\psi_{n}})^{SM}} \,  (\delta_n^1\delta_{o}^r \delta_{p}^s+ \delta_n^2\delta_{o}^t \delta_{p}^u)
+ 2 {\rm Re}[ \d  D^{\mathcal W}(k_n^2)]\right], \nn
&-&  \frac{16 \, \tilde C_{HW}}{\hat{g}_2^2\bar v_T^2} \int  {\it d ps}
|A^{\mathcal W^{\alpha \beta} \mathcal (W^{\gamma \delta})^\dagger}_{\psi_1^{rs},\psi_2^{tu}}|^2 \, g_{\alpha \gamma} \, (k_{ij} \cdot k_{kl} \,  g_{\beta \delta} - k_{kl}^\beta \, k_{ij}^\delta).
\eea
Here we have neglected interference effects due to SM phases in the propagator correction, and final state fermion masses.
The relevant inclusive phase space integrals can be evaluated to be
\bea
 \int  {\it d ps} \frac{k_{i} \cdot k_{l} \, k_{j} \cdot k_{k} }{|[k^2_{ij},\mathcal{W}]|^2 |[k^2_{kl},\mathcal{W}] |^2} \frac{8 (k^2_n - \hat{M}_W^2)}{\hat{v}_T^2}  &\equiv& N^\mathcal{W W}_2
 \simeq - 4.83 \times 10^{-7} , \\
  \int  {\it d ps} \frac{k_{i} \cdot k_{l} \, k_{j} \cdot k_{k} }{|[k^2_{ij},\mathcal{W}]|^2 |[k^2_{kl},\mathcal{W}] |^2} \, 2 \,{\rm Re}[ \d  D^{\mathcal W}(k^2_n)]
&\equiv&  N^\mathcal{W W}_3 \frac{\delta \Gamma_W}{\hat{\Gamma}_W} + N^\mathcal{W W}_4 \frac{\delta M_W^2}{\hat{M}_W^2}, \\
&\simeq&  - 1.22 \times 10^{-6} \,    \frac{\delta \Gamma_W}{\hat{\Gamma}_W} - 8.78 \times 10^{-6}  \frac{\delta M_W^2}{\hat{M}_W^2}, \nn
-
\int  {\it d ps} L_{k_i,k_j}^{\alpha \beta} L_{k_k,k_l}^{\gamma \delta} \frac{g_{\alpha \gamma}(g^{\beta \delta} k_{ij} \cdot k_{kl} -  k_{kl}^\beta k_{ij}^\delta)}{|[k^2_{ij},\mathcal{W}]|^2 |[k^2_{kl},\mathcal{W}] |^2 \hat{v}_T^2} &\equiv& N^\mathcal{W W}_5  \simeq
-8.15 \times 10^{-7} .
\eea
The shift in this inclusive partial decay width ($\delta \Gamma_{h \rightarrow  F_1({\psi_1^{rs},\psi_2^{tu}})}/ (\hat{\Gamma}(h \rightarrow F_1({\psi_1^{rs},\psi_2^{tu}}))_{SM}$) can then be defined as
\bea
 &\simeq&
\delta \Gamma(N^\mathcal{W W}_1) + \sum_{i=1,2} \frac{N^\mathcal{W W}_2}{N^\mathcal{W W}_1} \,  \frac{\tilde{C}_{\substack{H \psi_{i}\\ op}}^{(3)} (\delta_i^1\delta_{o}^r \delta_{p}^s+ \delta_i^2\delta_{o}^t \delta_{p}^u)}{\hat{g}_2^2 \, (g_{L,op}^{\mathcal W,\psi_{i}})^{SM}}
+ \frac{N^\mathcal{W W}_3}{N^\mathcal{W W}_1} \frac{\delta \Gamma_W}{\hat{\Gamma}_W} + \frac{N^\mathcal{W W}_4}{N^\mathcal{W W}_1}  \frac{\delta M^2_W}{\hat{M}^2_W}
+ \frac{N^\mathcal{W W}_5}{N^\mathcal{W W}_1} \frac{\tilde{C}_{H W}}{\hat{g}_2^2}, \nonumber \\
&\simeq &
\delta \Gamma(N^\mathcal{W W}_1) -0.38 \,   \sum_{i=1,2}  \frac{\tilde{C}_{\substack{H \psi_{i}\\ op}}^{(3)} (\delta_i^1\delta_{o}^r \delta_{p}^s+ \delta_i^2\delta_{o}^t \delta_{p}^u)}{\hat{g}_2^2 \, (g_{L,op}^{\mathcal W,\psi_{i}})^{SM}}
- 0.95 \frac{\delta \Gamma_W}{\hat{\Gamma}_W}
- 6.9 \frac{\delta M^2_W}{\hat{M}^2_W} -0.64 \frac{\tilde{C}_{H W}}{\hat{g}_2^2},
\eea
The first term in this expression can be obtained from rescaling the SM result, which is consistent with the approach in the $\kappa$ formalism.
When the population of phase space in the SM and the SMEFT due to an interaction is the same, the ratio of kinematic numbers is one.
The ratios of the kinematic numbers in the remaining terms give some intuition as to how the $\kappa$ formalism fails due to the decay kinematics being able to differ in the SMEFT, compared to the SM.
When measuring a decay channel, assumptions on SM like kinematics to define event rate acceptances is expected to require the introduction of further correction factors when the ratio of the kinematic numbers is very
far from one. The acceptance correction will be strongly dependent on the detailed experimental signal definition and is not determined in the calculation reported here.
Our results are intended to define and determine the theoretical inclusive total width and branching ratios in the SMEFT.

\subsubsection{\titlemath{$\delta \Gamma_{h \rightarrow \mathcal Z \, \mathcal Z^\star  \rightarrow \bar{\psi}^r_a \psi^r_a \, \bar{\psi}^s_b \psi^s_b}$}{h > ZZ* > ffff}}
A similar expression can be defined for $h \rightarrow \bar{\psi}^r_a \psi^r_a \, \bar{\psi}^s_b \psi^s_b$ through neutral currents (NC). There are several combinations of
intermediate states when considering neutral currents; we discuss each of the interference effects in turn.

\begin{figure}[h!]
\includegraphics[width=0.20\textwidth]{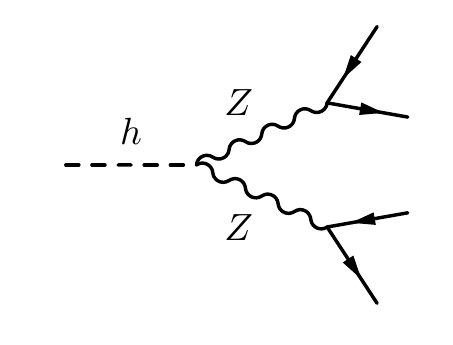}
\includegraphics[width=0.20\textwidth]{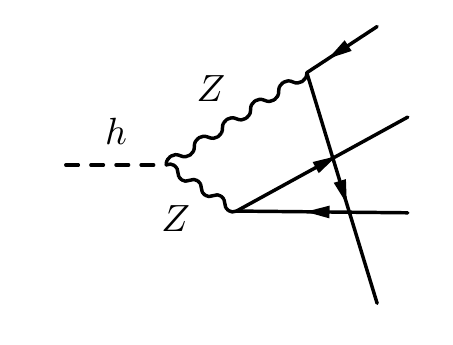}
\includegraphics[width=0.20\textwidth]{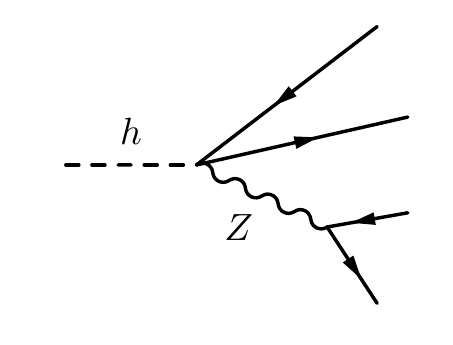}
\includegraphics[width=0.20\textwidth]{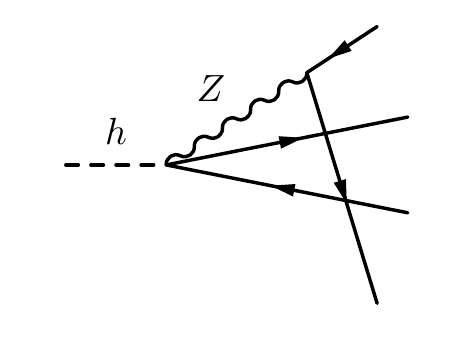}
\caption{Pure $Z$ neutral current contributions to $h \rightarrow 4 \psi$. The SM diagrams also represent $\mathcal{O}(\bar{v}_T^2/\Lambda^2)$ operator insertions perturbing the SM prediction through
diagrams with the same pole structure as the SM. \label{fig:pureZZ}} .
\end{figure}

We label the case where  one includes the effect of intermediate $\mathcal{Z}$ bosons only with the notation $h \rightarrow F_2({\psi_a^{rr},\psi_b^{ss}}) $.
The corresponding diagrams are shown in Fig.~\ref{fig:pureZZ}.

In the SM, this LO result  is
\bea\label{Zpairdecay}
|A_{{\psi_a^{r},\psi_b^{s}}}^{\mathcal Z \mathcal Z}|_{SM}^2 &=& |A^{\mathcal Z^{\alpha \beta} \mathcal Z^{\gamma \delta}}_{\psi_a^{rr},\psi_b^{ss}}|^2  g_{\alpha \gamma} g_{\beta \delta}, \nn
 |A^{\mathcal Z^{\alpha \beta} \mathcal Z^{\gamma \delta}}_{\psi_a^{rr},\psi_b^{ss}}|^2
&=&
\frac{\hat{g}_Z^4 \, \bar{g}_Z^4 \bar{v}^2_T}{4}  \left[T_{\psi_a^r}^{\mathcal Z^{\alpha} (\mathcal{Z}^\beta)^\dagger}(k^2_{ij}) \, T_{\psi_b^s}^{ \mathcal Z^{\gamma} (\mathcal Z^{\delta})^\dagger}(k^2_{kl}) (1- \frac{\delta_a^b\, \delta_r^s}{2})\right], \\
 &-& \frac{\hat{g}_Z^4 \, \bar{g}_Z^4 \bar{v}^2_T}{16} \delta_a^b \delta_r^s {\left[
T^{\mathcal Z^{\beta} Z^{\alpha} (Z^{\gamma})^\dagger (Z^{\delta})^\dagger}_{\psi_a^{r}  \, \psi_a^{r}
\, \psi_a^{r} \psi_a^r}(k^2_{ij},k^2_{kl},k^2_{jk},k^2_{li}) \,  + h.c.  \right]}. \nonumber
\eea
The second term in the expression above is complex. A relative sign in the two terms is due to Fermi statistics, and there are relative numerical factors
due to counting Wick contractions. Here $k_{ij,kl,jk,li}$ are the momentum carried by the $\mathcal{Z}$ propagators associated with the momenta of the final state spinors pairs $(\bar{u}(k_i), v(k_j)), (\bar{u}(k_k), v(k_l))$,
$(\bar{v}(k_j), u(k_k)),(\bar{v}(k_l), u(k_i))$.

It is useful to expand these results out explicitly obtaining
\bea\label{ZpairdecaySM}
|A_{{\psi_a^{r},\psi_b^{s}}}^{\mathcal Z \mathcal Z}|_{SM}^2 &=&
\frac{2 N_C^{\psi_a} N_C^{\psi_b} \,  \hat{g}_Z^4 \, \bar{g}_Z^4 \bar{v}_T^2}{|[k^2_{ij},\mathcal{Z}]|^2 |[k^2_{kl},\mathcal{Z}] |^2} k_{i} \cdot k_{k} \, k_{j} \cdot k_{l} \left[(\bar{g}_{+}^{\psi_a})^2  (\bar{g}_{+}^{\psi_b})^2 -(\bar{g}_{-}^{\psi_a})^2  (\bar{g}_{-}^{\psi_b})^2 \right](1- \frac{\delta_a^b \, \delta_r^s}{2} ), \nn
&+& \frac{2 N_C^{\psi_a} N_C^{\psi_b}  \hat{g}_Z^4 \, \bar{g}_Z^4 \bar{v}_T^2}{|[k^2_{ij},\mathcal{Z}]|^2 |[k^2_{kl},\mathcal{Z}] |^2} k_{i} \cdot k_{l} \, k_{j} \cdot k_{k} \left[ (\bar{g}_{+}^{\psi_a})^2  (\bar{g}_{+}^{\psi_b})^2 +(\bar{g}_{-}^{\psi_a})^2  (\bar{g}_{-}^{\psi_b})^2  \right] (1- \frac{\delta_a^b \, \delta_r^s}{2}), \nn
&+& \left[\frac{N_C^{\psi_a}   \hat{g}_Z^4 \, \bar{g}_Z^4 \bar{v}_T^2 \, \delta_{a}^b \, \delta_r^s }{[k^2_{ij},\mathcal{Z}] [k^2_{kl},\mathcal{Z}] [k^2_{jk},\mathcal{Z}]^\star [k^2_{li},\mathcal{Z}]^\star}
\, k_{i} \cdot k_{k}  \,k_{j} \cdot k_{l}  \left[|\bar{g}_{L}^{\psi_a}|^4 + |\bar{g}_{R}^{\psi_a}|^4 \right] + {\rm h.c.}\right].
\eea
Integrating over phase space we extract the kinematic numbers, one finds
\bea
N_1^{\mathcal{Z Z}} &=& \int  {\it d ps} \frac{k_{i} \cdot k_{k} \, k_{j} \cdot k_{l} }{|[k^2_{ij},\mathcal{Z}]|^2 |[k^2_{kl},\mathcal{Z}] |^2} =
\int  {\it d ps} \frac{k_{i} \cdot k_{l} \, k_{j} \cdot k_{k}  }{|[k^2_{ij},\mathcal{Z}]|^2 |[k^2_{kl},\mathcal{Z}] |^2} \simeq 1.74 \times 10^{-7},\\
N_2^{\mathcal{Z Z}}  &=& \int  {\it d ps} \frac{k_{i} \cdot k_{k}  \, k_{j} \cdot k_{l} }{[k^2_{ij},\mathcal{Z}] [k^2_{kl},\mathcal{Z}] [k^2_{jk},\mathcal{Z}]^\star [k^2_{li},\mathcal{Z}]^\star} + h.c. \simeq 6.84 \times 10^{-8}.
\eea
With these results the corresponding SM inclusive partial widths are constructed as
\bea
\Gamma(h \rightarrow F_2({\psi_a^{rr},\psi_b^{ss}}))_{SM} &=&  \Gamma^{\mathcal{ZZ}}_0  N_C^{\psi_a}  \left[
N_C^{\psi_b}  \, N_1^{\mathcal{Z Z}} \, (\bar{g}_{+}^{\psi_a})^2  (\bar{g}_{+}^{\psi_b})^2 (1- \frac{\delta_a^b \delta_r^s}{2})\right], \nonumber \\
&+& \frac{\Gamma^{\mathcal{ZZ}}_0}{4}  N_C^{\psi_a} \delta_{a}^b \, \delta_r^s \, N_2^{\mathcal{Z Z}} \,\left(|\bar{g}_{L}^{\psi_a}|^4 + |\bar{g}_{R}^{\psi_a}|^4\right).
\eea
where $\Gamma^{\mathcal{ZZ}}_0  = 32 \,  \,  \hat{g}_Z^4 \, \bar{M}_Z^4/(\hat{M}_h \bar{v}_T^2)$.
A subset of the SMEFT $\mathcal{L}^{(6)}$ corrections to this partial width directly follow as
\bea
\frac{\delta \Gamma(N_1^{\mathcal{Z Z}}, N_2^{\mathcal{Z Z}} )}{ \hat{\Gamma}^{\mathcal{ZZ}}_0  \, N_C^{\psi_a} } &=&
\, N_C^{\psi_b}  \, N_1^{\mathcal{Z Z}} \, \left[ \delta ( \bar{g}_{+}^{\psi_a})^2 \,   (\hat{g}_{+}^{\psi_b})^2 +  (\hat{g}_{+}^{\psi_a})^2  \delta ({g}_{+}^{\psi_b})^2 \right] (1- \frac{\delta_a^b \delta_r^s}{2}), \\
&+& 2  \,   \left(\tilde{C}_{HD} +  \ckin + 2 c_{\hat{\theta}} s_{\hat{\theta}} \tilde{C}_{HWB} - \frac{\delta m_Z^2}{\hat{M}^2_Z} - \frac{\delta G_F}{\sqrt{2}} \right) \, \frac{\Gamma(h \rightarrow F_2({\psi_a^{rr},\psi_b^{ss}}))_{SM}}{ \hat{\Gamma}^{\mathcal{ZZ}}_0  \, N_C^{\psi_a}},  \nonumber  \\
&+& \delta_{a}^b \, \delta_r^s  N_2^{\mathcal{Z Z}} \, \left( \delta{g}_{L}^{\psi_a} (\hat{g}_{L}^{\psi_a})^3+  \delta{g}_{R}^{\psi_a} (\hat{g}_{R}^{\psi_a})^3\right). \nonumber
\eea

In addition there are the perturbations to $2 \hat{M}_h \hat{\Gamma}(h \rightarrow F_2({\psi_a^{rr},\psi_b^{ss}}))$ of the form\footnote{Here we slightly abuse notation defining  $\hat{\Gamma}^{\mathcal{ZZ}}_0  = 32 \,  \,  \hat{g}_Z^4 \, \hat{M}_Z^4/(\hat{M}_h \hat{v}_T^2)$.}
\bea
&+&\hat{g}_Z^2 \, \bar{g}_Z^4 \bar{v}^2_T \! \! \! \!  \sum_{n= \{ij,kl\}} \frac{\delta \bar{{C}}_{H L/R}^{\psi_n}}{g_{L/R}^{\psi_{n},SM}}
\int  {\it d ps} \,
T_{\psi^{L/R,rr}_j, \psi^{L/R,rr}_j}^{\mathcal Z^{\alpha},(\mathcal Z^{\mu})^\dagger}
T_{\psi^{ss}_k, \psi^{ss}_k}^{\mathcal Z^{\mu},(\mathcal Z^{\alpha})^\dagger} [k^2_n,\mathcal{Z}], \nn
 &-& \frac{\hat{g}_Z^2 \, \bar{g}_Z^4 \bar{v}^2_T}{8} \delta_a^b  \frac{\delta \bar{C}_{H L/R}^{\psi_a}}{g_{L/R}^{\psi_{a},SM}}
\int  {\it d ps} \,
T_{\psi^{L/R}_{a,rr}, \psi^{L/R}_{a,rr} \psi^{L/R}_{a,rr}, \psi^{L/R}_{a,rr}}^{(\mathcal Z^{\alpha})^\dagger,(\mathcal Z^{\alpha})^\dagger \mathcal Z^{\beta},\mathcal Z^{\beta}} \, \,
([k^2_{jk},\mathcal{Z}]+[k^2_{li},\mathcal{Z}]), \nn
&+& \frac{\hat{g}_Z^4 \, \bar{g}_Z^4 \bar{v}^2_T}{4} \int  {\it d ps} \,  T_{\psi_a^r}^{\mathcal Z^{\alpha} (\mathcal{Z}^\beta)^\dagger} \, T_{\psi_b^t}^{ \mathcal Z^{\gamma} (\mathcal Z^{\delta})^\dagger}
 g_{\alpha \gamma} g_{\beta \delta}  \left[\d  D^{Z}(k^2_{ij})  + \d  D^{Z}(k^2_{kl})\right] (1- \frac{\delta_a^b\, \delta_r^s}{2}), \nn
&-& \frac{\hat{g}_Z^4 \, \bar{g}_Z^4 \bar{v}^2_T}{16}  \delta_a^b \int  {\it d ps} \,T^{\mathcal Z^{\beta} Z^{\alpha} (Z^{\gamma})^\dagger (Z^{\delta})^\dagger}_{\psi_a^{r}  \, \psi_a^{r}
\, \psi_a^{r} \psi_a^r}  g_{\alpha \gamma} g_{\beta \delta} \left(\d  D^{Z}(k^2_{ij})+ \d  D^{Z}(k^2_{kl})+ \d  D^{Z,\star}(k^2_{jk})+ \d  D^{Z,\star}(k^2_{li})\right), \nn
&-&  \frac{16 \, \tilde C_{\mathcal{Z} \mathcal{Z}}}{\hat{g}_Z^2\bar v_T^2} \int  {\it d ps}
|A^{\mathcal Z^{\alpha \beta} \mathcal (Z^{\gamma \delta})}_{\psi_a^{r},\psi_b^{s}}|^2 \, g_{\alpha \gamma} \, (k_{ij} \cdot k_{kl} \,  g_{\beta \delta} - k_{kl}^\beta \, k_{ij}^\delta) + h.c.,
\eea
where $\tilde C_{\mathcal{Z} \mathcal{Z}} = (c_{\hat{\theta}}^2 \tilde C_{HW} + s_{\hat{\theta}}^2\tilde  C_{HB} + c_{\hat{\theta}} \, s_{\hat{\theta}} \tilde C_{HWB})$.
This set of $\delta \Gamma(h \rightarrow F_2({\psi_a^{rr},\psi_b^{ss}}))$ perturbations numerically reduce to (neglecting fermion masses)
\bea
 &\simeq&   \hat{\Gamma}^{\mathcal{ZZ}}_0
\sum_{\{j,k\}=a,b} \frac{\delta \bar{C}_{H L/R}^{\psi_j}}{\hat{g}_Z^2}  N_C^a \,  \left[  (g_{L/R}^{\psi_{j},SM})  (g_{+}^{\psi_{k}})^2 N_C^b \, N_{3,L/R}^{\mathcal{Z Z}} {(1- \frac{\delta_a^b \delta_r^s}{2})}
+ \frac{(g_{L/R}^{\psi_{j},SM})^3}{4} \delta_{j}^k  \delta_{r}^s  N_{4,L/R}^{\mathcal{Z Z}} \right], \nn
 &+&   \hat{\Gamma}^{\mathcal{ZZ}}_0 \, N_C^{\psi_a} \left[
N_C^{\psi_b}  \, N_5^{\mathcal{Z Z}} \, (\bar{g}_{+}^{\psi_a})^2  (\bar{g}_{+}^{\psi_b})^2 (1- \frac{\delta_a^b\, \delta_r^s}{2})
+  \delta_{a}^b \, \delta_r^s  \, \frac{N_6^{\mathcal{Z Z}}}{4} \, \left[|\bar{g}_{L}^{\psi_a}|^4 + |\bar{g}_{R}^{\psi_a}|^4 \right]\right]  \frac{\delta \Gamma_Z}{\hat{\Gamma}_Z},
\nn
&+&   \hat{\Gamma}^{\mathcal{ZZ}}_0 \, N_C^{\psi_a}\left[
N_C^{\psi_b}  \, N_7^{\mathcal{Z Z}} \, (\bar{g}_{+}^{\psi_a})^2  (\bar{g}_{+}^{\psi_b})^2 (1- \frac{\delta_a^b\, \delta_r^s}{2})
+ \delta_{a}^b \, \delta_r^s \, \frac{N_8^{\mathcal{Z Z}}}{4} \, \left[|\bar{g}_{L}^{\psi_a}|^4 + |\bar{g}_{R}^{\psi_a}|^4 \right]\right]  \frac{\delta M^2_Z}{\hat{M}^2_Z},
\nn
&+&   \hat{\Gamma}^{\mathcal{ZZ}}_0 \, N_C^{\psi_a} \, \left[
N_C^{\psi_b}  \, N_9^{\mathcal{Z Z}} \, (\bar{g}_{+}^{\psi_a})^2  (\bar{g}_{+}^{\psi_b})^2 (1- \frac{\delta_a^b\, \delta_r^s}{2})
+  \delta_{a}^b \, \delta_r^s  \, \frac{N_{10}^{\mathcal{Z Z}}}{4} \, \left[|\bar{g}_{L}^{\psi_a}|^4 + |\bar{g}_{R}^{\psi_a}|^4 \right]\right]  \frac{\tilde{C}_{\mathcal{Z} \mathcal{Z}}}{\hat{g}_Z^2}. \nn
\eea

Here
\begin{align}
N_{3,L/R}^{\mathcal{Z Z}}  &\simeq -9.76 \times 10^{-8},  &\quad N_{4,L/R}^{\mathcal{Z Z}}  &\simeq -5.28 \times 10^{-8}, \\
N_{5}^{\mathcal{Z Z}}  &\simeq  -1.45 \times 10^{-7},  &\quad  N_{6}^{\mathcal{Z Z}}  &\simeq  -2.96 \times 10^{-9},\\
N_{7}^{\mathcal{Z Z}}  &\simeq  -1.37 \times 10^{-6},  & \quad  N_{8}^{\mathcal{Z Z}}  &\simeq  -3.79 \times 10^{-7}, \\
N_{9}^{\mathcal{Z Z}}  &\simeq  -9.55\times 10^{-8},  & \quad  N_{10}^{\mathcal{Z Z}}  &\simeq  -2.62 \times 10^{-8}.
\end{align}

\subsubsection{\titlemath{$\delta \Gamma_{h \rightarrow  \mathcal Z  \mathcal Z^\star  \times \mathcal V \, \mathcal V \rightarrow \bar{\psi}^r_a \psi^r_a \, \bar{\psi}^s_b \psi^s_b}$}{h > (ZZ* x VV*) > ffff}}
\label{VVZZsection}
\begin{figure}[h!]
\includegraphics[width=0.20\textwidth]{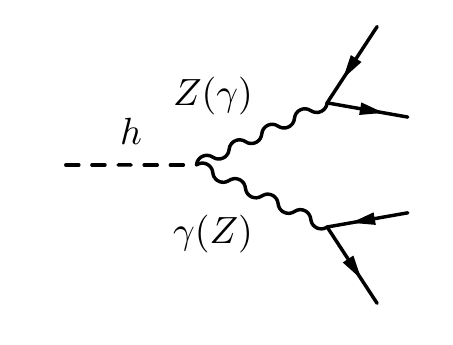}
\includegraphics[width=0.20\textwidth]{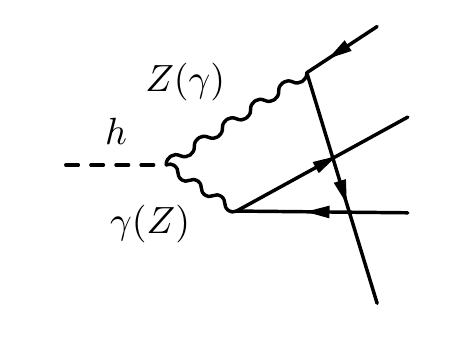}
\includegraphics[width=0.20\textwidth]{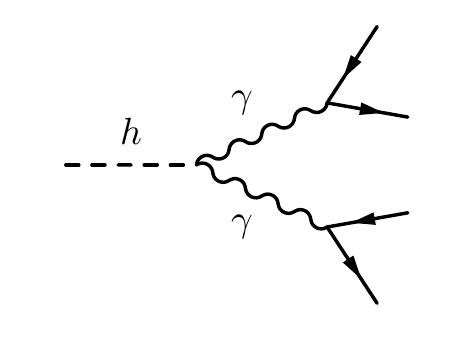}
\includegraphics[width=0.20\textwidth]{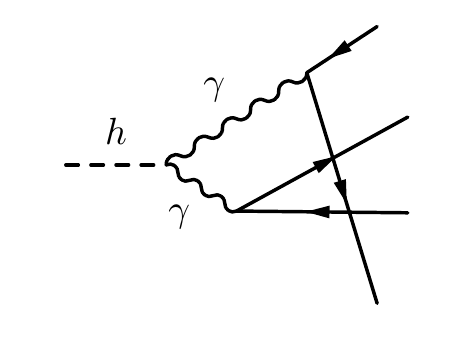}
\includegraphics[width=0.20\textwidth]{diagrams_H4f/diag_ZZ_direct.pdf}
\includegraphics[width=0.20\textwidth]{diagrams_H4f/diag_ZZ_cross.pdf}
\includegraphics[width=0.20\textwidth]{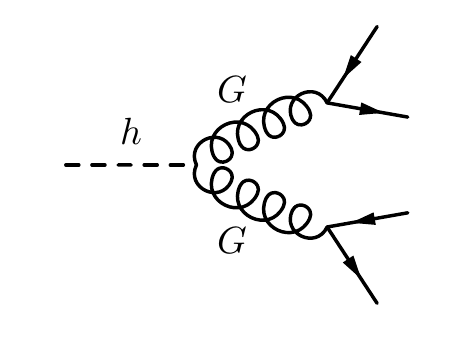}
\includegraphics[width=0.20\textwidth]{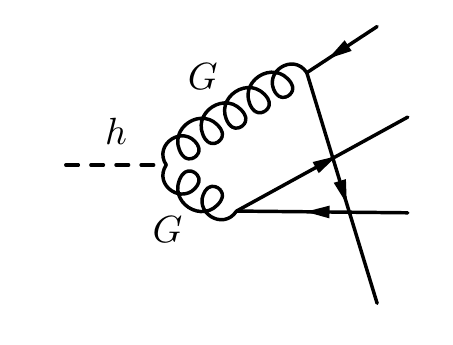}
\caption{Interference of $Z,A,G$ neutral current contributions to $h \rightarrow 4 \psi$.\label{fig:pureZA}} .
\end{figure}

In the SM, the amplitudes with $VV = \{\mathcal Z \, \mathcal A,\mathcal A \mathcal A, \mathcal G \mathcal G\}$ are loop suppressed.
This is not the case in the SMEFT in general \cite{Jenkins:2013fya}. This leads to a more significant breakdown of the narrow width approximation in the SMEFT. We include the tree level effects of these processes due to $\mathcal{L}^{(6)}$ interfering with the SM process through $\mathcal Z \mathcal Z^\star$
for a consistent LO SMEFT analysis. The corresponding diagrams are shown in Fig.~\ref{fig:pureZA} and we define
\bea\label{AZZZ}
\frac{|A^{\mathcal{ A Z Z Z}}_{\psi_a^{rr},\psi_b^{ss}}|^2}{{2 \hat{g}_Z^5 \, \hat{e}^3 \, \hat{v}^2_T (\hat{g}_2/\hat{g}_1)}} &=& Q_{\psi_a} C_{\mathcal{AZ}}
T_{\psi_a^{rr} \psi_a^{rr}}^{\mathcal A^{\alpha} (\mathcal Z^{\mu})^\dagger}\, T_{\psi_b^{ss} \psi_b^{ss}}^{\mathcal Z^{\beta} (\mathcal Z^{\mu})^\dagger}(1- \frac{\delta_a^b\, \delta_r^s}{2})
(P_{\mathcal A} \cdot P_{\mathcal Z} g^{\alpha \beta} - P_{\mathcal A}^\beta P_{\mathcal Z}^\alpha), \nn
&-& \frac{\delta_a^b\, \delta_r^s}{2} Q_{\psi_a} C_{\mathcal{AZ}} T^{\mathcal{A}^{\beta} \mathcal{Z}^{\alpha} (\mathcal{Z}^{\mu})^\dagger (\mathcal{Z}^{\mu})^\dagger}_{\psi^{rr}_a \, \psi^{rr}_a \, \psi^{rr}_a \, \psi^{rr}_a}
(P_{\mathcal A} \cdot P_{\mathcal Z} g^{\alpha \beta} - P_{\mathcal A}^\beta P_{\mathcal Z}^\alpha) + h.c.
\eea
Here the labeled momentum  $P_{\mathcal Z}$ is generated in the effective $h {\mathcal A} {\mathcal Z}$ vertex associated with $C_{H \mathcal{AZ}}$.
This interference effect in the SMEFT  with the SM neutral current mediated Higgs decay is given by
\bea
\delta \Gamma_{h \rightarrow F_3({\psi_a^{rr},\psi_b^{ss}})} &\simeq&
\frac{1}{2 \hat{M}_h} \int  {\it d ps}
|A^{\mathcal{ A Z ZZ}}_{\psi_a^r,\psi_b^s}|^2, \nonumber \\
 &=& {\frac{-\hat{g}_2}{\hat{g}_1}}  \frac{\hat{v}_T^2  \hat{g}_Z^5\, \hat{e}^3}{2 \hat{M}_h} \, \tilde{C}_{\mathcal A \mathcal Z} \, (\hat{g}_{V}^{\psi_a,SM}) \,  (\hat{g}_{+}^{\psi_b})^2 \, Q_{\psi_a}  \, N_C^{\psi_a}  \, N_C^{\psi_b} \, N_1^{\mathcal{ A Z Z Z}} \, (1- \frac{\delta_a^b\, \delta_r^s}{2}), \nonumber \\
&{-}&  {\frac{\hat{g}_2}{\hat{g}_1}}  \frac{\hat{v}_T^2  \hat{g}_Z^5\, \hat{e}^3}{2 \hat{M}_h} \, \tilde{C}_{\mathcal A \mathcal Z} \, (\hat{g}_{V}^{\psi_b,SM}) \,  (\hat{g}_{+}^{\psi_a})^2 \, Q_{\psi_b}  \, N_C^{\psi_b}  \, N_C^{\psi_a} \, N_1^{\mathcal{ A Z Z Z}} \, (1- \frac{\delta_a^b\, \delta_r^s}{2}), \nonumber \\
&{-}&  {\frac{\hat{g}_2}{\hat{g}_1}} \frac{\hat{v}_T^2  \hat{g}_Z^5 \, \hat{e}^3}{2 \hat{M}_h} \, \tilde{C}_{\mathcal A \mathcal Z} \,  \delta_{a}^b  \delta_{r}^s  \left[(\hat{g}_L^{\psi_a,SM})^3+ (\hat{g}_R^{\psi_a,SM})^3\right]  Q_{\psi_a}  \, N_C^{\psi_a} \, N_2^{\mathcal{ A Z Z Z}}.
\eea
The kinematic numbers can be approximated as
\begin{align}
N_1^{\mathcal{A Z Z Z}} &\simeq 2.7 \times 10^{-6}, & \quad
&N_2^{\mathcal{A Z Z Z}} \simeq 1.0 \times 10^{-7}.
\end{align}
We also define the following expression for interference with  $\mathcal A \mathcal A$ with the SM neutral currents
\bea\label{ZZAA}
\frac{|A^{\mathcal{ Z Z A A}}_{\psi_a^{rr},\psi_b^{ss}}|^2}{-2 \hat{g}_Z^4 \, \hat{e}^4 } &=& Q_{\psi_a} \, Q_{\psi_b} \tilde C_{\mathcal{AA}}
\left[T_{\psi_a^{rr} \psi_a^{rr}}^{\mathcal Z^{\mu} (\mathcal A^{\alpha})^\dagger}\, T_{\psi_b^{ss} \psi_b^{ss}}^{\mathcal Z^{\mu} (\mathcal A^{\beta})^\dagger} (1- \frac{\delta_a^b\, \delta_r^s}{2})\right](P^1_{\mathcal A} \cdot P^2_{\mathcal A} g^{\alpha \beta} - P^{1,\beta}_{\mathcal A} P^{2,\alpha}_{\mathcal A}), \nn
&-&  \frac{\delta_a^b\, \delta_r^s}{2} Q_{\psi_a} \, Q_{\psi_b} \tilde C_{\mathcal{AA}}
\left[T^{\mathcal{Z}^{\mu} \mathcal{Z}^{\mu} (\mathcal{A}^{\alpha})^\dagger (\mathcal{A}^{\beta})^\dagger}_{\psi^{rr}_a \, \psi^{rr}_a \, \psi^{rr}_a \, \psi^{rr}_a} \right]
(P^1_{\mathcal A} \cdot P^2_{\mathcal A} g^{\alpha \beta} - P^{1,\beta}_{\mathcal A} P^{2,\alpha}_{\mathcal A}) + h.c.
\eea

This result contributes to a partial width as
\bea
\delta \Gamma_{h \rightarrow F_4(\psi_a^{rr},\psi_b^{ss})} &\simeq&
\frac{1}{2 \hat{M}_h} \int  {\it d ps}
|A^{\mathcal{ Z Z A A}}_{\psi_a^r,\psi_b^s}|^2, \nonumber \\
 &=& \frac{\hat{v}_T^2  \hat{g}_Z^4 \, \hat{e}^4  }{2 \hat{M}_h} \tilde{C}_{\mathcal A \mathcal A} \, \left[(\hat{g}_V^{\psi_a,SM}) \, Q_{\psi_a} \, N_C^{\psi_a} \, (\hat{g}_V^{\psi_b,SM}) \,  N_C^{\psi_b} \,  Q_{\psi_b}  \right] \, (1- \frac{\delta_a^b\, \delta_r^s}{2}) \,
\, N_1^{\mathcal{Z Z A A}}, \nonumber \\
&+& \frac{\hat{v}_T^2  \hat{g}_Z^4 \, \hat{e}^4  }{2 \hat{M}_h} \delta_{a}^b  \delta_{r}^s \tilde{C}_{\mathcal A \mathcal A} \, (\hat{g}_+^{\psi_a,SM})^2 \, N_C^{\psi_a} Q_{\psi_a}^2  \, N_2^{\mathcal{Z Z A A}}.
\eea
where the numerical results can be approximated as
\begin{align}
N_1^{\mathcal{Z Z A A}} &\simeq -1.9 \times 10^{-6}, &  \quad
&N_2^{\mathcal{Z Z A A}} \simeq -1.9 \times 10^{-7}.
\end{align}

The operator $\mathcal{Q}_{HG} = H^\dagger H G_{\mu \nu}^A  G_A^{\mu \nu}$ has a tree level interference contribution to
$h \rightarrow  F_5(\psi_a^{rr},\psi_b^{ss})$ for quark final states. Due to the $\rm SU(3)$ generator of the gluon coupling to fermions, only the single trace form is present, and the result is
\bea\label{ZZGG}
\frac{|A^{\mathcal{ Z Z G G}}_{\psi_a^{rr},\psi_b^{ss}}|^2}{-2 \hat{g}_Z^4 \, \hat{g}_s^2} &=&
-  \frac{\delta_a^b\, \delta_r^s}{8}  C_{HG}
\left[T^{\mathcal{Z}^{\mu} \mathcal{Z}^{\mu} (\mathcal{G}^{\alpha})^\dagger (\mathcal{G}^{\beta})^\dagger}_{\psi^{rr}_a \, \psi^{rr}_a \, \psi^{rr}_a \, \psi^{rr}_a} \right]
(P^1_{\mathcal G} \cdot P^2_{\mathcal G} g^{\alpha \beta} - P^{1,\beta}_{\mathcal G} P^{2,\alpha}_{\mathcal G}) + h.c.
\eea
which contributes to a $h \rightarrow   F_5(\psi_a^{rr},\psi_b^{ss})$ partial width as
\bea
\delta \Gamma_{h \rightarrow F_5(\psi_a^{rr},\psi_b^{ss})} &\simeq&
\frac{1}{2 \hat{M}_h} \int  {\it d ps}
|A^{\mathcal{ Z Z G G}}_{\psi_a^r,\psi_b^s}|^2, \nonumber \\
 &=&  \frac{\hat{v}_T^2  \hat{g}_Z^4 \, \hat{g}_s^2  }{2 \hat{M}_h} \delta_{a}^b  \delta_{r}^s \tilde{C}_{H G} \, (\hat{g}_+^{\psi_a,SM})^2  \, N_2^{\mathcal{Z Z GG}}.
\eea
The numerical results can be approximated as $ N_2^{\mathcal{Z Z G G}} \simeq -7.6 \times 10^{-7} = 4 \, N_2^{\mathcal{Z Z A A}}$.

\subsubsection{IR behavior when interfering with tree level photon exchange}
The numerical evaluation of the four body phase space integrations in the cases with intermediate photons are more challenging than the remaining numerical evaluations.
All the kinematic numbers are extracted with a direct numerical evaluation with the Vegas Monte Carlo integration algorithm and the CUBA package \cite{Hahn:2004fe} and cross-checked both with the RAMBO phase space generator and with the numbers extracted from massless simulations in \MG\ with {\tt SMEFTsim}.
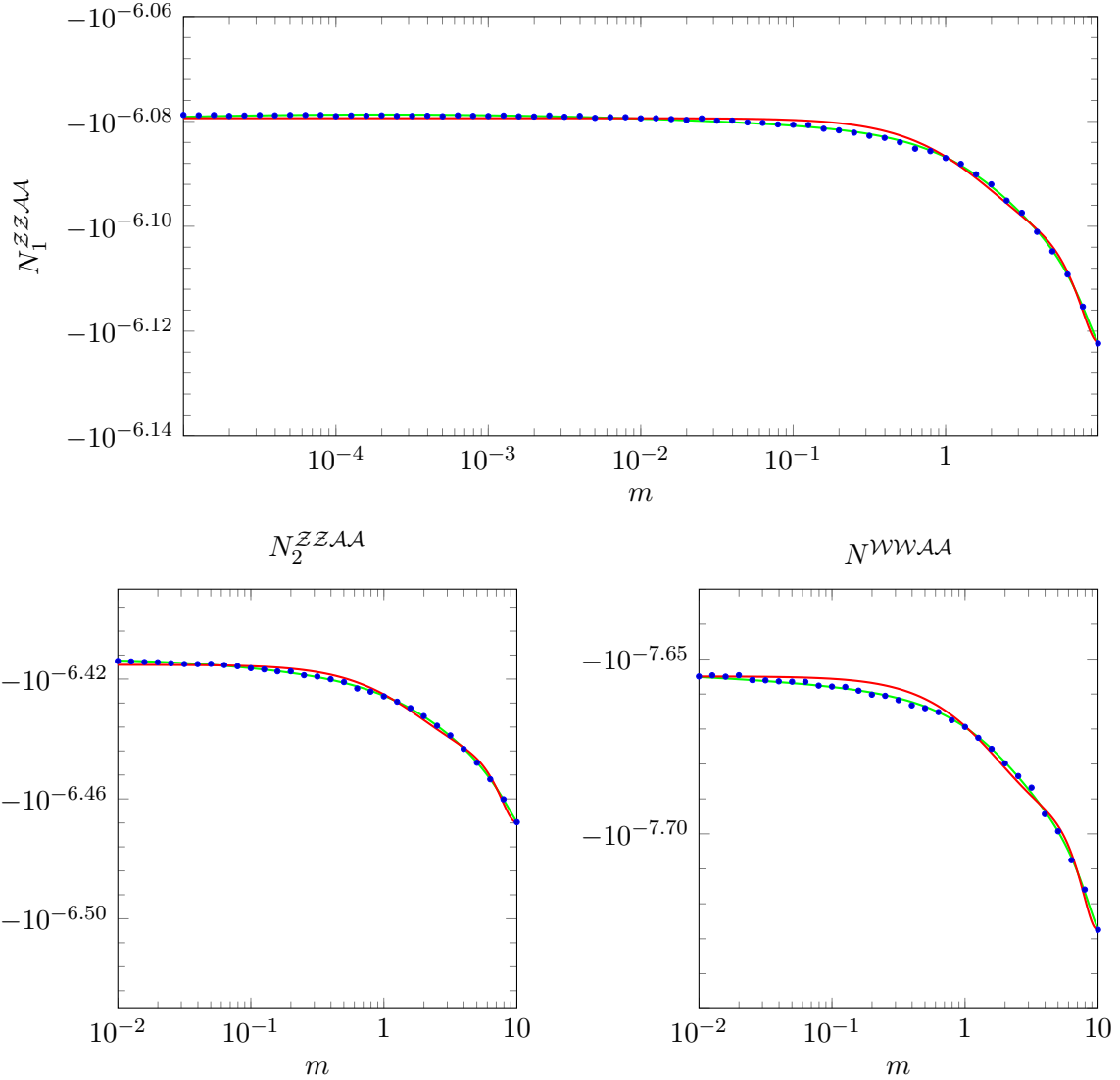
\begin{figure}
\begin{tabular}{cc}
\multicolumn{2}{c}{
\begin{tikzpicture}\begin{semilogxaxis}[
    width=.9\textwidth,
    height=7.275cm,
    minor x tick num = 3,
    minor y tick num = 4,
    ymin = -6.14,ymax = -6.06,
    xmin = .00001,xmax = 10,
    xlabel={$m$},
    xtick = {.00001,.0001,.001,.01,.1,1,10},
    xticklabels = {,$10^{-4}$,$10^{-3}$,$10^{-2}$,$10^{-1}$,$1$,},
    ylabel={$N_1^{\mathcal{ZZAA}}$},
    ytick = {-6.06,-6.08,-6.10,-6.12,-6.14},
    yticklabels = {$-10^{-6.06}$,$-10^{-6.08}$,$-10^{-6.10}$,$-10^{-6.12}$,$-10^{-6.14}$}
        ]
    \addplot+ [blue,only marks,mark size=1] table {data/ZZAA1results_percent1.txt};
    \addplot+ [green,mark=none,thick] table {data/ZZAA1wlogs.txt};
    \addplot+ [red,mark=none,thick] table {data/ZZAA1IRfinite.txt};
\end{semilogxaxis}\end{tikzpicture}
}\\
\begin{tikzpicture}\begin{semilogxaxis}[
    height=7.275cm,
    width=.45\textwidth,
    minor x tick num = 3,
    minor y tick num = 4,
    ymin = -6.53,ymax = -6.39,
    xmin = .01,xmax = 10,
    xlabel={$m$},
    xtick = {.01,.1,1,10},
    xticklabels = {$10^{-2}$,$10^{-1}$,$1$,$10$},
    title={$N_2^{\mathcal{ZZAA}}$},
    ytick = {-6.54,-6.50,-6.46,-6.42,-6.38},
    yticklabels = {,$-10^{-6.50}$,$-10^{-6.46}$,$-10^{-6.42}$,$-10^{-6.38}$}
        ]
    \addplot+ [blue,only marks,mark size=1] table {data/ZZAA2results_percent1.txt};
    \addplot+ [green,mark=none,thick] table {data/ZZAA2wlogs.txt};
    \addplot+ [red,mark=none,thick] table {data/ZZAA2IRfinite.txt};
\end{semilogxaxis}\end{tikzpicture}
&
\begin{tikzpicture}\begin{semilogxaxis}[
    height=7.275cm,
    width=.45\textwidth,
    minor x tick num = 3,
    minor y tick num = 4,
    ymin = -7.75,ymax = -7.63,
    xmin = .01,xmax = 10,
    xlabel={$m$},
    xtick = {.01,.1,1,10},
    xticklabels = {$10^{-2}$,$10^{-1}$,$1$,$10$},
    title={$N^{\mathcal{WWAA}}$},
    ytick = {-7.75,-7.70,-7.65},
    yticklabels = {,$-10^{-7.70}$,$-10^{-7.65}$}
        ]
    \addplot+ [blue,only marks,mark size=1] table {data/WWAAresults_percent1.txt};
    \addplot+ [green,mark=none,thick] table {data/WWAAwlogs.txt};
    \addplot+ [red,mark=none,thick] table {data/WWAAIRfinite.txt};
\end{semilogxaxis}\end{tikzpicture}
\end{tabular}
\caption{Top: $N_1^{\mathcal{ZZAA}}$, Bottom Left: $N_2^{\mathcal{ZZAA}}$, Bottom Right: $N^{WWAA}$, as functions of the fermions' mass $m$. The green line includes the IR divergent $\log(m/M_H)$ and $\log(m/M_H)^2$ dependence of Eq.~\eqref{eq:empiricalfit} while the red line neglects these contributions. The top plot shows a larger mass range to demonstrate the approximately constant behavior for lower $m$, this behavior is observed for $N_{1,2}^{\mathcal{ZZAA}}$ as well, but is cut off from the plots to better show the mass dependence and quality of the fit including the IR divergent contributions.
}\label{fig:empiricalfermion}
\end{figure}
The numerical integration in Vegas was evaluated using both massless and massive phase space boundaries and validated with two different phase space variable sets and numerical methodologies.
In the case of extracting an interference effect with a double photon pole, i.e. the results giving $N_1^{\mathcal{Z Z A A}}, N_2^{\mathcal{Z Z A A}}$, RAMBO did not converge with sufficient numerical accuracy to afford a cross-check of results and the \MG\ simulation was found to be subject to significant numerical uncertainties in the massless fermions case. Retaining fermion masses overcomes the latter issue, and allowed us to confirm the Vegas results.

The reason these results are numerically challenging to determine is due to the IR behavior of the corresponding phase space in the massless fermion limit.
The phase space volume in part is
\bea
\int_{0}^{m_h^2/2} d \kappa_{12} \int_{0}^{(m_h - \sqrt{2 \kappa^2_{12}})^2/2} \hspace{-2cm} d \kappa_{34}  \hspace{1cm} \frac{1}{(\kappa^2_{12} + i \epsilon)} \frac{1}{(\kappa^2_{34} + i \epsilon)} \cdots
\eea
where the photon invariant masses are $\kappa^2_{12},\kappa^2_{34}$.
A logarithmic dependence on the final state fermion masses results  when integrating the phase space. We believe this is due to soft and collinear emissions of the final state fermions.
For example, consider the massless fermion limit. The boundaries of the phase space volume are defined by
\bea
- \kappa_{12}^2 &<& 0, \\
2 \kappa^2_{12} \, \kappa^2_{23} \, \kappa_{13} &<& 0, \\
(\kappa_{13} \, \kappa_{24} - \kappa_{14} \kappa_{23})^2 &<& 0
\eea
in this case. (See Appendix A for details on the phase space integration.) The collinear momentum configuration $\kappa_{12} \rightarrow 0$ while $\kappa_{34} \rightarrow 0$ on the phase space boundary leads to fermion mass dependent logarithmic behavior.
As does the case where $\kappa_{12} \rightarrow 0$, while $\kappa_{14} = \kappa_{13}$ and $\kappa_{24} = \kappa_{23}$. These momentum configurations are also allowed when fermion masses are included in the final states,
but the presence of such masses softens the logarithmic singularity into logarithmic and dilog dependences on the fermion masses. An empirical fit to the dependence on the fermion masses in the result is shown
in Fig.~\ref{fig:empiricalfermion}. The functional form fit to was
\begin{eqnarray}
f(m)&=&c_1\log\frac{m^2}{M_H^2}+c_2 \frac{m^2}{M_H^2}\log\frac{m^2}{M_H^2}+c_3\log^2\left[\frac{m^2}{M_H^2}\right]+c_4\frac{m^2}{M_H^2}\log^2 \left[\frac{m^2}{M_H^2}\right]
\nonumber\\
&+&c_5\, {\rm Li}_2\frac{m^2}{M_H^2}+c_6\frac{m^2}{M_H^2}{\rm Li}_2\frac{m^2}{M_H^2}\label{eq:empiricalfit}
\end{eqnarray}
with free parameters $c_i$.A constant term was also included and determined in the fit. This expression should not be understood to imply that the massless limit is formally divergent, as cancellations can occur between
the logarithmic and polylogarithmic terms shown. The massless limit is show in Fig.\ref{fig:empiricalfermion}, and is empirically found to be finite in our numerical fit.

These fermion mass effects are numerically small enough to be neglected in the LO analysis included here, so long as an appropriate theoretical error is included in the corresponding theoretical predictions.
In the case of the decay through $\mathcal{A} \mathcal{Z}$ the IR limit is sufficiently regulated by the mass of the $\mathcal{Z}$ to further soften the logarithmic behavior.

It is important to note the interplay of these regions of phase space, where fermion masses regulate IR behavior in this manner, also coincide with
the final state photon being reconstructed in the detector, not the experimental case where the photon has converted to two distinct final state fermions.
As such, the regulation of phase space is practically cut off by detector effects and the signal definition, in addition to fermion masses,  when this particular decay is studied experimentally.
We stress that the results in Sections~\ref{AAsection},~\ref{ZAsection} are not a double counting even in this collinear limit. The interference effects
in each case are with distinct processes, at tree or the loop level in the SM.

\subsubsection{\titlemath{$\delta \Gamma_{h \rightarrow \mathcal W \mathcal W^\star \times \mathcal V \mathcal V  \rightarrow \bar{\psi}^s_a \psi^s_a \, \bar{\psi}^r_b \psi^r_b}$}{h > (WW* x VV*) > ffff}}

\begin{figure}[ht!]
\includegraphics[width=0.20\textwidth]{diagrams_H4f/diag_ZZ_direct.pdf}
\includegraphics[width=0.20\textwidth]{diagrams_H4f/diag_ZZ_cross.pdf}
\includegraphics[width=0.20\textwidth]{diagrams_H4f/diag_AA_direct.pdf}
\includegraphics[width=0.20\textwidth]{diagrams_H4f/diag_AA_cross.pdf}
\includegraphics[width=0.20\textwidth]{diagrams_H4f/diag_ZA_direct.pdf}
\includegraphics[width=0.20\textwidth]{diagrams_H4f/diag_ZA_cross.pdf}
\includegraphics[width=0.20\textwidth]{diagrams_H4f/diag_WW_direct.pdf}
\includegraphics[width=0.20\textwidth]{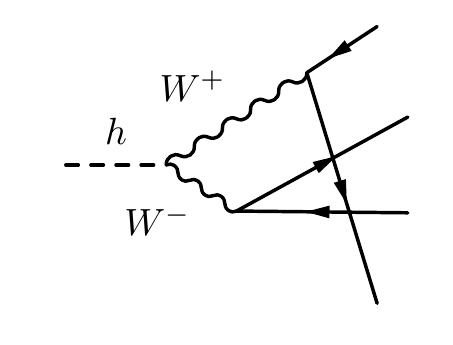}
\includegraphics[width=0.20\textwidth]{diagrams_H4f/diag_ZE_direct.pdf}
\includegraphics[width=0.20\textwidth]{diagrams_H4f/diag_ZE_cross.pdf}
\includegraphics[width=0.20\textwidth]{diagrams_H4f/diag_WpE_direct.pdf}
\includegraphics[width=0.20\textwidth]{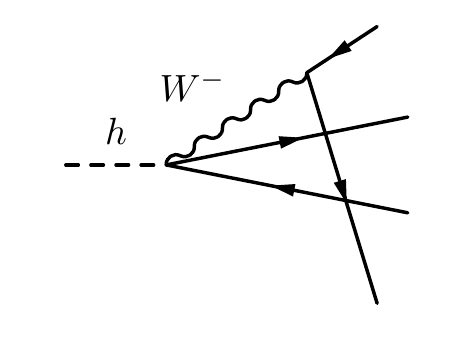}
\includegraphics[width=0.20\textwidth]{diagrams_H4f/diag_GG_direct.pdf}
\includegraphics[width=0.20\textwidth]{diagrams_H4f/diag_GG_cross.pdf}
\caption{Neutral-charged current interference contributions to $h \rightarrow 4 \psi$. The SM diagrams also represent $\mathcal{O}(\bar{v}_T^2/\Lambda^2)$ operator insertions perturbing the SM prediction
with the same pole structure as the SM.\label{fig:pureWV}}
\end{figure}

There is also a contribution due to the interference of the charged and neutral currents, where $\mathcal V = \{\mathcal{Z},\mathcal{A}\}$ in this subsection.
The corresponding diagrams are shown in  Fig.~\ref{fig:pureWV}.

In the SM, as the couplings to $\mathcal{ZA},\mathcal{AA},\mathcal{GG}$  are loop suppressed, and the LO expression is given by\footnote{The presence of a minus sign again follows from Fermi statistics, see Ref.~\cite{Bredenstein:2006rh}.}
\bea
|A_{SM}^{\mathcal{WWVV}}|^2 &=& -\frac{\hat{g}_2^2 \bar{g}_2^2 \hat{g}_Z^2 \bar{g}_Z^2  \bar{v}^2_T}{8}  T^{ \mathcal{W}_{\mu} (\mathcal{Z}^{\nu})^\dagger  (\mathcal{W}^{\mu}) (\mathcal{Z}^{\nu})^\dagger}_{\psi_1^{rs} \psi_{a,L}^{ss}  (\psi_1^{rs})^\dagger \psi_{b,L}^{rr}} (k_{ij}^2,k_{jk}^2,k_{kl}^2,k_{li}^2)  + h.c. \\
&=& 2 \hat{g}_2^2 \bar{g}_2^2 \hat{g}_Z^2 \bar{g}_Z^2  \bar{v}^2_T  N_C^{\psi_1} |(\bar g_{L,rs}^{W_\pm,\psi_1})|^2  (\bar g_{L,ss}^{\psi_a}) (\bar g_{L,rr}^{\psi_b})
\frac{k_{i} \cdot k_{k} k_{j} \cdot k_{l}}{[k^2_{ij},\mathcal{W}][k^2_{kl},\mathcal{W}][k^2_{jk},\mathcal{Z}]^\star [k^2_{li},\mathcal{Z}]^\star} + h.c. \nonumber
\eea
leading to the SM result
\bea
\Gamma_{h \rightarrow F_6({\psi_a^{rr},\psi_b^{ss}})} &=& \Gamma_{\psi_1^{rs}, \psi_a^r,\psi_b^s}^{\mathcal{W} \mathcal{Z}}   N^{\mathcal{W} \mathcal{V}}_1, \quad
 \Gamma_{\psi_1^{rs}, \psi_a^r,\psi_b^s}^{\mathcal{W} \mathcal{Z}} = \frac{16 \, \hat{g}_2^2  \hat{g}_Z^2 \, N_C^{\psi_1} \hat{M}_{W}^2 \hat{M}_{Z}^2}{\hat{M}_h \, \hat{v}_T^2} |(\bar g_{L,rs}^{W_\pm,\psi_1})|^2  (\bar g_{L,ss}^{\psi_a}) (\bar g_{L,rr}^{\psi_b}) , \nn
\eea
with the kinematic number
\bea
N^{\mathcal{W} \mathcal{V}}_1 = \int  {\it d ps} \frac{k_{i} \cdot k_{k} k_{j} \cdot k_{l}}{[k^2_{ij},\mathcal{W}][k^2_{kl},\mathcal{W}][k^2_{jk},\mathcal{Z}]^\star [k^2_{li},\mathcal{Z}]^\star}  + h.c. \simeq 1.33 \times 10^{-7}.
\eea

The SMEFT corrections to charged-neutral current interference are defined as $\delta \Gamma_{h \rightarrow F_5({\psi_a^{ss},\psi_b^{rr}})}$ and is given by
\bea
&\,&\delta \Gamma_{\psi_1^{rs}, \psi_a^r,\psi_b^s}^{\mathcal{W} \mathcal{Z}}   N^{\mathcal{W} \mathcal{V}}_1
+  \frac{4  \, C_{\substack{H \psi_{1}\\ rs}}^{(3)}  \Gamma_{\psi_1^{rs}, \psi_a^r,\psi_b^s}^{\mathcal{W} \mathcal{Z}} }{\hat{g}_2^2 \, (\bar g_{L,rs}^{\mathcal{W},\psi_1})} \int  {\it d ps} \,\frac{k_{i} \cdot k_{k} k_{j} \cdot k_{l}  \left([k^2_{ij}, \mathcal{W}]+[k^2_{kl}, \mathcal{W}]\right)}{D[k^2_{ij},k^2_{kl},k^2_{jk},k^2_{li}]} ,
\nonumber \\
&+&  \frac{4}{\hat{g}_Z^2} \int  {\it d ps} \, \frac{k_{i} \cdot k_{k} k_{j} \cdot k_{l} \,  \Gamma_{\psi_1^{rs}, \psi_a^r,\psi_b^s}^{\mathcal{W} \mathcal{Z}}}
{D[k^2_{ij},k^2_{kl},k^2_{jk},k^2_{li}]}
\left(\frac{\delta C_{\substack{H L\\ rr}}^{\psi_a}}{(\bar g_{L,rr}^{\psi_a})} [k^2_{jk}, \mathcal{Z}^\dagger]+\frac{\delta C_{\substack{H L\\ ss}}^{\psi_b}}{(\bar g_{L,ss}^{\psi_b})}[k^2_{li}, \mathcal{Z}^\dagger]\right), \nonumber\\
&+& \int  {\it d ps} \, \frac{k_{i} \cdot k_{k} k_{j} \cdot k_{l} \,  \Gamma_{\psi_1^{rs}, \psi_a^r,\psi_b^s}^{\mathcal{W} \mathcal{Z}} \left(\d  D^{W}(k^2_{ij})+ \d  D^{Z,\star}(k^2_{jk})+ \d  D^{W}(k^2_{jk})+ \d  D^{Z,\star}(k^2_{li})\right)}
{D[k^2_{ij},k^2_{kl},k^2_{jk},k^2_{li}]},
\eea
and also
\bea
&+& \frac{{8} \hat{g}_2^2 \, \hat{g}_Z^2}{2 \hat{M}_h^2} \left(
 \hat{g}_Z^2 \, \tilde{C}_{HW} \int  {\it d ps} T^{ \mathcal{W}_{\alpha} (\mathcal{Z}^{\nu})^\dagger  (\mathcal{W}^{\beta}) (\mathcal{Z}^{\nu})^\dagger}_{\psi_1^{rs} \psi_{a,L}^{ss}  (\psi_1^{rs})^\dagger \psi_{b,L}^{rr}}
+  \hat{g}_2^2 \, \tilde C_{\mathcal{Z} \mathcal{Z}} \int  {\it d ps} T^{ \mathcal{W}_{\mu} (\mathcal{Z}^{\alpha})^\dagger  (\mathcal{W}^{\mu}) (\mathcal{Z}^{\beta})^\dagger}_{\psi_1^{rs} \psi_{a,L}^{ss}  (\psi_1^{rs})^\dagger \psi_{b,L}^{rr}} \right) K_{\alpha \beta}, \nn
&+& \frac{{4 } \hat{g}_2^4 \, {\hat{e}^2}}{2 \hat{M}_h^2} Q_{\psi_b} \left(
{\hat{e}^2}\, Q_{\psi_a} \, {2} \, \tilde{C}_{\mathcal{A} \mathcal{A}} \int  {\it d ps} T^{ \mathcal{W}_{\mu} (\mathcal{A}^{\alpha})  (\mathcal{W}^{\mu}) (\mathcal{A}^{\beta})}_{\psi_1^{rs} \psi_{a,L}^{ss}  (\psi_1^{rs})^\dagger \psi_{b,L}^{rr}}
{- \frac{\hat{g}_2}{\hat{g}_1}} \, \hat{g}_Z^2 \, \tilde{C}_{\mathcal{A} \mathcal{Z}}  \int  {\it d ps} T^{ \mathcal{W}_{\mu} (\mathcal{A}^{\alpha})  (\mathcal{W}^{\mu}) (\mathcal{Z}^{\beta})^\dagger}_{\psi_1^{rs} \psi_{b,L}^{ss}  (\psi_1^{rs})^\dagger \psi_{a,L}^{rr}}
  \right) K_{\alpha \beta},  \nonumber \\
&+&  \frac{{2} \hat{g}_2^4 \, \hat{g}_s^2}{2 \hat{M}_h^2}\,  \tilde{C}_{HG} \int  {\it d ps} T^{ \mathcal{W}_{\mu} (\mathcal{G}^{\alpha})  (\mathcal{W}^{\mu}) (\mathcal{G}^{\beta})}_{\psi_1^{rs} \psi_{a,L}^{ss}  (\psi_1^{rs})^\dagger \psi_{b,L}^{rr}} K_{\alpha \beta}
+ \frac{{4 }  \hat{g}_2^4 {\hat{e}}^2 \hat{g}_Z^2}{2 \hat{M}_h^2} \, Q_{\psi_a}  \, \tilde{C}_{\mathcal{A} \mathcal{Z}} \int  {\it d ps} T^{ \mathcal{W}_{\mu} (\mathcal{A}^{\alpha})  (\mathcal{W}^{\mu}) (\mathcal{Z}^{\beta})^\dagger}_{\psi_1^{rs} \psi_{a,L}^{ss}  (\psi_1^{rs})^\dagger \psi_{b,L}^{rr}}  K_{\alpha \beta} + h.c \nonumber
\eea
where $K^{\alpha \beta} =  (k_{jk} \cdot k_{il} g^{\alpha \beta} - k_{li}^{\alpha} k_{jk}^{\beta})$, $D[k^2_{ij},k^2_{kl},k^2_{jk},k^2_{li}] = [k^2_{ij},\mathcal{W}][k^2_{kl},\mathcal{W}][k^2_{jk},\mathcal{Z}]^\star [k^2_{li},\mathcal{Z}]^\star$
and
\bea
\frac{\delta \hat{\Gamma}_{\psi_1^{rs}, \psi_a^r,\psi_b^s}^{\mathcal{W} \mathcal{Z}}}{\hat{\Gamma}_{\psi_1^{rs}, \psi_a^r,\psi_b^s}^{\mathcal{W} \mathcal{Z}}}  &=&
\left[- \frac{ \delta m_{\mathcal{Z}}^2}{\hat{M}^2_Z} + \frac{ \delta M_W^2}{\hat{M}^2_W} - \sqrt{2} \delta G_F + 2 \ckin + \tilde{C}_{HD}  + 2 c_{\hat{\theta}} s_{\hat{\theta}} \tilde{C}_{HWB}\right], \nn
&+&  \left[ 2 \frac{{\rm Re}[ \d  (g_L^{\mathcal W^+,\psi_{1}})_{rs}]}{(g_L^{\mathcal W^+,\psi_{1}})_{rs}^{SM}} + \frac{\delta g_{L,ss}^{\psi_a}}{\hat{g}_{L,ss}^{\psi_a,SM}}
+ \frac{\delta g_{L,rr}^{\psi_b}}{\hat g_{L,rr}^{\psi_a,SM}} \right].
\eea
This expression numerically reduces to
\bea
\frac{\delta \Gamma_{h \rightarrow F_6({\psi_a^{ss},\psi_b^{rr}})}}{\hat{\Gamma}_{\psi_1^{rs}, \psi_a^r,\psi_b^s}^{\mathcal{W} \mathcal{Z}}}
&\simeq& \frac{\delta \hat{\Gamma}_{\psi_1^{rs}, \psi_a^r,\psi_b^s}^{\mathcal{W} \mathcal{Z}}}{\hat{\Gamma}_{\psi_1^{rs}, \psi_a^r,\psi_b^s}^{\mathcal{W} \mathcal{Z}}}    N^{\mathcal{W} \mathcal{V}}_1
+ \frac{\tilde{C}_{\substack{H \psi_{1}\\ rs}}^{(3)} }{ (\bar g_{L,rs}^{\mathcal{W},\psi_1})}  \frac{N^{\mathcal{W} \mathcal{V}}_2}{\hat{g}_2^2}
+ \left[\frac{\delta \tilde{C}_{\substack{H L\\ rr}}^{\psi_a}}{(\bar g_{L,rr}^{\psi_a})} + \frac{\delta \tilde{C}_{\substack{H L\\ ss}}^{\psi_b}}{(\bar g_{L,ss}^{\psi_b})}\right]   \frac{N^{\mathcal{W} \mathcal{V}}_3}{ \hat{g}_Z^2}, \nonumber \\
&+&
\frac{\tilde{C}_{H W}}{\hat{g}_2^2} \,  N^{\mathcal{W} \mathcal{V}}_4
+ \frac{\tilde{C}_{\mathcal{Z} \mathcal{Z}}}{\hat{g}_Z^2} \,   N^{\mathcal{W} \mathcal{V}}_5
+ \frac{ \hat{e}^4 \, Q_{\psi_a}  Q_{\psi_b}}{\hat{g}_Z^4 \,  (\bar g_{L,ss}^{\psi_a}) (\bar g_{L,rr}^{\psi_b})}\,  \tilde{C}_{\mathcal{A} \mathcal{A}} \,  N^{\mathcal{W} \mathcal{V}}_6, \nn
&+& \frac{\hat{g}_s^2}{\hat{g}_Z^4 \,  (\bar g_{L,ss}^{\psi_a}) (\bar g_{L,rr}^{\psi_b})} \, \tilde{C}_{H G} \,  N^{\mathcal{W} \mathcal{V}}_7
- {\frac{\hat{g}_2}{\hat{g}_1} \frac{ \hat{e}^3 }{\hat{g}_Z^3}} \, \left[\frac{Q_{\psi_a}}{\bar g_{L,ss}^{\psi_a}}+  \frac{Q_{\psi_b}}{\bar g_{L,ss}^{\psi_b}} \right]\,  \tilde{C}_{\mathcal{A} \mathcal{Z}} \,  N^{\mathcal{W} \mathcal{V}}_8,  \nonumber \\
&+&  \left[N^{\mathcal{W} \mathcal{V}}_9  \frac{\delta \Gamma_Z}{\hat{\Gamma}_Z} +  N^{\mathcal{W} \mathcal{V}}_{10}  \frac{\delta \Gamma_W}{\hat{\Gamma}_W}\right]
+ \left[N^{\mathcal{W} \mathcal{V}}_{11}  \frac{\delta M_Z^2}{\hat{M}_Z^2} +  N^{\mathcal{W} \mathcal{V}}_{12}  \frac{\delta M_W^2}{\hat{M}_W^2} \right].
\eea
With the (inclusive) kinematic numbers
\begin{align}
N^{\mathcal{W} \mathcal{V}}_2 &\simeq -7.21  \times 10^{-8}, & \quad  N^{\mathcal{W} \mathcal{V}}_3 &\simeq -5.01  \times 10^{-8}, \nn
N^{\mathcal{W} \mathcal{V}}_4 &\simeq -2.8  \times 10^{-8}, & \quad  N^{\mathcal{W} \mathcal{V}}_5 &\simeq -2.6  \times 10^{-8}, \nn
N^{\mathcal{W} \mathcal{V}}_6 &= \frac{N^{\mathcal{W} \mathcal{V}}_7}{4} \simeq  - 1.8 \times 10^{-7} , & \quad  N^{\mathcal{W} \mathcal{V}}_8 &\simeq 5.2 \times 10^{-8}, \nn
N^{\mathcal{W} \mathcal{V}}_9 &\simeq - 1.42 \times 10^{-11}, & \quad   N^{\mathcal{W} \mathcal{V}}_{10} &\simeq -1.37 \times 10^{-10}, \nn
N^{\mathcal{W} \mathcal{V}}_{11} &\simeq 4.77 \times 10^{-10}, & \quad
N^{\mathcal{W} \mathcal{V}}_{12} &\simeq 7.01  \times 10^{-10}.
\end{align}
The kinematic numbers $N^{\mathcal{W} \mathcal{V}}_{6,7,8}$ weak logarithmic dependence on the final state masses is neglected
here.

\section{Numerical results and analysis of the contributions to \titlemath{$h\to 4f$}{h > 4f}}\label{sec:Pheno}

Taking into account all of these results, the total Higgs width combining these decays is given by
\bea
\Gamma_{h,full}^{SMEFT} = \Gamma_h^{SM} + \hspace{-0.5cm}\sum_{\psi=\{\substack{u,c,d,s, \\ b,e,\mu,\tau} \}} \hspace{-0.3cm} \delta \Gamma_{h \rightarrow \bar{\psi} \psi} + \delta \Gamma_{h \rightarrow \mathcal{A} \mathcal{A}}
+ \delta \Gamma_{h \rightarrow \mathcal{Z} \mathcal{A}} + \delta \Gamma_{h \rightarrow gg} + \hspace{-0.2cm}\sum_{\psi_{1,2,3,4}} \hspace{-0.1cm} \delta \Gamma_{h \rightarrow \bar{\psi}_1 \psi_2 \bar{\psi}_3 \psi_4}, \nonumber \\
\eea
where $\sum_{\psi_{1,2,3,4}}$ indicates a sum over all possible final state fermions kinematically allowed. Due to the experimental definition of $\mathcal{A} \mathcal{A}, \mathcal{Z} \mathcal{A},gg$
final states, there is no double counting.
For reference, the total SM Higgs width is~\cite{deFlorian:2016spz}
\begin{equation}
\Gamma_{h,full}^{SM} = 4.100~{\rm MeV}.
\end{equation}
These corrections lead to branching ratio modifications of the Higgs decaying to a set of final states $S$. We define this  branching ratio in the SMEFT as
\bea
{\rm Br}^{SMEFT}_{h \rightarrow S} = {\rm Br}^{SM}_{h \rightarrow S}\left[1 + \frac{\delta \Gamma_{h \rightarrow S}}{\Gamma^{SM}_{h \rightarrow S}} - \frac{\sum_S \delta \Gamma_{h \rightarrow S}}{\sum_S\Gamma^{SM}_{h \rightarrow S}} \right].
\eea
The SMEFT branching ratio defined in this way retains the leading order interference effect of $\mathcal{A}^{(6)}(h\rightarrow S)$ interfering with $\mathcal{A}^{SM}(h\rightarrow S)$.
The SM Higgs has suppressions by small Yukawa couplings $Y_b$ in dominant SM decays at leading order in perturbation theory, and phenomenologically important contributions due to one loop decays.
Retaining the leading order $\mathcal{A}^{SM}\times \mathcal{A}^{(6),\star}(h\rightarrow S)$ SMEFT effects retains a subset of Yukawa coupling suppressed, and $\mathcal{O}(1/16 \pi^2 \, \Lambda^2)$ corrections.
Obviously this encourages developing the SMEFT to include higher order corrections in time, to retain a full set of terms at each mixed order in perturbation theory. As such results are not completely available at this time
we perform a LO analysis in this work retaining the leading $\mathcal{A}^{SM}\times \mathcal{A}^{(6),\star}(h\rightarrow S)$ contributions in each case.

The expressions derived in the previous sections allow to infer the relative SMEFT correction to each partial Higgs decay width:
\begin{equation}\label{eq:PWshift}
\frac{\d\Gamma_{h\to S}}{\Gamma_{h\to S}^{SM,{\rm tree}}} = 1+ \sum_i a_i^{(S)}\, \tilde{C}_i \,
\end{equation}
where $a_i^{(S)}$ are input scheme-dependent functions of the SM parameters. The expression in Eq.~\eqref{eq:PWshift} represents the leading relative SMEFT correction for each channel: in a realistic numerical analysis, it can be assigned to the most accurate prediction available for $\Gamma_{h\to S}^{SM}$, leading to the numerical estimate
\begin{equation}\label{eq.general_W_parameterization}
\Gamma_{h\to S}^{SMEFT} = \Gamma_{h\to S}^{SM}\left[1+  \sum_i a_i^{(S)}\, \tilde{C}_i\right]\,.
\end{equation}
The numerical values of the coefficients $a_i^{(S)}$ found for all the decay channels considered are reported in Tables~\ref{tab:results_23body_MW} - \ref{tab:results_4f_a} in Appendix~\ref{app:tables}, with the numerical inputs reported in Table~\ref{tab:inputs}. Note that the fermion masses $M_{b,c,\tau}$ were used for the $h\to\bar ff$ channels but were neglected in the $h\to4f$ estimates.
The CKM matrix is always taken to be the unit matrix, thereby omitting flavor changing channels.
Finally, the top quark mass is relevant for the numerical evaluation of the SM Higgs couplings to $gg,\,Z\g,\,\g\g$ (see Sec.~\ref{AAsection} - \ref{ZAsection}).
In this section we refer only to results obtained with the $\{\hat M_W, \hat M_Z, \hat G_F, \hat M_h\}$ input parameter scheme for concreteness. We find the main considerations illustrated here to be also valid for the $\{\hat \alpha_{ew}, \hat M_Z, \hat G_F, \hat M_h\}$ input schemes result.

The SM predictions for 2-body decays (see e.g. Table~\ref{tab:results_23body_MW}) are provided by the LHC Higgs Cross Section Working Group~\cite{deFlorian:2016spz,LHCHXSWG-tables}.
The SM predictions for the $4f$ channels (e.g. Table~\ref{tab:results_4f_MW}) are extracted with {\tt Prophecy4f 2.0}~\cite{Bredenstein:2006rh,Bredenstein:2006nk,Bredenstein:2006ha} using Monte Carlo settings consistent with the Working Group recommendations~\cite{deFlorian:2016spz}.
\begin{table}[t]\centering
\renewcommand{\arraystretch}{1.2}
 \begin{tabular}{crlc}\hline
  $\hat{M}_{W}$ & 80.365 &GeV &  \cite{Aaltonen:2013iut} \\
  $\hat\alpha_{ew}(M_Z)$ & 1/127.950& & \cite{Olive:2016xmw}\\\hline
  $\hat{M}_Z$ &  91.1876 &GeV & \cite{Z-Pole,Olive:2016xmw,Mohr:2012tt} \\
  $\hat{G}_F$ & 1.1663787 $\cdot 10^{-5}$ &GeV$^{-2}$ &  \cite{Olive:2016xmw,Mohr:2012tt} \\

  $\hat{M}_h$ & 125.09 &GeV & \cite{Aad:2015zhl} \\
  $\hat{\alpha}_s(\hat m_Z)$& 0.1181 & &\cite{Olive:2016xmw}\\
  \hline
  $\hat{M}_t$&     173.21   &GeV & \cite{Olive:2016xmw}\\
  $\hat{M}_b$&     4.18 &GeV & \cite{Olive:2016xmw}\\
  $\hat{M}_c$&     1.28  &GeV & \cite{Olive:2016xmw}\\
  $\hat{M}_\tau$&  1.77686  &GeV & \cite{Olive:2016xmw}\\
  \hline
  \end{tabular}
\caption{Numerical central values of the relevant SM parameters used as inputs for the estimate of the leading SMEFT corrections. Only one among the values of $\hat M_W$ and $\hat\alpha_{ew}$ is used as input, depending on the scheme adopted. All the other parameters are common to the two input schemes considered.}\label{tab:inputs}
\end{table}

The dependence on the Wilson coefficients has been cross-checked with \MG\ with the UFO model {\tt SMEFTsim\_A\_U35\_MwScheme\_UFO\_v2.1}, generating the interference contribution to the partial widths for 5 values of each Wilson coefficient and extracting the corresponding $a_i^{(S)}$ via a linear interpolation. Agreement to $1\%$ or better was found between the theoretical prediction and Monte Carlo result for all $a_i^{(S)}$, when corrections from the $W,Z$ propagators are neglected.

The dependence of the total inclusive width on the $\mathcal{L}^{(6)}$ Wilson coefficients of the SMEFT is found to be
\bea
\begin{aligned}
\frac{\d\Gamma_{h,full}^{SMEFT}}{\Gamma_h^{SM}}
\simeq \,  1  &
- 1.50    \,\tilde{C}_{HB}
- 1.21    \,\tilde{C}_{HW}
+ 1.21    \,\tilde{C}_{HWB}
+ 50.6    \,\tilde{C}_{HG}
\\ &
+ 1.83    \,\tilde{C}_{H\square}
- 0.43    \,\tilde{C}_{HD}
+ 1.17    \,\tilde{C}_{ll}'
\\&
- 7.85  \,  \hat{Y}_{\substack{u \\ cc}} \, \re\tilde{C}_{uH}
- 48.5 \,  \hat{Y}_{\substack{d \\ bb}} \, \re\tilde{C}_{dH}
- 12.3   \,  \hat{Y}_{\substack{\ell \\ \tau \tau}} \, \re\tilde{C}_{eH}
\\&
+ 0.002   \,\tilde{C}_{Hq}^{(1)}
+ 0.06    \,\tilde{C}_{Hq}^{(3)}
+ 0.001 \,\tilde{C}_{Hu}
- 0.0007  \,\tilde{C}_{Hd}
\\&
- 0.0009   \,\tilde{C}_{Hl}^{(1)}
- 2.32    \,\tilde{C}_{Hl}^{(3)}
- 0.0006   \,\tilde{C}_{He},
\end{aligned}
\eea
using the $\{\hat M_W,\hat M_Z,\hat G_F,\hat M_h\}$ input scheme. Here we  have pulled out the explicit Yukawa factor from the Wilson coefficient, consistent with the $\rm U(3)^5$ limit considered.
In the remaining results the Yukawa factor is included in the numerical $a_i^{(S)}$ reported.

Using the $\{\hat \alpha_{ew},\hat M_Z,\hat G_F,\hat M_h\}$ input scheme we find analogously
\bea
\begin{aligned}
\frac{\d\Gamma_{h,full}^{SMEFT}}{\Gamma_h^{SM}}
\simeq \,  1  &
- 1.40    \,\tilde{C}_{HB}
- 1.22    \,\tilde{C}_{HW}
+ 2.89    \,\tilde{C}_{HWB}
+ 50.6    \,\tilde{C}_{HG}
\\ &
+ 1.83    \,\tilde{C}_{H\square}
+ 0.34    \,\tilde{C}_{HD}
+ 0.70    \,\tilde{C}_{ll}'
\\&
- 7.85  \,  \hat{Y}_{\substack{u \\ cc}} \, \re\tilde{C}_{uH}
- 48.5 \,  \hat{Y}_{\substack{d \\ bb}} \, \re\tilde{C}_{dH}
- 12.3   \,  \hat{Y}_{\substack{\ell \\ \tau \tau}} \, \re\tilde{C}_{eH}
\\&
+ 0.002   \,\tilde{C}_{Hq}^{(1)}
+ 0.06    \,\tilde{C}_{Hq}^{(3)}
+ 0.001 \,\tilde{C}_{Hu}
- 0.0008  \,\tilde{C}_{Hd}
\\&
- 0.0008   \,\tilde{C}_{Hl}^{(1)}
- 1.38    \,\tilde{C}_{Hl}^{(3)}
- 0.0007   \,\tilde{C}_{He}.
\end{aligned}
\eea

It is interesting to examine the impact of different contributions to the final result and in particular of contributions that were previously neglected, to our knowledge, in the estimate of SMEFT corrections to $h\to 4f$. In the SM, these decays are well-described in a narrow-width approximation for the $W,Z$ bosons, that gives
\begin{equation}
\Gamma_{h\to \bar\psi_a\psi_a \bar\psi_b \psi_b }^{NC,nw.} =
\Gamma_{h\to Z Z^*,Z^*\to\bar\psi_a\psi_a}  {\rm Br}_{Z\to  \bar\psi_b\psi_b}
+
\Gamma_{h\to Z Z^*,Z^*\to \bar\psi_b\psi_b}  {\rm Br}_{Z\to  \bar\psi_a\psi_a}
\end{equation}
for channels proceeding through NC,
and analogously for charged currents\footnote{For channels that allow both neutral and charged current contractions, the inclusive width is the sum of two $h\to ZZ^*$ and two $h\to WW^*$ terms.}.
The same approach is usually  generalized to the SMEFT case, leading to estimates of the form
\begin{equation}\label{eq.narrow_width_shifts}
\begin{aligned}
\Gamma_{h\to\bar\psi_a\psi_a\bar\psi_b\psi_b}^{NC,nw.\; SMEFT} &=
\Gamma_{h\to Z Z^*,Z^*\to\bar\psi_a\psi_a}^{SM}  {\rm Br}_{Z\to  \bar\psi_b\psi_b}^{SM}
\left[
1+\frac{\d\Gamma_{h\to ZZ^*,Z^*\to \bar\psi_a\psi_a}}{\Gamma_{h\to ZZ^*,Z^*\to \bar\psi_a\psi_a}^{SM}} +\frac{\d\Gamma_{Z\to \bar\psi_b\psi_b}}{\Gamma_{Z\to\bar\psi_b\psi_b}^{SM}} 
\right]
\\
&
+\Gamma_{h\to Z Z^*,Z^*\to\bar\psi_b\psi_b}^{SM}  {\rm Br}_{Z\to  \bar\psi_a\psi_a}^{SM}
\left[
1+\frac{\d\Gamma_{h\to ZZ^*,Z^*\to \bar\psi_b\psi_b}}{\Gamma_{h\to ZZ^*,Z^*\to \bar\psi_b\psi_b}^{SM}} +\frac{\d\Gamma_{Z\to \bar\psi_a\psi_a}}{\Gamma_{Z\to\bar\psi_a\psi_a}^{SM}} 
\right]
\\
&
- \Gamma_{h\to\bar\psi_a\psi_a\bar\psi_b\psi_b}^{NC,nw.\; SM} \frac{\d\Gamma_{Z,full}}{\Gamma_{Z,full}^{SM}}\,.
\end{aligned}
\end{equation}
The implementation of the narrow-width approximation in this context is not unique, as there is some arbitrariness in the choice of the contributions included in each term. However, the following classes of terms are often omitted in this approach:
\begin{enumerate}
\item Diagrams with intermediate off-shell photons.\\
Contributions containing the $Z\gamma$ interaction are compatible with the narrow-width assumption for NC, and could therefore be included, while $\g\g$-mediated diagrams are always missed in this approximation.

\item Interference terms between NC and CC contributions, that are not compatible with the amplitude factorization into $(h\to \bar\psi\psi V) \times (V\to \bar\psi\psi)$.

\item Interference terms between ZZ diagrams with different current contractions in channels with 2 indistinguishable fermion pairs ($\bar \psi_a \psi_a \bar\psi'_a\psi'_a$ vs $ \bar \psi_a \psi'_a \bar\psi'_a\psi_a$).

\item Propagator corrections for the off-shell boson.
\end{enumerate}
In the following we isolate and quantify the impact of each of these terms.

\subsection{Photon-mediated diagrams}

\begin{table}[t]\centering
\renewcommand{\arraystretch}{1.2}
\hspace*{-10mm}
\begin{tabular}{>{$}c<{$}|*3{>{$}c<{$}}|*3{>{$}c<{$}}|*3{>{$}c<{$}}}
\hline
\multirow{2}{*}{$h\to S$}& \multicolumn{3}{c|}{$\tilde{C}_{HW}$}& \multicolumn{3}{c|}{$\tilde{C}_{HB}$}& \multicolumn{3}{c}{$\tilde{C}_{HWB}$}\\\cline{2-10}
& Z\g& \g\g& WW,ZZ& Z\g& \g\g& WW,ZZ& Z\g& \g\g& WW,ZZ\\
\hline
\ell^+_p\ell^-_p\ell^+_r\ell^-_r     &\bf 1.04 & -0.009 & -0.78 &\bf -1.04 & -0.03 & -0.22 &\bf -0.70 & 0.02 & 0.30 \\
\ell^+_p\ell^-_p\bar\nu_r\nu_r       & 0.52 &          & -0.78 &\bf -0.52 &   & -0.22 &\bf -0.35 &   & -0.06 \\
\bar u_pu_p\bar u_ru_r               &\bf 2.26 & -0.04 & -0.78 &\bf -2.26 & -0.15 & -0.22 &\bf\bf -1.51 & 0.08 & 1.13 \\
\bar d_pd_p\bar d_rd_r               &\bf 1.53 & -0.02 & -0.78 &\bf -1.53 & -0.07 & -0.22 &\bf -1.02 & 0.04 & 0.63 \\
\bar u_pu_p\bar d_rd_r               &\bf 1.89 & -0.03 & -0.78 &\bf -1.89 & -0.10 & -0.22 &\bf -1.26 & 0.05 & 0.88 \\
\ell^+_p\ell^-_p \bar u_{p,r}u_{p,r} &\bf 1.65 & -0.02 & -0.78 &\bf -1.65 & -0.07 & -0.22 &\bf -1.10 & 0.04 & 0.71 \\
\ell^+_p\ell^-_p \bar d_{p,r}d_{p,r} &\bf 1.29 & -0.01 & -0.78 &\bf -1.29 & -0.05 & -0.22 & -0.86 & 0.02 & 0.46 \\
\nu_p\nu_p\bar u_{p,r}u_{p,r}        &\bf 1.13 &   & -0.78 &\bf -1.13 &   & -0.22 &\bf -0.75 &   & 0.36 \\
\nu_p\nu_p \bar d_{p,r}d_{p,r}       &\bf 0.76 &   & -0.78 &\bf -0.76 &   & -0.22 &\bf -0.51 &   & 0.11 \\
\hline
\ell^+_p\ell^-_p\ell^+_p\ell^-_p     &\bf 1.06 & -0.29 & -0.75 &\bf -1.06 &\bf -1.01 & -0.22 &\bf -0.70 &\bf 0.54 & 0.43 \\
\bar u_pu_p\bar u_pu_p               &\bf 2.23 & -0.08 & -0.77 &\bf -2.23 &\bf -0.27 & -0.22 &\bf -1.49 & 0.15 & 1.20 \\
\bar d_pd_p\bar d_pd_p               &\bf 1.48 & -0.03 & -0.76 &\bf -1.48 & -0.09 & -0.22 &\bf -0.99 & 0.05 & 0.65 \\
\hline
\bar u_pd_p\bar d_pu_p               & 0.06 & 0.001 & -1.47 &\bf -0.06 & 0.004 & -0.008 &\bf -0.04 & -0.002 & 0.02\\
\ell_p^+\nu_p\bar\nu_p \ell_p^-      & -0.02 &   & -1.49 &\bf 0.02 &   & -0.007 & 0.01 &   & -0.07 \\
\hline
\end{tabular}
\caption{Contribution to $a_{HW}^{(S)},\,a_{HB}^{(S)},\,a_{HWB}^{(S)}$ from $Z\gamma$, $\g\g$ and $WW+ZZ$ mediated diagrams, using the $\{\hat M_W,\hat M_Z,\hat G_F,\hat M_h\}$ input scheme. The channels in the three blocks admit NC only with a unique current contraction, NC with two possible contractions and  both NC and CC. We distinguish channels with same- or different- flavor fermion pairs ($p\neq r$).
The double subscript $p,r$ indicates that both same and different flavor-currents are included.  The most significant contributions are highlighted in bold.}\label{tab:photon_contrib}
\end{table}

As mentioned previously, due to its coupling to $Z\g$ and $\g\g$ the Higgs boson can decay to 4 fermions via electromagnetic currents, in addition to the weak ones. In the SM this effect is negligible due to the $hZ\g$, $h\g\g$ effective couplings being loop suppressed
 (this is essentially an accidental suppression due to the $d \leq 4$ operator mass dimensions of the SM, for a related discussion see Ref.\cite{Jenkins:2013fya}). In the SMEFT, in contrast, these interactions formally arise at tree-level together with the leading corrections to the $hZZ$, $hWW$ couplings. This is the prime reason that the narrow width approximation fails more dramatically in the SMEFT compared to the SM.

The calculation presented in this work includes for the first time the interference terms
\begin{equation}
\frac{1}{\Gamma_{h\to S}^{SM,tree}}\dfrac{1}{2\hat M_h}\int dps\; |\mathcal{A}^{\cal ZZZA}|^2 +  |\mathcal{A}^{\cal ZZAA}|^2 +  |\mathcal{A}^{\cal WWZA}|^2 + |\mathcal{A}^{\cal WWAA}|^2,
\end{equation}
which are proportional to either $C_{\cal AZ}$ or $C_{\cal AA}$ and therefore affect the dependence on $C_{HW},$ $C_{HB},$ $C_{HWB}$.
Table~\ref{tab:photon_contrib} shows the numerical contribution of these diagrams to the coefficients $a_i^{(S)}$ in the linearized SMEFT expressions compared to the contributions from $WW$ and $ZZ$ diagrams\footnote{For comparison, the quantities in Table~\ref{tab:results_4f_MW} are given by the sum of these three contributions, plus the corrections from the $W,Z$ propagators.}.

It is immediate to see that the photon contribution to these quantities is significant, especially for the $Z\gamma$ terms that exceed in absolute value the $ZZ,WW$ contributions in most channels. In several cases, the $Z\gamma$ contribution flips the overall sign in the $C_i$ dependence compared to the one when only including $ZZ,WW$ currents. The photon effect is largest for channels with NC only, and involving the up quark, due to a color factor and electromagnetic charge enhancement. Channels allowing both NC and CC decay are largely dominated by the CC diagrams, so both $ZZ$ and photon contributions are suppressed.

\subsection{NC-CC interference terms}
The channels $h\to \ell^+\ell^-\bar\nu \nu$, $h\to \bar u u \bar d d $ with 4 fermions of the same generation admit both CC and NC diagrams. When assuming narrow $W$ or $Z$ bosons, one usually sums over the 4 configurations in which either a $Z$ or a $W$ is nearly on-shell. By construction this calculation neglects interference terms between diagrams mediated by $W$ and $Z$.
\begin{table}[t]\centering
\renewcommand{\arraystretch}{1.2}
\hspace*{-10mm}
\begin{tabular}{>{$}c<{$}|*9{>{$}c<{$}}}
\hline
h\to e^+ e^- \bar\nu_e\nu_e&
\tilde{C}_{HW} & \tilde{C}_{HB}& C_{HWB}& \tilde{C}_{HD}& \tilde{C}_{H\square} & \tilde{C}_{Hl}^{(1)}& \tilde{C}_{Hl}^{(3)}& \tilde{C}_{He} & \tilde{C}_{ll}'
\\\hline

|\mathcal{A}_{ZZ}|^2&-0.04 & -0.01 & -0.003 & 0.09 & -0.008 & 0.009 & -0.08 & -0.08 & 0.14 \\
|\mathcal{A}_{WW}|^2& -1.49 &   &   & 2.00 & -0.50 &   & -2.00 &   & 3.00 \\
\mathcal{A}_{ZZ}\cdot \mathcal{A}_{WW}& 0.04 &\bf  0.004 & \bf -0.07 & -0.10 & -0.04 & \bf-0.04 & 0.06 &   & -0.14 \\
\hline
\mathcal{A}_{ZZ}\cdot \mathcal{A}_{HZff} &   &   &   &   &   & -0.005 & -0.10 & 0.04 &   \\
\mathcal{A}_{WW}\cdot \mathcal{A}_{HWff} &   &   &   &   &   &   & -1.77 &   &   \\
\mathcal{A}_{ZZ(WW)}\cdot \mathcal{A}_{HW(Z)ff}
&   &   &   &   &   & \bf 0.03 & 0.15 &   &   \\
\hline
\end{tabular}
\caption{
Contribution to $a_{i}^{(ee\nu\nu)}$ from different interference terms. First block: $ZZ$ and $WW$ mediated diagrams. Second block: diagrams involving contact operators. Corrections to the $W,Z$ propagators are omitted in this table. The most significant contributions are highlighted in bold.
}\label{tab:ZZWW_contrib_eevv}
\end{table}
\begin{table}[h!]\centering
\renewcommand{\arraystretch}{1.2}
\hspace*{-24mm}
\begin{tabular}{>{$}c<{$}|*{11}{>{$}c<{$}}}
\hline
h\to \bar uu \bar dd&
\tilde{C}_{HW} & \tilde{C}_{HB}& \tilde{C}_{HWB}& \tilde{C}_{HD}& \tilde{C}_{H\square} & \tilde{C}_{Hl}^{(3)}& \tilde{C}_{Hq}^{(1)}& \tilde{C}_{Hq}^{(3)}& \tilde{C}_{Hu} & \tilde{C}_{Hd} & \tilde{C}_{ll}'
\\\hline
|\mathcal{A}_{ZZ}|^2&                      -0.03 & -0.009 & 0.03 & 0.08 & 0.03 & -0.24 & -0.005 & 0.19 & 0.04 & -0.02 & 0.12 \\
|\mathcal{A}_{WW}|^2&                       -1.45 &   &   & 1.95 & -0.49 & -5.86 &   & 3.91 &   &   & 2.93 \\
\mathcal{A}_{ZZ}\cdot \mathcal{A}_{WW}&     0.012 & 0.001 &\bf -0.02 & -0.03 & -0.009 & 0.10 & 0.004 & -0.08 &   &   & -0.05 \\
\hline
\mathcal{A}_{ZZ}\cdot \mathcal{A}_{HZff}  &    &   &   &   &   &   & 0.003 & -0.09 & -0.02 & 0.008 &   \\
\mathcal{A}_{WW}\cdot \mathcal{A}_{HWff}  &    &   &   &   &   &   &   & -1.72 &   &   &   \\
\mathcal{A}_{ZZ(WW)}\cdot \mathcal{A}_{HW(Z)ff}&
 &   &   &   &   &   & -0.003 & 0.05 &   &   &   \\
\hline
\end{tabular}
\caption{Contribution to $a_{i}^{(\bar uu\bar dd)}$ from different interference terms. First block: $ZZ$ and $WW$ mediated diagrams. Second block: diagrams involving contact operators. Corrections to the $W,Z$ propagators are omitted in this table. The most significant contributions are highlighted in bold.
}\label{tab:ZZWW_contrib_uudd}
\end{table}

Tables~\ref{tab:ZZWW_contrib_eevv},~\ref{tab:ZZWW_contrib_uudd} show a breakdown of the contributions to the quantities $a_i^{(S)}$ for the relevant operators $\mathcal{O}_i$ from different interference terms in the squared amplitude, obtained with the full computation.
The first three rows report the contribution from $ZZ$ and $WW$ mediated diagrams, while the last three rows indicate the contributions from contact interactions $HVff$. All numbers are normalized to the corresponding SM tree level partial width, that contains both $ZZ$, $WW$ and interference terms.
This table omits the contributions from the $W,Z$ propagator corrections as well as contributions from photon diagrams.

As the SM partial width for both channels is dominated by the $WW$ diagram, corrections to the latter are generally more important than corrections to the $ZZ$ topology. The interference between the two gives significant contributions to the dependence on $C_{HWB}$ and on $C_{Hl}^{(1)}$ in the leptonic case. The latter is due to an accidental cancellation between the corresponding charged lepton and neutrino corrections in $ZZ$ diagrams that does not occur in the interference with $W$ currents.

\subsection{Interference between NC diagrams with different current contractions}
\begin{figure}[b!]
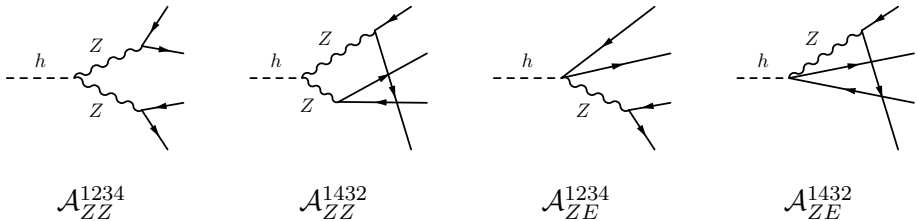
\centering
\includegraphics[width=0.2\textwidth]{diagrams_H4f/diag_ZZ_direct.pdf}
\includegraphics[width=0.2\textwidth]{diagrams_H4f/diag_ZZ_cross.pdf}
\includegraphics[width=0.2\textwidth]{diagrams_H4f/diag_ZE_direct.pdf}
\includegraphics[width=0.2\textwidth]{diagrams_H4f/diag_ZE_cross.pdf}
$ \mathcal{A}_{ZZ}^{1234}
\hspace*{2.3cm}
\mathcal{A}_{ZZ}^{1432}
\hspace*{2.3cm}
\mathcal{A}_{ZE}^{1234}
\hspace*{2.3cm}
\mathcal{A}_{ZE}^{1432}$
\caption{Current contractions allowed in the presence of 2 pairs of identical particles in final state.}\label{fig:ZZ_contractions}
\end{figure}

Decays with 2 pairs of identical particles in final state admit 2 independent neutral-current contractions, depicted in Figure~\ref{fig:ZZ_contractions} for the $ZZ$ and contact-term cases. The same contractions are allowed for photon mediated diagrams.

In the squared amplitude, the ``direct'' products $\mathcal{A}^{1234}\mathcal{A}^{1234,\dag}$, $\mathcal{A}^{1432}\mathcal{A}^{1432,\dag}$ are related by relabeling of the final states and give therefore identical results, while the ``crossed'' interference $\mathcal{A}^{1234}\mathcal{A}^{1432,\dag}$  provides an independent contribution, that is neglected in the narrow width approximation.

In the complete calculation, the ``crossed'' interference terms are found to be most relevant in the $h\to \ell^+\ell^-\ell^+\ell^-$ channel, particularly for diagrams involving the photon. Table~\ref{tab:crossedNC_contrib} shows a comparison of the contribution to $a_i^{(eeee)}$ from direct and crossed amplitude products for $Z$, $\g$ and contact diagrams independently.

For the remaining channels $h\to \bar\nu\nu\bar\nu\nu,\bar uu\bar uu,\bar dd\bar dd$ we find that the size of $\mathcal{A}^{1234}\mathcal{A}^{1432,\dag}$ contributions is generally smaller, ranging between a few \% and 20 \% of the corresponding ``direct'' contribution. The different behavior is due to two numerical effects: on one hand, all ``crossed'' contributions in the quarks case are suppressed by a factor $N_c=3$ compared to the ``direct'' ones. In addition, the photon contributions are further reduced by factors of $|Q_q|<|Q_e|=1$.

\begin{table}[t]\centering
\renewcommand{\arraystretch}{1.2}
\begin{tabular}{>{$}l<{$}|*9{>{$}c<{$}}}
\hline
h\to \ell^+\ell^-\ell^+\ell^- & \tilde{C}_{HW} & \tilde{C}_{HB} & \tilde{C}_{HWB} & \tilde{C}_{H\square} & \tilde{C}_{HD} & \tilde{C}_{Hl}^{(1)} & \tilde{C}_{Hl}^{(3)} & \tilde{C}_{He} & \tilde{C}_{ll}' \\
\hline
\mathcal{A}_{ZZ}\mathcal{A}^\dag_{ZZ} \text{ direct} & -0.70 & -0.20 & 0.27 & 1.81 & 0.15 & 3.97 & -1.47 & -3.20 & 2.72 \\
\mathcal{A}_{ZZ}\mathcal{A}^\dag_{ZZ} \text{ cross} & -0.05 & -0.01 &\bf 0.16 & 0.19 &\bf 0.13 & 0.47 & -0.08 & -0.25 & 0.28 \\
\mathcal{A}_{ZZ}\mathcal{A}^\dag_{ZE}\text{ direct} &   &   &   &   &   & -2.03 & -2.03 & 1.63 &   \\
\mathcal{A}_{ZZ}\mathcal{A}^\dag_{ZE}\text{ cross} &   &   &   &   &   & -0.33 & -0.33 & 0.17 &   \\
\mathcal{A}_{ZZ}\mathcal{A}^\dag_{AA,ZA}\text{ direct} & 0.94 & -0.98 & -0.62 &   &   &   &   &   &   \\
\mathcal{A}_{ZZ}\mathcal{A}^\dag_{AA,ZA}\text{ cross} &\bf -0.17 &\bf -1.09 &\bf 0.45 &   &   &   &   &   &   \\
\hline
\end{tabular}
\caption{Contribution to $a_{i}^{(eeee)}$ from the interference of $ZZ$ diagrams with $ZZ$, contact and photon diagrams, for ``direct'' ($\mathcal{A}^{1234}\mathcal{A}^{1234,\dag}$)  and ``crossed'' ($\mathcal{A}^{1234}\mathcal{A}^{1432,\dag}$) current contractions. The most significant contributions are highlighted in bold.}\label{tab:crossedNC_contrib}
\end{table}

\subsection{Propagator corrections to the off-shell boson}\label{propcorrections}

Finally, the complete calculation allows to extract the exact dependence on the $W,Z$ propagator corrections. In the narrow $V$ width approximation, neglecting for simplicity the off-shell boson's contribution, one would just have (see also Eq.~\eqref{eq.narrow_width_shifts}):
\begin{equation}
\frac{\Gamma_{h\to VV^*\to 4f}^{SMEFT}}{\Gamma_{h\to VV^*\to ff}^{SM}} = 1-\frac{\d\Gamma_V}{\Gamma_V^{SM}} + \dots
\end{equation}
Once all the contributions are taken into account, the coefficient of $\d\Gamma_V/\Gamma_V^{SM}$ in this expression generally deviates from $-1$.
Table~\ref{tab:width_contrib} shows the values obtained in this work. We use the $\{\hat M_W,\hat M_Z,\hat G_F,\hat M_h\}$ input scheme, so the only relevant corrections are due to shifts in the width of $W,Z$, as $\d M_Z=\d M_W=0$.

For completeness, we also report here the numerical expression of the width shifts in terms of Wilson coefficients, in the same scheme:
\begin{align}\label{eq:dGammaZ_MW}
\frac{\d\Gamma_Z}{\Gamma_Z^{SM}} &=
 0.46 \tilde{C}_{HWB}
- 0.07 \tilde{C}_{HD}
- 0.18 \tilde{C}_{Hl}^{(1)}
- 1.37 \tilde{C}_{Hl}^{(3)}
- 0.18 \tilde{C}_{He}
\nn
&
+ 0.47 \tilde{C}_{Hq}^{(1)}
+ 1.61 \tilde{C}_{Hq}^{(3)}
+ 0.24 \tilde{C}_{Hu}
-0.18 \tilde{C}_{Hd}
 + \tilde{C}_{ll}',
\\
\frac{\d\Gamma_W}{\Gamma_W^{SM}} &= \frac{4}{3}\left(\tilde{C}_{Hq}^{(3)}-\tilde{C}_{Hl}^{(3)}\right)+\tilde{C}_{ll}'.\label{eq:dGammaW_MW}
\end{align}

\begin{table}[t]\centering
\renewcommand{\arraystretch}{1.2}
\begin{tabular}{*2{|>{$}l<{$}|>{$}c<{$}>{$}c<{$}}|}
\hline
h\to S& \d\Gamma_Z/\Gamma_Z^{SM} & \d\Gamma_W / \Gamma_W^{SM} &
h\to S& \d\Gamma_Z/\Gamma_Z^{SM} & \d\Gamma_W / \Gamma_W^{SM}
\\\hline
\ell^+_p\ell^-_p \ell^+_r\ell^-_r& -0.82 &                  &\ell^+_p \ell^-_p \ell^+_p \ell^-_p& -0.74 &
\\
\bar \nu_p\nu_p \bar \nu_r\nu_r& -0.82 &                    &\bar \nu_p \nu_p \bar \nu_p \nu_p& -0.68   &
\\
\ell^+_p \ell^-_p \bar \nu_r\nu_r& -0.82 &                  &\bar u_p u_p \bar u_p u_p& -0.78  &
\\
\bar u_p u_p \bar u_r u_r& -0.82 &                          &\bar d_p d_p \bar d_p d_p& -0.77   &
\\
\bar d_p d_p \bar d_r d_r& -0.82 &                          &\ell^+_p \nu_p \bar\nu_r \ell^-_r+\hc& & -0.92
\\
\bar u_p u_p \bar d_r d_r& -0.82 &                          &\bar u_p d_p \bar d_r u_r +\hc& & -0.92
\\
\ell^+_p \ell^-_p \bar u_{p,r} u_{p,r}& -0.82 &             &\ell^+_p \nu_p \bar u_{p,r} d_{p,r}+\hc& & -0.92
\\
\ell^+_p \ell^-_p \bar d_{p,r }d_{p,r}& -0.82 &             &\bar u_p u_p \bar d_p d_p& -0.03 & -0.89
\\
\bar \nu_p \nu_p \bar u_{p,r} u_{p,r}& -0.82 &              &\ell^+_p \ell^-_p \bar \nu_p \nu_p& -0.04 &  -0.91
\\
\bar \nu_p \nu_p \bar d_{p,r} d_{p,r}& -0.82 &       & & &
\\\hline
\end{tabular}
\caption{Coefficients of $\d\Gamma_V/\Gamma_V^{SM}$ appearing in the relative SMEFT correction $\Gamma_{h\to S}^{SMEFT} / \Gamma_{h\to S}^{SM}$, using the $\{\hat M_W,\hat M_Z,\hat G_F,\hat M_h\}$ input scheme.  We distinguish channels with same- or different- flavor fermion pairs ($p\neq r$).
The double subscript $p,r$ indicates that both same and different flavor-currents are included. }
\label{tab:width_contrib}
\end{table}

\subsection{Summary of the impact of various contributions to \titlemath{$h\to 4f$}{h > 4f}}
In this section we have examined the impact of various classes of terms in the squared amplitude to the final SMEFT calculation for $h\to 4f$, and in particular those that are usually omitted in narrow $W,Z$-width calculations.

We find that the largest among the latter contributions are those from photon-mediated diagrams. These have a very significant impact on the determination of the dependence on the Wilson coefficients $C_{HW},\,C_{HB},\,C_{HWB}$ in the $h\to 4f$ partial widths. This effect can be a few times larger in absolute value compared to the contribution from $ZZ,WW$ diagrams only and is most relevant for channels proceeding via NC.

The accurate estimate of the corrections due to $W,Z$ propagator shifts is also found to be important, as it leads to a $\mathcal{O}(20-30)\%$ difference in the dependence on $\d\Gamma_V/\Gamma_V^{SM}$ with respect to the naive narrow-width estimate.

The interference among NC and CC diagrams, when present, is found to affect significantly the $C_{HWB}$ dependence, as well as that on $C_{HB}$ and $C_{Hl}^{(1)}$ in the leptonic channels. Its contribution is subleading (between a few \% and $\mathcal{O}(15)\%$) for all other parameters.

Finally, the interference between two different NC contractions contributes only to $\mathcal{O}(10)\%$ or less of the dependence on all Wilson coefficients, with the exception of the $\ell^+\ell^-\ell^+\ell^-$ channel, where the ``crossed'' photon diagrams effect is unsuppressed.

\section{Conclusions}\label{conclusions}
We have calculated and presented the Higgs width in the SMEFT for a set of two and four body Higgs decays. Our results are presented in a manner
that more than one input parameter scheme can be used.
The resulting dependence on the Wilson coefficients in the Higgs width, and branching ratios, is significantly different than the partial results in the literature,
and significantly different than various results obtained using the narrow width approximation.  The main reason for this difference is more naive narrow width approaches miss large interference
effects which introduce a leading dependence on Wilson coefficients in the SMEFT in some final states.

The numerical size of the corrections we have determined, in the perturbation of the Higgs width depends upon the Wilson coefficients.
Determining constraints on the Wilson coefficients in a consistent global SMEFT analysis is an active pursuit in the theoretical
community. Some combinations of the parameters
perturbing the Higgs width are significantly constrained by EWPD and diboson production data. We leave
a combined analysis of the constraints inferred on physics beyond the SM, combining EWPD, diboson data, and Higgs data
to future publications.

These results we have presented allow the inclusive branching ratios and total width
of the Higgs, constructed from the processes reported here, to be determined without a Monte Carlo generation of phase space being performed for each Wilson coefficient
value chosen.\footnote{A future version of this work will include a numerical code of our results consistent with SMEFTsim conventions and inputs.}

\newpage
\begin{acknowledgments}
We acknowledge generous support from the Villum Fonden and partial support by the Danish National Research Foundation (DNRF91) through the Discovery centre.
We thank M. D\"uhrssen, C. Hays, A. Manohar, D. McGady, G. Passarino, M. Pellen and D. Straub for useful discussions and comments and O. Mattelaer for support with \MG.
\end{acknowledgments}

\appendix

\section{Tables of numerical results}\label{app:tables}
In this appendix we report tables that summarize  the SM partial width and relative SMEFT corrections for all the Higgs decay channels considered in this work.

We parameterize each partial width as in Eq.~\eqref{eq.general_W_parameterization}, with the SM result taken to be the current best estimate, as provided by the LHC Higgs Cross Section Working Group~\cite{deFlorian:2016spz,LHCHXSWG-tables}. The SMEFT corrections are tabulated reporting the values of the coefficients $a_i^{(S)}$ for each channel $S$ and $\mathcal{L}^{(6)}$ coefficient $C_i$. These are determined directly from our tree-level calculation and have been cross-checked with \MG\ and the {\tt SMEFTsim} packages.
We give results both in the $\{\hat M_W,\hat M_Z,\hat G_F,\hat M_h\}$ (Tab.~\ref{tab:results_23body_MW},~\ref{tab:results_4f_MW})  and in the $\{\hat\alpha_{ew},\hat M_Z,\hat G_F,\hat M_h\}$ (Tab.~\ref{tab:results_23body_a},~\ref{tab:results_4f_a}) input schemes.

We note that the scheme dependence is particularly large for the coefficients $C_{HWB}$, $C_{HD}$, $C_{Hl}^{(3)}$ and $C_{ll}'$ and stronger in the 4-fermion decay channels that are dominantly mediated by charged currents.
These discrepancies are mostly due to the different definition of the weak mixing angle (or equivalently, of the weak gauge couplings $g_1$, $g_2$) and of $M_W$ in the two schemes. Numerically, for the $\delta s_\theta^2$ correction we find:
\begin{align}
\delta s_\theta^2 &= -0.39 \tilde C_{HD} - 0.42 \tilde C_{HWB} & &(\hat M_W \text{ scheme})
\\
\delta s_\theta^2 &= 0.17 \tilde C_{HD} + 0.79 \tilde C_{HWB} + 0.76 \tilde C_{Hl}^{(3)} - 0.34 \tilde C_{ll}' & &(\hat \alpha_{ew} \text{ scheme})
\end{align}
As $\delta s_\theta^2$ enters directly the $Z$ couplings to fermions, the large numerical difference between these two results directly propagates to the 4-fermions partial widths mediated by neutral currents.

The total decay width of the $Z$ boson has also a significantly different dependence on these 4 parameters in the two schemes. When $\hat \a_{ew}$ is an input, one has numerically
\begin{equation}
\begin{aligned}
\frac{\d\Gamma_Z}{\Gamma_Z^{SM}} &= -0.82 \tilde C_{HWB} - 0.67 \tilde C_{HD} - 0.19 \tilde C_{Hl}^{(1)} - 2.06 \tilde C_{Hl}^{(3)} - 0.19 \tilde C_{He}\\
&\, +0.47\tilde C_{Hq}^{(1)} + 1.61 \tilde C_{Hq}^{(3)} + 0.26 \tilde C_{Hu} - 0.19 \tilde C_{Hd} + 1.35 \tilde C_{ll}'
\end{aligned}
\end{equation}
which can be compared to the result for the $\hat M_W$ input scheme in Eq.~\eqref{eq:dGammaZ_MW}.

The shift in $M_W$ (see Eq.~\eqref{eq:dMW}), on the other hand, has a very significant impact on the predictions for the total $W$ width and for the 4-fermion Higgs decays proceedings via charged currents. In the $\hat \alpha_{ew}$ scheme one has
\begin{equation}
\frac{\d\Gamma_W}{\Gamma_W^{SM}} = -3.97 \tilde C_{HWB} - 1.80 \tilde C_{HD} - 3.52 \tilde C_{Hl}^{(3)} + 1.33 \tilde C_{Hq}^{(3)}  + 2.10 \tilde C_{ll}'\,.
\end{equation}
Comparing this result to Eq.~\ref{eq:dGammaW_MW}, one finds that the dependence on $C_{HWB}$ and $C_{HD}$ is present only in the $\hat\alpha_{ew}$ scheme, and at the same time corrections due to $C_{Hl}^{(3)}$ and $C_{ll}'$ are very scheme-dependent.

These effects are all reflected in the tables presented in this appendix.

\vskip 2cm
\begin{table}[h!]
\renewcommand{\arraystretch}{1.2}
\hspace*{-15mm}\begin{tabular}{l|c|*{11}{>{$}c<{$}}}
\hline
$h\to S$& $\Gamma_{h\to S}^{SM}$ (MeV)
&\tilde{C}_{HW}& \tilde{C}_{HB}& \tilde{C}_{HWB}& \tilde{C}_{HG}& \tilde{C}_{HD}& \tilde{C}_{H\square}& \tilde{C}_{Hl}^{(3)}& \tilde{C}_{ll}'& \re\tilde{C}_{dH}& \re\tilde{C}_{uH}& \re\tilde{C}_{eH}
\\\hline
$\bar bb$&        2.38 &   &   &   &   & -0.5 & 2 & -2  & 1 & -2 &   &   \\
$\bar cc$&        0.12 &   &   &   &   & -0.5 & 2 & -2  & 1 &   &   & -2 \\
$\tau^+\tau^-$&   0.26 &   &   &   &   & -0.5 & 2 & -2  & 1 &   & -2 &   \\
$gg$&             0.33 &   &   &   & 619 &   &   &   &    &   &   &   \\
$Z\gamma$&        6.32 $\cdot 10^{-3}$ & -243 & 243 & 162 &   &    &   &   &   &   &   &   \\
$\gamma\gamma$&   9.31 $\cdot 10^{-3}$ & -231 & -805 & 431 &   &   &    &   &   &   &   &   \\
\hline
\end{tabular}
\caption{Partial SM Higgs decay width and coefficients $a_i^{(S)}$ in the relative SMEFT correction for 2-body decay channels, using the $\{\hat M_W,\hat M_Z,\hat G_F,\hat M_h\}$ input scheme and including all contributions. The SM values are taken from the tables provided by the LHCHXSWG and include higher order corrections~\cite{deFlorian:2016spz,LHCHXSWG-tables}. }\label{tab:results_23body_MW}
\end{table}

\begin{table}[h!]
\renewcommand{\arraystretch}{1.2}
\hspace*{-15mm}\begin{tabular}{l|c|*{11}{>{$}c<{$}}}
\hline
$h\to S$& $\Gamma_{h\to S}^{SM}$ (MeV)
&\tilde{C}_{HW}& \tilde{C}_{HB}& \tilde{C}_{HWB}& \tilde{C}_{HG}& \tilde{C}_{HD}& \tilde{C}_{H\square}& \tilde{C}_{Hl}^{(3)}& \tilde{C}_{ll}'& \re\tilde{C}_{dH}& \re\tilde{C}_{uH}& \re\tilde{C}_{eH}
\\\hline
$\bar bb$&        2.38 &   &   &   &   & -0.5 & 2 & -2  & 1 & -2 &   &   \\
$\bar cc$&        0.12 &   &   &   &   & -0.5 & 2 & -2  & 1 &   &   & -2 \\
$\tau^+\tau^-$&   0.26 &   &   &   &   & -0.5 & 2 & -2  & 1 &   & -2 &   \\
$gg$&             0.33 &   &   &   & 619 &   &   &   &    &   &   &   \\
$Z\gamma$&        6.32 $\cdot 10^{-3}$ & -246 & 246 & 155 &   &    &   &   &   &   &   &   \\
$\gamma\gamma$&   9.31 $\cdot 10^{-3}$ & -233 & -765 & 422 &   &   &    &   &   &   &   &   \\
\hline
\end{tabular}
\caption{Partial SM Higgs decay width and coefficients $a_i^{(S)}$ in the relative SMEFT correction for 2-body decay channels, using the $\{\hat \alpha_{ew},\hat M_Z,\hat G_F,\hat M_h\}$ input scheme and including all contributions. The SM values are taken from the tables provided by the LHCHXSWG and include higher order corrections~\cite{deFlorian:2016spz,LHCHXSWG-tables}. }\label{tab:results_23body_a}
\end{table}

\clearpage
\begin{landscape}
\thispagestyle{empty}
\begin{table}[h!]
\renewcommand{\arraystretch}{1.3}
\hspace*{-3cm}
\scalebox{.95}{
\begin{tabular}{l>{$\times$ }l|c|*{15}{>{$}c<{$}}}
\hline
\multicolumn{2}{c|}{$h\to S$}& $\Gamma_{h\to S}^{SM}$ (MeV)
 &\tilde{C}_{HW}& \tilde{C}_{HB}& \tilde{C}_{HWB}& \tilde{C}_{HG}& \tilde{C}_{HD}& \tilde{C}_{H\square}& \tilde{C}_{Hl}^{(1)}& \tilde{C}_{Hl}^{(3)}& \tilde{C}_{He}& \tilde{C}_{Hq}^{(1)}& \tilde{C}_{Hq}^{(3)}& \tilde{C}_{Hu}& \tilde{C}_{Hd}& \tilde{C}_{ll}'&
\\\hline
$\ell^+_p\ell^-_p \ell^+_r\ell^-_r$& 3 &           0.0007 & 0.26 & -1.30 & -0.76 & & 0.23 & 2 & 2.30 & -2.71 & -1.58 & -0.39 & -1.34 & -0.20 & 0.15 & 2.17
\\
$\bar \nu_p\nu_p \bar \nu_r\nu_r$& 3 &             0.003 & -0.78 & -0.22 & -0.80 & & -0.44 & 2 & -1.81 & -2.90 & 0.15 & -0.39 & -1.34 & -0.20 & 0.15 & 2.17
\\
$\ell^+_p \ell^-_p \bar \nu_r\nu_r$& 6 &           0.003 & -0.25 & -0.75 & -0.79 & & -0.11 & 2 & 0.24 & -2.81 & -0.71 & -0.39 & -1.34 & -0.20 & 0.15 & 2.17
\\
$\bar u_p u_p \bar u_r u_r$& 1 &                   0.003 & 1.44 & -2.63 & -0.68 & & 1.00 & 2 & 0.15 & -4.86 & 0.15 & -2.76 & 1.02 & 0.80 & 0.15 & 2.17
\\
$\bar d_p d_p \bar d_r d_r$& 3 &                   0.015 & 0.73 & -1.82 & -0.73 & & 0.53 & 2 & 0.15 & -4.86 & 0.15 & 1.84 & 0.89 & -0.20 & -0.24 & 2.17
\\
$\bar u_p u_p \bar d_r d_r$& 4 &                   0.016 & 1.09 & -2.22 & -0.71 & & 0.77 & 2 & 0.15 & -4.86 & 0.15 & -0.46 & 0.96 & 0.30 & -0.04 & 2.17
\\
$\ell^+_p \ell^-_p \bar u_{p,r} u_{p,r}$& 6 &      0.005 & 0.86 & -1.94 & -0.73 & & 0.62 & 2 & 1.23 & -3.79 & -0.71 & -1.57 & -0.16 & 0.30 & 0.15 & 2.17
\\
$\ell^+_p \ell^-_p \bar d_{p,r }d_{p,r}$& 9 &      0.010 & 0.50 & -1.55 & -0.75 & & 0.38 & 2 & 1.23 & -3.79 & -0.71 & 0.73 & -0.23 & -0.20 & -0.04 & 2.17
\\
$\bar \nu_p \nu_p \bar u_{p,r} u_{p,r}$& 6 &       0.010 & 0.35 & -1.35 & -0.78 & & 0.28 & 2 & -0.83 & -3.89 & 0.15 & -1.57 & -0.16 & 0.30 & 0.15 & 2.17
\\
$\bar \nu_p \nu_p \bar d_{p,r} d_{p,r}$& 9 &       0.020 & -0.01 & -0.99 & -0.78 & & 0.05 & 2 & -0.83 & -3.89 & 0.15 & 0.73 & -0.23 & -0.20 & -0.04 & 2.17
\\
$\ell^+_p \ell^-_p \ell^+_p \ell^-_p$& 3 &         0.0004 & 0.01 & -2.28 & -0.08 & & 0.34 & 2 & 2.23 & -2.87 & -1.50 & -0.36 & -1.23 & -0.18 & 0.14 & 2.24
\\
$\bar \nu_p \nu_p \bar \nu_p \nu_p$& 3 &           0.002 & -0.74 & -0.21 & -0.72 & & -0.45 & 2 & -1.70 & -3.21 & 0.13 & -0.33 & -1.13 & -0.17 & 0.13 & 2.30
\\
$\bar u_p u_p \bar u_p u_p$& 2 &                   0.003 & 1.39 & -2.72 & -0.51 & -8.03& 1.06 & 2 & 0.14 & -4.91 & 0.14 & -2.70 & 1.05 & 0.77 & 0.14 & 2.20
\\
$\bar d_p d_p \bar d_p d_p$& 3 &                   0.008 & 0.69 & -1.79 & -0.65 & -6.18 & 0.55 & 2 & 0.14 & -4.92 & 0.14 & 1.82 & 0.92 & -0.19 & -0.22 & 2.21
\\
\hline
$\ell^+_p \nu_p \bar\nu_r \ell^-_r+\hc$& 3  &      0.06 & -1.49 &   &   & &-0.5 & 2 &   & -2.50 &   &   & -1.26 &   &   & 2.05
\\
$\bar u_p d_p \bar d_r u_r +\hc$& 1 &              0.20 & -1.49 &   &   & & -0.5 & 2 &   & -4.74 &   &   & 0.97 &   &   & 2.05
\\
$\ell^+_p \nu_p \bar u_{p,r} d_{p,r}+\hc$& 6&      0.39 & -1.49 &   &   & & -0.5 & 2 &   & -3.62 &   &   & -0.14 &   &   & 2.05
\\
\hline
$\bar u_p u_p \bar d_p d_p$& 2 &                   0.20 & -1.41 & -0.07 & -0.04 & -0.98& -0.47 & 2 & 0.006 & -4.72 & 0.006 & -0.02 & 0.96 & 0.01 & -0.002 & 2.04
\\
$\ell^+_p \ell^-_p \bar \nu_p \nu_p$& 3 &          0.03 & -1.51 & 0.01 & -0.07 & & -0.55 & 2 & -0.0005 & -2.43 & -0.03 & -0.02 & -1.32 & -0.009 & 0.007 & 2.02
\\
\hline\hline
\multicolumn{2}{l|}{tot. $h\to 4 f$}& 0.985 &
-1.27 & -0.17 & -0.08 & -0.28 & -0.40 & 2 & -0.004 & -4.03 & -0.003 & 0.009 & 0.24 & 0.004 & -0.003 & 2.06
\\\hline
\end{tabular}}
\caption{
Partial SM Higgs decay width and coefficients $a_i^{(S)}$ in the relative SMEFT correction for each 4-fermion decay channel, using the $\{\hat M_W,\hat M_Z,\hat G_F,\hat M_h\}$ input scheme, neglecting fermion masses and including all contributions. We distinguish channels with same- or different- flavor fermion pairs ($p\neq r$).
The double subscript $p,r$ indicates that both same and different flavor-currents are included.
$\Gamma_{h\to S}^{SM}$ are estimated with {\tt Prophecy4f 2.0} and all the allowed flavor combinations are summed over: the multiplicities are indicated in the first column. Channels of the first, second and third block proceed via neutral currents only, charged currents only and neutral + charged currents respectively. The last row gives the total $\Gamma_{h\to 4f}^{SMEFT}$.  }\label{tab:results_4f_MW}
\end{table}

\clearpage
\thispagestyle{empty}
\begin{table}[h!]
\renewcommand{\arraystretch}{1.3}
\hspace*{-3.5cm}
\scalebox{.98}{
\begin{tabular}{l>{$\times$ }l|c|*{15}{>{$}c<{$}}}
\hline
\multicolumn{2}{c|}{$h\to S$}& $\Gamma_{h\to S}^{SM}$ (MeV)
 &\tilde{C}_{HW}& \tilde{C}_{HB}& \tilde{C}_{HWB}& \tilde{C}_{HG}& \tilde{C}_{HD}& \tilde{C}_{H\square}& \tilde{C}_{Hl}^{(1)}& \tilde{C}_{Hl}^{(3)}& \tilde{C}_{He}& \tilde{C}_{Hq}^{(1)}& \tilde{C}_{Hq}^{(3)}& \tilde{C}_{Hu}& \tilde{C}_{Hd}& \tilde{C}_{ll}'&
\\\hline
$\ell^+_p\ell^-_p \ell^+_r\ell^-_r$& 3 &           0.0007 & -0.11& -0.90& -0.98&  & -0.12& 2& 2.24& -2.92& -1.66& -0.39& -1.33& -0.21& 0.16& 2.24  \\
$\bar \nu_p\nu_p \bar \nu_r\nu_r$& 3 &             0.003 & -0.77& -0.23& 0.25&  & 0.06& 2& -1.80& -2.34& 0.16& -0.39& -1.33& -0.21& 0.16& 1.89
\\
$\ell^+_p \ell^-_p \bar \nu_r\nu_r$& 6 &           0.003 & -0.44& -0.56& -0.37&  & -0.03& 2& 0.22& -2.63& -0.75& -0.39& -1.33& -0.21& 0.16& 2.07
\\
$\bar u_p u_p \bar u_r u_r$& 1 &                   0.003 & 1.41& -2.58& -3.86&  & -0.54& 2& 0.16& -6.67& 0.16& -2.76& 1.03& 0.86& 0.16& 3.07
\\
$\bar d_p d_p \bar d_r d_r$& 3 &                   0.015 & 0.78& -1.87& -2.67&  & -0.36& 2& 0.16& -5.97& 0.16& 1.85& 0.92& -0.21& -0.26& 2.73
\\
$\bar u_p u_p \bar d_r d_r$& 4 &                   0.016 & 1.10& -2.22& -3.27&  & -0.45& 2& 0.16& -6.32& 0.16& -0.45& 0.98& 0.32& -0.05& 2.90
\\
$\ell^+_p \ell^-_p \bar u_{p,r} u_{p,r}$& 6 &      0.005 & 0.66& -1.71& -2.44&  & -0.33& 2& 1.20& -4.79& -0.75& -1.57& -0.15& 0.32& 0.16& 2.66
\\
$\ell^+_p \ell^-_p \bar d_{p,r }d_{p,r}$& 9 &      0.010 & 0.34& -1.37& -1.84&  & -0.24& 2& 1.20& -4.45& -0.75& 0.73& -0.20& -0.21& -0.05& 2.48
\\
$\bar \nu_p \nu_p \bar u_{p,r} u_{p,r}$& 6 &       0.010 & 0.34& -1.34& -1.84&  & -0.24& 2& -0.82& -4.50& 0.16& -1.57& -0.15& 0.32& 0.16& 2.48
\\
$\bar \nu_p \nu_p \bar d_{p,r} d_{p,r}$& 9 &       0.020& 0.02& -1.02 &-1.23& & -0.15& 2& -0.82& -4.16& 0.16& 0.73& -0.20& -0.21& -0.05& 2.31
\\
$\ell^+_p \ell^-_p \ell^+_p \ell^-_p$& 3 &         0.0004 & -0.39& -1.92& -0.63&  & -0.20& 2& 2.16& -3.26& -1.60& -0.36& -1.21& -0.19& 0.14& 2.40
\\
$\bar \nu_p \nu_p \bar \nu_p \nu_p$& 3 &           0.002 & -0.73& -0.22& 0.17&  & -0.03& 2& -1.70& -2.73& 0.13& -0.33& -1.12& -0.18& 0.13& 2.06
\\
$\bar u_p u_p \bar u_p u_p$& 2 &                   0.003 & 1.35& -2.69& -3.91& -8.20& -0.58& 2& 0.15& -6.84& 0.15& -2.71& 1.06& 0.83& 0.15& 3.17
\\
$\bar d_p d_p \bar d_p d_p$& 3 &                   0.008 & 0.74& -1.84& -2.69& -6.27& -0.40& 2& 0.15& -6.09& 0.15& 1.83& 0.94& -0.20& -0.24& 2.80
\\
\hline
$\ell^+_p \nu_p \bar\nu_r \ell^-_r+\hc$& 3  &      0.06 & -1.51& & 8.09&  & 3.16& 2&  & 1.91&  &  & -1.24& & & -0.16
\\
$\bar u_p d_p \bar d_r u_r +\hc$& 1 &              0.20 & -1.51& & 8.09&  & 3.16& 2&  & -0.30&  &  & 0.97&  &  & -0.16
\\
$\ell^+_p \nu_p \bar u_{p,r} d_{p,r}+\hc$& 6&      0.39 & -1.51& & 8.09&  & 3.16& 2&  & 0.81&  &  & -0.13&  &  & -0.16
\\
\hline
$\bar u_p u_p \bar d_p d_p$& 2 &                   0.20 & -1.42& -0.07& 7.79& -1.01& 3.08& 2& 0.006& -0.42& 0.006& -0.02& 0.97& 0.01& -0.002& -0.10
\\
$\ell^+_p \ell^-_p \bar \nu_p \nu_p$& 3 &          0.03 & -1.53& 0.02& 8.18& & 3.19& 2& -0.002& 2.08& -0.04& -0.02& -1.30& -0.01& 0.008& -0.24
\\
\hline\hline
\multicolumn{2}{l|}{tot. $h\to 4 f$}& 0.985 &
-1.29 & -0.17 & 7.00 &  -0.28 & 2.80 & 2 & -0.004 & -0.15 & -0.003 & 0.009 & 0.25 & 0.005 & -0.003 & 0.13
\\\hline
\end{tabular}}
\caption{
Partial SM Higgs decay width and coefficients $a_i^{(S)}$ in the relative SMEFT correction for each 4-fermion decay channel, using the $\{\hat \alpha_{ew},\hat M_Z,\hat G_F,\hat M_h\}$ input scheme, neglecting fermion masses and including all contributions. We distinguish channels with same- or different- flavor fermion pairs ($p\neq r$).
The double subscript $p,r$ indicates that both same and different flavor-currents are included.
$\Gamma_{h\to S}^{SM}$ are estimated with {\tt Prophecy4f 2.0} and all the allowed flavor combinations are summed over: the multiplicities are indicated in the first column. Channels of the first, second and third block proceed via neutral currents only, charged currents only and neutral + charged currents respectively. The last row gives the total $\Gamma_{h\to 4f}^{SMEFT}$.  }\label{tab:results_4f_a}
\end{table}

\end{landscape}

\section{Four body phase space integrations}

\subsection{Analytic results}
Integrating four body phase space is a formally solved problem. Executing such integrations in the SMEFT still presents technical challenges.
Our interest in the four body phase space volume is to describe the decays of the form $h \rightarrow  \bar{\psi} \psi \, \bar{\psi} \psi$.
When directly numerically integrating this phase space volume, we use the approach in Ref.~\cite{Byckling:1971vca}, which relies on Ref.~\cite{RevModPhys.36.595}.

It is helpful to transform the phase space integral to an integration over the set of independent Lorentz invariants $\kappa_{ij}$, the scalar product of the two four vectors $k_i$ and $k_j$, instead of angular variables
which are not Lorentz invariant. There are five independent invariants of the form $\{\kappa_{12}, \kappa_{13}, \kappa_{14}, \kappa_{23}, \kappa_{24}, \kappa_{34}\}$ that are present in four body decays,
subject  to the momentum conservation condition
\bea
m_h^2 = \sum_i m_i^2 + 2 \sum_{i < j} \kappa_{ij}.
\eea
When an index is repeated, we use the convention that $\kappa_i^2 = \kappa_{ii}$
Although closely related in the massless limit the notation $\kappa_{ij}$ and $k_{ij}^2$ are {\it distinct}. The massless limit relationship between the quantities is $k_{ij}^2 = 2 \, \kappa_{ij}$.

The phase space volume in these variables \cite{RevModPhys.36.595} is
\bea
\int  {\it d ps}  &=& \int (2 \pi)^4 \delta^4(P_h -\sum_i k_i) \prod_{k_i} \frac{d^3 k_i}{(2 \pi)^3 2 E_{k_i}}, \nn
 &=& \frac{1}{2^8 \, m_h^2 \, \pi^{6} \, \sqrt{- {\rm Det} M_4}}  \int  \delta^4(\sum_{i < j} \kappa_{ij}   - (m_h^2 - \sum_i m_i^2)/2) \prod_{i < j} d (\kappa_{ij}).
\eea
where the determinant is on the real symmetric matrix constructed of the Lorentz invariants
\bea
M_4 =  \left(\begin{array}{cccc}
\kappa_{1}^2  & \kappa_{12} & \kappa_{13} & \kappa_{14} \\
\kappa_{21}  & \kappa_{2}^2 & \kappa_{23} & \kappa_{24} \\
\kappa_{31}  & \kappa_{32} & \kappa_{3}^2 & \kappa_{34} \\
\kappa_{41}  & \kappa_{42} & \kappa_{43} & \kappa_{4}^2 \\
\end{array}\right).
\eea
The momentum configuration is physical so long as the matrix $M_4$ has one positive and three negative eigenvalues \cite{RevModPhys.36.595,Byckling:1971vca}.
Imposing this condition on the momentum is aided by performing a Gram-Schmidt diagonalization of the momentum vectors.
The basis vectors of the Lorentz space of the $\kappa_{ij}$ can be chosen to be independent. This is easily done by imposing the condition that one vector is time-like and three are space-like.
Then the physical momentum configurations defining the phase space are defined by the simultaneous set of conditions
\begin{align}
\kappa_1^2 &>0, \nn
\kappa_1^2 \, \kappa_2^2 - \kappa_{12}^2 &< 0, \nn
\kappa_1^2 \, \kappa_2^2 \, \kappa_3^2 - \kappa_1^2 \, \kappa^2_{23}  - \kappa_2^2 \, \kappa_{13}^2 - \kappa_3^2  \, \kappa_{12}^2 + 2 \, \kappa_{12} \, \kappa_{23} \, \kappa_{13}  &<0,
\end{align}
and
\bea
&\,&\kappa_1^2 \, \kappa_2^2 \, \kappa_3^2  \, \kappa_4^2 - \kappa_1^2 \, \kappa_2^2 \,  \kappa_{34}^2 - \kappa_1^2  \, \kappa_{23}^2 \,   \kappa_4^2 + 2 \, \kappa_1^2 \,
\kappa_{23} \,  \kappa_{24} \, \kappa_{34}  - \kappa_1^2 \, \kappa_3^2 \,\kappa_{24}^2 - \kappa_{12}^2 \, \kappa_3^2 \, \kappa_4^2  + \kappa_{12}^2 \, \kappa_{34}^2 \nn
&+& 2 \, \kappa_{12} \,  \kappa_{13}\, \left[\kappa_{23}  \, \kappa_4^2 - \kappa_{24}\,  \kappa_{34} \right]
-2 \kappa_{14} \, (\kappa_{12} \,  \kappa_{23} \, \kappa_{34}  - \kappa_{12}\,  \kappa_{24} \, \kappa_3^2 - \kappa_{13}  \, \kappa_2^2  \, \kappa_{34} + \kappa_{13} \, \kappa_{23} \, \kappa_{24}),  \nn
&+& \kappa_{13}^2 \,  \left[\kappa_{24}^2- \kappa_2^2\,  \kappa_4^2\right] + \kappa_{14}^2 \, \left(\kappa_{23}^2- \kappa_2^2 \, \kappa_3^2 \right) <0
\eea
In the limit that all final state masses are taken to vanish $k_i^2 \rightarrow 0$ and these conditions can be simultaneously solved to
give the phase space volume:
\subsubsection{Region 1}
\begin{equation}
\begin{array}{rcl}
0&\le \kappa_{12}\le&\frac{m_h^2}{2}\\
\\
0&\le \kappa_{34}\le&\frac{1}{2}(m_h - \sqrt{2} \kappa_{12})^2)\\
\\
0&\le \kappa_{13}\le&\frac{1}{4}\left(m_h^2-2 \kappa_{12}-2 \kappa_{34}\right)-\frac{m_h^2}{4}\beta( \kappa_{12},\kappa_{34},\frac{m_h^2}{2})\\
\\
\frac{1}{4}(m_h^2-2\kappa_{12}-4\kappa_{13}-2\kappa_{34})-\frac{m_h^2}{4}\beta(\kappa_{12},\kappa_{34},\frac{m_h^2}{2})&\le \kappa_{14}\le&\frac{1}{4}(m_h^2-2k_{12}-4\kappa_{13}-2\kappa_{34})+\frac{m_h^2}{4}\beta(k_{12},\kappa_{34},\frac{m_h^2}{2})\\
\\
\frac{1}{2(k_{13}+\kappa_{14})^2}\left[A-2\sqrt{B}\right]&\le \kappa_{23}\le&\frac{1}{2(\kappa_{13}+\kappa_{14})^2}\left[A+2\sqrt{B}\right]
\end{array}
\end{equation}

\subsubsection{Region 2}
\begin{equation}
\begin{array}{rcl}
0&\le \kappa_{12}\le&\frac{m_h^2}{2}\\
\\
0&\le \kappa_{34}\le&\frac{1}{2}(m_h - \sqrt{2} \kappa_{12})^2)\\
\\
\frac{1}{4}\left(m_h^2-2\kappa_{12}-2\kappa_{34}\right)-\frac{m_h^2}{4}\beta(\kappa_{12},\kappa_{34},\frac{m_h^2}{2})&\le \kappa_{13}\le&\frac{1}{4}\left(m_h^2-2\kappa_{12}-2\kappa_{34}\right)+\frac{m_h^2}{4}\beta(\kappa_{12},\kappa_{34},\frac{m_h^2}{2})\\
\\
0&\le \kappa_{14}\le&\frac{1}{4}\left(m_h^2-2\kappa_{12}-2\kappa_{34}-4 \kappa_{13}\right)+\frac{m_h^2}{4}\beta(\kappa_{12},\kappa_{34},\frac{m_h^2}{2})\\
\\
\frac{1}{2(\kappa_{13}+\kappa_{14})^2}\left[A-2\sqrt{B}\right]&\le \kappa_{23}\le&\frac{1}{2(\kappa_{13}+\kappa_{14})^2}\left[A+2\sqrt{B}\right]
\end{array}
\end{equation}
where,
\begin{equation}
\begin{array}{rcl}
\beta(a,b,c)&=&\sqrt{1-\frac{2(a+b)}{c}+\frac{(a-b)^2}{c^2}}\\
\\
A&=&\kappa_{13}(\kappa_{13}+\kappa_{14})\left[m_h^2-2(\kappa_{12}+\kappa_{13}+\kappa_{14})]-2\kappa_{34}[\kappa_{12}(\kappa_{13}-\kappa_{14})+\kappa_{13}(\kappa_{13}+\kappa_{14})\right]\\
\\
B&=&2\kappa_{12} \kappa_{13} \kappa_{14}\kappa_{34}\left[m_h^2 (\kappa_{13}+\kappa_{14})-2 (\kappa_{12}+\kappa_{13}+\kappa_{14})(\kappa_{13}+\kappa_{14}+\kappa_{34})\right]
\end{array}
\end{equation}

Retaining final state masses is numerically required when the double photon pole is present in some interference cases.
The conditions above can be directly imposed on a numerical integration over the $\kappa_{ij}$ variable set in this case, modifying the allowed phase space volume further.

\providecommand{\href}[2]{#2}\begingroup\raggedright\endgroup

\end{document}